\documentclass[aps,prd,superscriptaddress,showkeys,showpacs,a4paper]{revtex4}

\usepackage[english]{babel}
\usepackage{amsmath,bm}
\usepackage{amssymb}
\usepackage{enumitem}
\usepackage{textcomp}
\usepackage{graphicx}
\usepackage{dsfont}
\usepackage{color}
\usepackage{hyperref}
\usepackage{mathrsfs}
\usepackage[normalem]{ulem}

\usepackage{placeins}

\graphicspath{{Figures/}}

\begin{document}
	
	\title{Gradient expansion formalism for magnetogenesis in the kinetic coupling model}
	
	\author{O.~O.~Sobol}
	\email{oleksandr.sobol@epfl.ch}
	\affiliation{Institute of Physics, Laboratory for Particle Physics and Cosmology (LPPC), \'{E}cole Polytechnique F\'{e}d\'{e}rale de Lausanne (EPFL), CH-1015 Lausanne, Switzerland}
	\affiliation{Physics Faculty, Taras Shevchenko National University of Kyiv, 64/13, Volodymyrska Street, 01601 Kyiv, Ukraine}
	
	\author{A.~V.~Lysenko}
	\affiliation{Physics Faculty, Taras Shevchenko National University of Kyiv, 64/13, Volodymyrska Street, 01601 Kyiv, Ukraine}
	
	\author{E.~V.~Gorbar}
	\affiliation{Physics Faculty, Taras Shevchenko National University of Kyiv, 64/13, Volodymyrska Street, 01601 Kyiv, Ukraine}
	\affiliation{Bogolyubov Institute for Theoretical Physics, 14-b, Metrologichna Street, 03680 Kyiv, Ukraine}
	
	\author{S.~I.~Vilchinskii}
	\affiliation{Institute of Physics, Laboratory for Particle Physics and Cosmology (LPPC), \'{E}cole Polytechnique F\'{e}d\'{e}rale de Lausanne (EPFL), CH-1015 Lausanne, Switzerland}
	\affiliation{Physics Faculty, Taras Shevchenko National University of Kyiv, 64/13, Volodymyrska Street, 01601 Kyiv, Ukraine}
	
	\date{\today}
	
	\begin{abstract}
		In order to describe magnetogenesis during inflation in the kinetic coupling model, we utilize a gradient expansion which is based on the fact that only long-wavelength (superhorizon) modes undergo amplification. For this purpose, we introduce a set of functions (bilinear combinations of electromagnetic fields with an arbitrary number of spatial curls) satisfying an infinite chain of equations. Apart from the usual mode enhancement due to interaction with the inflaton, these equations also take into account the fact that the number of relevant modes constantly grows during inflation. Truncating this chain, we show that even with a relatively small number of equations, it is possible to describe the electric and magnetic energy densities with a few percent accuracy during the whole inflation stage. We arrive at this conclusion for different types of coupling functions (increasing, decreasing, and nonmonotonic) in the regime with strong backreaction and its absence.
	\end{abstract}

\maketitle

\section{Introduction}
\label{sec-intro}

Analysis of the gamma-ray observations from distant blazars indicates the presence of magnetic fields in intergalactic voids of our Universe~\cite{Tavecchio:2010,Ando:2010,Neronov:2010,Dolag:2010,Dermer:2011,Taylor:2011,Caprini:2015}. The very large coherence length of these fields, measured in megaparsecs, strongly favors their cosmological nature (see Refs.~\cite{Kronberg:1994,Grasso:2001,Widrow:2002,Giovannini:2004,Kandus:2011,Durrer:2013,Subramanian:2016} for a review of different magnetogenesis models). As is well known, one of the most important successes of inflationary models is the explanation of the cosmic microwave background (CMB) power spectra and the observed large-scale structure of the Universe which arose as a result of the evolution of primordial quantum perturbations generated during inflation~\cite{Harrison:1970, Zeldovich:1972, Chibisov:1982} (for a review, see Refs.~\cite{Mukhanov:1992,Durrer:book}). It is natural to assume that primordial magnetic fields were also formed at that epoch. In such a case, primordial magnetic fields are very important observables, because they, in addition to the abundance of chemical elements, the CMB radiation, and the large-scale structure of galaxies and their clusters, can be a unique source of information about physical processes in the early Universe at high energies, which are unattainable in the laboratory.

One of the key requirements necessary for generating magnetic fields during inflation is the violation of conformal invariance in the electromagnetic sector~\cite{Parker:1968}. This can be done through the coupling of the electromagnetic field to the inflaton field or to the gravitational sector~\cite{Turner:1988,Ratra:1992,Garretson:1992,Dolgov:1993}. One of the most popular models in the literature is the kinetic coupling model, where the inflaton couples to the square of the electromagnetic field tensor. It was first proposed by Ratra~\cite{Ratra:1992} and revisited many times in the literature~\cite{Giovannini:2001,Bamba:2004,Martin:2008,Demozzi:2009,Kanno:2009,Fujita:2012,Fujita:2013,Ferreira:2013,Ferreira:2014,Fujita:2016,Vilchinskii:2017,Sharma:2017b,Shtanov:2018,Sobol:2018,Shtanov:2020,Talebian:2020}.

In the standard approach to the kinetic coupling model, one works with separate Fourier modes of the electromagnetic field which evolve in a given background of the inflaton field and the scale factor (see, e.g., Refs.~\cite{Martin:2008,Demozzi:2009}). This approximation is valid only in the weak field regime and is based on the two facts: (i)~the equation of motion for the mode function is a linear homogeneous equation [see Eq.~(\ref{eq-mode-physical}) below], and (ii)~the kinetic coupling to the inflaton does not mix different modes. Sufficiently strong generated electromagnetic fields cause the backreaction on the Universe expansion and the inflaton evolution \cite{Kanno:2009,Sobol:2018}. This in turn affects the evolution of all electromagnetic modes. Thus, the system enters a strongly nonlinear regime. Moreover, some other strong-field effects may become relevant---e.g., the Schwinger pair production, which also makes all electromagnetic modes coupled to each other~\cite{Kobayashi:2014}. In these cases, one has to consider all relevant modes simultaneously. This means that one should solve the system of all mode equations together with the background equations that is numerically very costly \cite{Fujita:2015,Notari:2016}.
Alternatively, one may use a lattice simulation. However, during inflation the spacetime is expanding exponentially and one should take the exponentially small lattice spacing at the beginning in order to get satisfactory accuracy at the end. That is why the lattice simulations are successfully used to describe only the last few $e$-foldings of inflation and preheating (see, e.g., Ref.~\cite{Cuissa:2018} for the case of axial coupling). In the present paper, we utilize a different approach based on the gradient expansion formalism. We work with vacuum expectation values of an infinite set of bilinear electromagnetic functions with an arbitrary number of spatial curls. Thus, the coordinate dependence of electromagnetic fields is traded for an infinite set of spatially homogeneous quantities. The same method was previously applied in the axial coupling model by three of us in Ref.~\cite{Sobol:2019}. Its extension to the case of the kinetic coupling model provides the main motivation for the present paper.

This paper is organized as follows: In Sec.~\ref{sec-basics}, we formulate the problem and apply the gradient expansion in order to derive an infinite set of equations for the bilinear electromagnetic quantities. Further, in Sec.~\ref{sec-mode}, we consider the evolution of the electromagnetic mode function and separate the physically relevant modes which contribute to the generated electromagnetic field. Then, in Sec.~\ref{sec-boundary}, we derive the boundary terms which take into account the variable number of physically relevant modes. Finally, in Sec.~\ref{sec-numerical}, we discuss our truncation scheme for the infinite chain of equations and then provide the numerical results obtained from such a truncated system. Section~\ref{sec-concl} is devoted to conclusions. Appendixes~\ref{app-envelope} and \ref{app-cylindric} provide some supplemental mathematical material which is used in the main text for the derivation of the boundary terms.
Throughout the work, we use the natural units and set $\hbar=c=1$.

\section{Gradient expansion formalism}
\label{sec-basics}

The action describing the electromagnetic field $A_{\mu}$ kinetically coupled to the inflaton field $\phi$ reads as
\begin{equation}
\label{action}
S=\int d^{4}x \sqrt{-g}\left[\frac{1}{2}g^{\mu\nu}\partial_{\mu}\phi\partial_{\nu}\phi-V(\phi)-\frac{1}{4}f^{2}(\phi)g^{\mu\alpha}g^{\nu\beta}F_{\mu\nu}F_{\alpha\beta}\right],
\end{equation}
where $g={\rm det}\,g_{\mu\nu}$, $g_{\mu\nu}$ is the metric tensor, $V(\phi)$ is the effective potential of the inflaton, $f(\phi)$ is the kinetic coupling function, and $F_{\mu\nu}=\partial_{\mu}A_{\nu}-\partial_{\nu}A_{\mu}$ is the electromagnetic field tensor.

The equation of motion for the inflaton field has the form
\begin{equation}
\label{KGF}
\frac{1}{\sqrt{-g}}\partial_{\mu}\left[\sqrt{-g}g^{\mu\nu}\partial_{\nu}\phi\right]+\frac{dV}{d\phi}=-\frac{1}{2}f\frac{df}{d\phi}F_{\mu\nu}F^{\mu\nu},
\end{equation}
while the Maxwell equations read as
\begin{equation}
\label{Maxwell}
\frac{1}{\sqrt{-g}}\partial_{\mu}\left[\sqrt{-g}g^{\mu\alpha}g^{\nu\beta}f^{2}(\phi)F_{\alpha\beta}\right]=0.
\end{equation}
They are supplemented by the Bianchi identities
\begin{equation}
\frac{1}{\sqrt{-g}}\partial_{\mu}\left[\sqrt{-g}\tilde{F}^{\mu\nu}\right]=0,
\end{equation}
where $\tilde{F}^{\mu\nu}=\frac{1}{2}\eta^{\mu\nu\lambda\rho}F_{\lambda\rho}$ is the dual electromagnetic tensor, $\eta^{\mu\nu\lambda\rho}=\varepsilon^{\mu\nu\lambda\rho}/\sqrt{-g}$, and $\varepsilon^{\mu\nu\lambda\rho}$ is the totally antisymmetric Levi-Civita symbol with $\varepsilon^{0123}=+1$.

Throughout this paper, we work in a spatially flat Friedmann-Lema\^{i}tre-Robertson-Walker (FLRW) universe whose metric is expressed in terms of the cosmic time as follows:
\begin{equation}
\label{metric}
g_{\mu\nu}={\rm diag}\,(1,\,-a^{2},\,-a^{2},\,-a^{2}), \quad \sqrt{-g}=a^{3}.
\end{equation}
Here $a(t)$ is the scale factor which describes the Universe expansion.

It is convenient to use the Coulomb gauge for the electromagnetic field $A_{\mu}=(0,\,\mathbf{A})$, where ${\rm div\,}\mathbf{A}=0$. Then, electric 
and magnetic fields are defined as follows:
\begin{equation}
\label{fields-E-and-B}
\mathbf{E}\equiv -\frac{1}{a}\dot{\mathbf{A}}, \qquad \mathbf{B}\equiv \frac{1}{a^{2}} {\rm rot\,}\mathbf{A},
\end{equation}
where the overdot denotes a derivative with respect to the cosmic time.  The components of the electromagnetic tensors are expressed through the 
electric and magnetic fields as usual:
\begin{equation}
\label{EM-tensors}
F^{0i}=\frac{1}{a} E^{i},\quad F_{ij}=a^{2}\varepsilon_{ijk}B^{k},
\quad \tilde{F}^{0i}=\frac{1}{a}B^{i}, \quad \tilde{F}_{ij}=-a^{2}\varepsilon_{ijk}E^{k},
\end{equation}
where $\varepsilon_{ijk}$ is the three-dimensional Levi-Civita symbol and indices $i,\,j,\,k$ on the right-hand side denote the components of
3-vectors. We would like to note that the electric and magnetic fields defined in Eqs.~(\ref{fields-E-and-B}) and (\ref{EM-tensors}) are physical fields measured by a comoving observer.

Then, the Maxwell equations for the electric and magnetic fields take the form
\begin{eqnarray}
\label{Maxwell-E}
&&\dot{\mathbf{E}}+2H\mathbf{E}+2\frac{\dot{f}}{f}\mathbf{E}-\frac{1}{a}{\rm rot\,}\mathbf{B}=0,
\\
\label{Maxwell-B}
&&\dot{\mathbf{B}}+2H\mathbf{B}+\frac{1}{a}{\rm rot\,}\mathbf{E}=0,\\
&&{\rm div\,}\mathbf{B}=0,\quad {\rm div\,}\mathbf{E}=0.
\label{constraints}
\end{eqnarray}

It is conventional to assume that the inflaton field is spatially homogeneous. In this case, we have
\begin{equation}
\label{KGF-2}
\ddot{\phi}+3H\dot{\phi}+\frac{dV}{d\phi}=f(\phi)\frac{df(\phi)}{d\phi}(\mathbf{E}^{2}-\mathbf{B}^{2}).
\end{equation}

The evolution of the Universe is driven by the total energy density of all fields, which can be calculated as the $00$-component of the stress-energy tensor. It is defined as usual:
\begin{equation}
T_{\mu\nu}=\frac{1}{\sqrt{-g}}\frac{\delta S}{\delta g^{\mu\nu}}=\partial_{\mu}\phi\partial_{\nu}\phi-f^{2}(\phi)g^{\alpha\beta}F_{\mu\alpha}F_{\nu\beta}-g_{\mu\nu} \left[\frac{1}{2}\partial_{\alpha}\phi\partial^{\alpha}\phi-V(\phi)-\frac{1}{4}f^{2}(\phi)F_{\alpha\beta}F^{\alpha\beta}\right].
\end{equation}

For a spatially homogeneous inflaton field, the energy density reads as
\begin{equation}
\label{energy-density}
\rho=\left[\frac{1}{2}\dot{\phi}^{2}+V(\phi)\right]+\frac{1}{2}f^{2}(\mathbf{E}^{2}+\mathbf{B}^{2})=\rho_{\rm inf}+\rho_{\rm EM}.
\end{equation}

Then, the Friedmann equation governing the evolution of the scale factor takes the form
\begin{eqnarray}
\label{Friedmann}
H^{2}=\left(\frac{\dot{a}}{a}\right)^{2}=\frac{1}{3M_{p}^{2}}\left(\rho_{\rm inf}+\rho_{\rm EM}\right).
\end{eqnarray}

Observable classical electromagnetic fields are created during inflation due to the enhancement of vacuum fluctuations of the quantum electromagnetic field. As we will see below, when the Universe expands quasiexponentially during inflation, the wavelength of electromagnetic modes becomes much larger than the observable region. Such long-wavelength modes can be treated classically. Nevertheless, their quantum nature implies that the corresponding electric and magnetic fields are chaotically oriented in different regions of the Universe. Therefore, it is not convenient to work with vector quantities such as $\mathbf{E}$ and $\mathbf{B}$. Instead, we introduce a set of bilinear functions of the electromagnetic field,
\begin{eqnarray}
\label{EE}
\mathscr{E}^{(n)}&\equiv&\frac{f^{2}(\phi)}{a^{n}}\langle \mathbf{E}\cdot {\rm rot}^{n}\mathbf{E}\rangle,\\
\label{EB}
\mathscr{G}^{(n)}&\equiv&-\frac{f^{2}(\phi)}{2a^{n}}\langle \mathbf{E}\cdot {\rm rot}^{n}\mathbf{B}+ {\rm rot}^{n}\mathbf{B}\cdot \mathbf{E}\rangle,\\
\label{BB}
\mathscr{B}^{(n)}&\equiv&\frac{f^{2}(\phi)}{a^{n}}\langle \mathbf{B}\cdot {\rm rot}^{n}\mathbf{B}\rangle,
\end{eqnarray}
where $\langle \cdots \rangle$ denotes the vacuum expectation value. Using Eqs.~(\ref{Maxwell-E}) and (\ref{Maxwell-B}), we derive the system of coupled equations for these quantities:
\begin{eqnarray}
\dot{\mathscr{E}}^{(n)}+(n+4)H\mathscr{E}^{(n)}+2\frac{\dot{f}}{f}\mathscr{E}^{(n)}+2\mathscr{G}^{(n+1)}&=&0,\label{eq-EE}\\
\dot{\mathscr{B}}^{(n)}+(n+4)H\mathscr{B}^{(n)}-2\frac{\dot{f}}{f}\mathscr{B}^{(n)}-2\mathscr{G}^{(n+1)}&=&0,\label{eq-BB}\\
\dot{\mathscr{G}}^{(n)}+(n+4)H\mathscr{G}^{(n)}+\mathscr{B}^{(n+1)}-\mathscr{E}^{(n+1)}&=&0.\label{eq-EB}
\end{eqnarray}
These equations describe the evolution of quadratic electromagnetic correlators [Eqs.~(\ref{EE})--(\ref{BB})] which contain the contributions from \textit{all} Fourier modes of the electromagnetic field. As we will see in Sec.~\ref{sec-boundary}, a naive computation of the vacuum expectation values gives the UV-divergent results for them. This is a usual situation in quantum field theory, because the quantum fluctuations of all possible modes are always present in physical vacuum, and their total energy is infinite. Therefore, in order to define physically relevant quantities, we should omit the modes corresponding to pure vacuum fluctuations. The separation of such modes is performed in Sec.~\ref{sec-mode}. The number of physically relevant modes changes in time during inflation, and this leads to an additional time dependence of electromagnetic quantities which can be described by means of boundary terms in the equations of motion. These boundary terms are derived in Sec.~\ref{sec-boundary}.

\section{Evolution of mode functions}
\label{sec-mode}

In this section, we consider the evolution of mode functions of the electromagnetic field. For this purpose, we decompose the electromagnetic field operator (in the Coulomb gauge) over the set of creation and annihilation operators:
\begin{equation}
\label{quant-operator}
\hat{\mathbf{A}}(t,\mathbf{x})=\int\frac{d^{3}\mathbf{k}}{(2\pi)^{3/2}f(t)}\!\!\sum_{\lambda=\pm}\left\{ \boldsymbol{\varepsilon}_{\lambda}(\mathbf{k})\hat{b}_{\lambda,\mathbf{k}}\mathcal{A}_{\lambda}(t,\mathbf{k})e^{i\mathbf{k}\cdot\mathbf{x}}+\boldsymbol{\varepsilon}^{*}_{\lambda}(\mathbf{k})\hat{b}^{\dagger}_{\lambda,\mathbf{k}}\mathcal{A}_{\lambda}^{*}(t,\mathbf{k})e^{-i\mathbf{k}\cdot\mathbf{x}}\right\},
\end{equation}
where $\mathcal{A}_{\lambda}(t,\mathbf{k})$ is the mode function with the wave vector (momentum) $\mathbf{k}$ and polarization $\lambda$; $\boldsymbol{\varepsilon}_{\lambda}(\mathbf{k})$ are two circular polarization 3-vectors, which have the following properties:
\begin{equation}
\mathbf{k}\cdot\boldsymbol{\varepsilon}_{\lambda}(\mathbf{k})=0,\quad \boldsymbol{\varepsilon}^{*}_{\lambda}(\mathbf{k})=\boldsymbol{\varepsilon}_{-\lambda}(\mathbf{k}), \quad [i\mathbf{k}\times\boldsymbol{\varepsilon}_{\lambda}(\mathbf{k})]=\lambda k \boldsymbol{\varepsilon}_{\lambda}(\mathbf{k}), \quad \boldsymbol{\varepsilon}^{*}_{\lambda}(\mathbf{k})\cdot\boldsymbol{\varepsilon}_{\lambda'}(\mathbf{k})=\delta_{\lambda\lambda'}. 
\end{equation}
The creation and annihilation operators have the standard commutation relations
\begin{equation}
[\hat{b}_{\lambda,\mathbf{k}},\,\hat{b}^{\dagger}_{\lambda',\mathbf{k}'}]=\delta_{\lambda\lambda'}\delta^{(3)}(\mathbf{k}-\mathbf{k}').
\end{equation}

Our goal now is to describe the evolution of the mode function $\mathcal{A}_{\lambda}(t,\mathbf{k})$. Substituting decomposition (\ref{quant-operator}) into Eq.~(\ref{Maxwell-E}), we find the following equation:
\begin{equation}
\label{eq-mode-physical}
\ddot{\mathcal{A}}_{\lambda}(t,\mathbf{k})+H\dot{\mathcal{A}}_{\lambda}(t,\mathbf{k})+\left[\frac{k^{2}}{a^{2}}-H\frac{\dot{f}}{f}-\frac{\ddot{f}}{f} \right]\mathcal{A}_{\lambda}(t,\mathbf{k})=0
\end{equation}
or in the conformal time $d\eta=dt/a(t)$,
\begin{equation}
\label{eq-mode-conformal}
\mathcal{A}_{\lambda}''(\eta,\mathbf{k})+\left[k^{2}-\frac{f''}{f}\right]\mathcal{A}_{\lambda}(\eta,\mathbf{k})=0.
\end{equation}
Here, primes denote derivatives with respect to the conformal time. Initially, both polarizations are present equally in vacuum state. Moreover, from Eqs.~(\ref{eq-mode-physical}) or (\ref{eq-mode-conformal}), we conclude that mode functions of different polarizations satisfy the same equations of motion. Consequently, the mode functions of both polarizations would be identical at any moment of time. Therefore, in what follows, we will omit the polarization index $\lambda$ in the mode function.

There are two completely different terms in the square brackets in Eq.~(\ref{eq-mode-conformal}), and depending on their relative magnitude, two completely different regimes of evolution can take place. The transition between these two regimes occurs when both terms are equally important. By analogy with the description of primordial perturbations, we will refer to this moment as the \textit{horizon crossing}.

We assume that the mode $k$ crosses (exits) the horizon (not to be confused with the Hubble horizon) if the equality
\begin{equation}
\label{horizon-crossing}
k^{2}=\left|\frac{f''(\eta)}{f}\right|=a^{2}H^{2}\left|\frac{\dot{f}}{fH}+\frac{\ddot{f}}{f H^{2}}\right|
\end{equation}
is achieved \textit{for the first time} during inflation. For a general coupling function $f$, the right-hand side of Eq.~(\ref{horizon-crossing}) may be a nonmonotonic function of time---that is why it is important to take only the first moment of time at which Eq.~(\ref{horizon-crossing}) is satisfied. 

The wave number of the last mode which has crossed the horizon before a certain moment of time $t$ is determined by the maximal value of the right-hand side of Eq.~(\ref{horizon-crossing}) in the interval $[0,\ t]$. This can be represented as follows:
\begin{equation}
\label{horizon-crossing-mode}
k_{h}(t)\equiv {\rm Env}\left\{a(t)H(t)\sqrt{|f_{1}(t)+f_{2}(t)|}\right\},
\end{equation}
where $f_{1,2}$ are the first two functions from the set
\begin{equation}
f_{n}\equiv \frac{1}{f H^{n}}\frac{d^{n}}{dt^{n}}f, \qquad n\in \mathds{N},
\end{equation}
and 
\begin{equation}
{\rm Env}\{F(t)\}\equiv \underset{t'\in[0,\,t]}{{\rm max}}\,F(t')
\end{equation}
stands for the \textit{upper monotonic envelope of a function} $F(t)$. Its definition, properties, and examples are given in Appendix~\ref{app-envelope}.

For modes deep inside the horizon, $k^{2}\gg k_{h}^{2}$, the Bunch-Davies boundary condition~\cite{Bunch:1978} is imposed:
\begin{equation}
\label{init-BD}
\mathcal{A}(\eta,\mathbf{k})=\frac{1}{\sqrt{2k}}e^{-i k\eta}, \quad -k\eta\to \infty.
\end{equation}
This solution is realized in the absence of the kinetic coupling and describes vacuum fluctuations of the electromagnetic field. The influence of the coupling function becomes important only after the mode exits the horizon.

Our aim is to determine the mode function at the moment of horizon crossing. For this purpose, we will use an approach similar to the Wentzel-Kramers-Brillouin method near the classical turning point. Namely, we will approximate the exact time dependence of the second term in brackets in Eq.~(\ref{eq-mode-conformal}) by some simple ansatz which is accurate near the moment of time when a given mode crosses the horizon and admits an exact analytical solution to Eq.~(\ref{eq-mode-conformal}). Unknown integration constants will be fixed by matching this solution with the Bunch-Davies mode function [Eq.~(\ref{init-BD})] deep inside the horizon.

The essence of our approximation is the following: In general, $f_{n}$'s are smooth functions and vary much slower than the scale factor during inflation. Indeed, they depend only on the inflaton, its time derivatives, and the Hubble parameter, which are changing very slowly in the slow-roll regime. In the vicinity of the horizon crossing, we exploit this fact and replace $f_{1,2}(t)$ with their (constant) values at the moment of horizon crossing, $f_{1,2}(t_{h})$. Moreover, since the conformal time can be approximately expressed as $\eta=-1/(aH)$ during inflation, Eq.~(\ref{eq-mode-conformal}) can be written as
\begin{equation}
\label{eq-mode-conformal-horizon-crossing}
\mathcal{A}''(\eta,\mathbf{k})+\left[k^{2}-\frac{f_{1}(t_{h})+f_{2}(t_{h})}{\eta^{2}}\right]\mathcal{A}(\eta,\mathbf{k})=0.
\end{equation}
This equation is accurate in the vicinity of the horizon crossing and admits an exact analytical solution, which is expressed in terms of cylindric functions as follows (see Appendix~\ref{app-cylindric} for details):
\begin{equation}
\label{solution-general-cylindric}
\mathcal{A}(\eta,\mathbf{k})=\sqrt{-k\eta}\left[D_{1}J_{\alpha}(-k\eta)+D_{2}Y_{\alpha}(-k\eta)\right].
\end{equation}
Here $J_{\alpha}$ and $Y_{\alpha}$ are the Bessel and Neumann functions, respectively, and the index $\alpha$ equals
\begin{equation}
\alpha=\sqrt{\frac{1}{4}+f_{1}(t_{h})+f_{2}(t_{h})}.
\end{equation}
In order to determine $D_{1,2}$, we have to match the solution in Eq.~(\ref{solution-general-cylindric}) with the Bunch-Davies boundary condition. Using the properties of the cylindric functions listed in Appendix~\ref{app-cylindric}, we finally get the solution valid in the vicinity of the horizon crossing:
\begin{equation}
\label{solution-horizon-crossing}
\mathcal{A}(\eta,\mathbf{k})=\frac{e^{i\left(\frac{\alpha \pi}{2}+\frac{\pi}{4}\right)}}{\sqrt{2k}}\sqrt{\frac{\pi}{2}(-k\eta)}H^{(1)}_{\alpha}(-k\eta),
\end{equation}
where $H^{(1)}_{\alpha}=J_{\alpha}+iY_{\alpha}$ is the Hankel function of the first kind. For further convenience, let us find also the time derivative of this solution:
\begin{eqnarray}
\mathcal{D}(\eta,\mathbf{k})\!\!\!\!&\equiv&\!\!\!\frac{f}{k}\frac{\partial}{\partial\eta}\frac{\mathcal{A}(\eta,\mathbf{k})}{f}=\nonumber\\
\!\!\!\!&=&\!\!\!\frac{e^{i\left(\frac{\alpha \pi}{2}+\frac{\pi}{4}\right)}}{2\sqrt{2k}}\sqrt{\frac{\pi}{2}(-k\eta)}\left\{\left(1-\frac{\frac{1}{2}+f_{1}(\eta)}{\alpha}\right)H^{(1)}_{\alpha+1}(-k\eta)-\left(1+\frac{\frac{1}{2}+f_{1}(\eta)}{\alpha}\right)H^{(1)}_{\alpha-1}(-k\eta)\right\},\label{D-function-horizon-crossing}
\end{eqnarray}
where we have used properties of the Hankel function considered in Appendix~\ref{app-cylindric}.

Finally, far outside the horizon $k\ll k_{h}(t)$, the term $k^2/a^2$ in Eq.~(\ref{eq-mode-physical}) can be neglected, and the latter can be rewritten in the form
\begin{equation}
\frac{d}{dt}\left[a(t)f^{2}(t)\frac{d}{dt}\left(\frac{\mathcal{A}}{f}\right)\right]=0.
\end{equation}
Its general solution is
\begin{equation}
\label{outside-horizon}
\mathcal{A}(t,\mathbf{k})=C_{1}(k)f(t)+C_{2}(k)f(t)\int_{t_{h}}^{t}\frac{dt'}{a(t')f^{2}(t')}, \quad \mathcal{D}(t,\mathbf{k})=\frac{C_{2}(k)}{k f(t)},
\end{equation}
where $C_{1,2}(k)$ can be determined by matching the solution (\ref{outside-horizon}) with Eqs.~(\ref{solution-horizon-crossing}) and (\ref{D-function-horizon-crossing}) valid at the horizon crossing. The resulting expressions are the following:
\begin{eqnarray}
C_{1}(k)&=&\frac{e^{i\left(\frac{\alpha \pi}{2}+\frac{\pi}{4}\right)}}{\sqrt{2k}f(t_{h})}\sqrt{\frac{\pi}{2}\frac{k}{a(t_{h})H(t_{h})}}H^{(1)}_{\alpha}\bigg(\frac{k}{a(t_{h})H(t_{h})}\bigg),\\
C_{2}(k)&=&\frac{k f(t_{h})}{2\sqrt{2k}}e^{i\left(\frac{\alpha \pi}{2}+\frac{\pi}{4}\right)}\sqrt{\frac{\pi}{2}\frac{k}{a(t_{h})H(t_{h})}}\times\nonumber\\
&\times&\left\{\left(1-\frac{1/2+f_{1}}{\alpha}\right)H^{(1)}_{\alpha+1}\bigg(\frac{k}{a(t_{h})H(t_{h})}\bigg)-\left(1+\frac{1/2+f_{1}}{\alpha}\right)H^{(1)}_{\alpha-1}\bigg(\frac{k}{a(t_{h})H(t_{h})}\bigg)\right\}.
\end{eqnarray}
Here $t_{h}=t_{h}(k)$ is the moment of time when the mode with wave number $k$ crosses the horizon; the values of $f_{1,2}$ and $\alpha$ should be determined also at this moment of time.

\begin{figure}[ht!]
	\centering
	\includegraphics[width=0.4\textwidth]{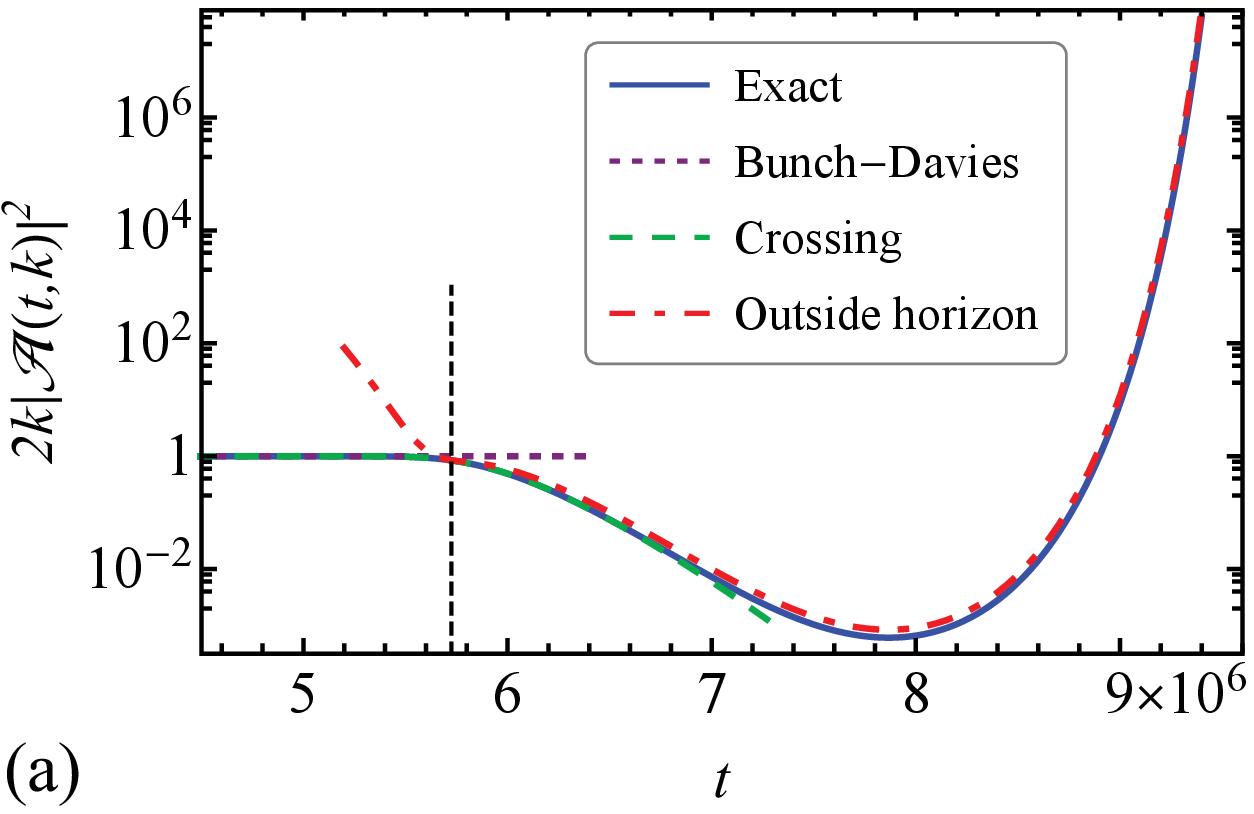}\hspace*{0.5cm}
	\includegraphics[width=0.4\textwidth]{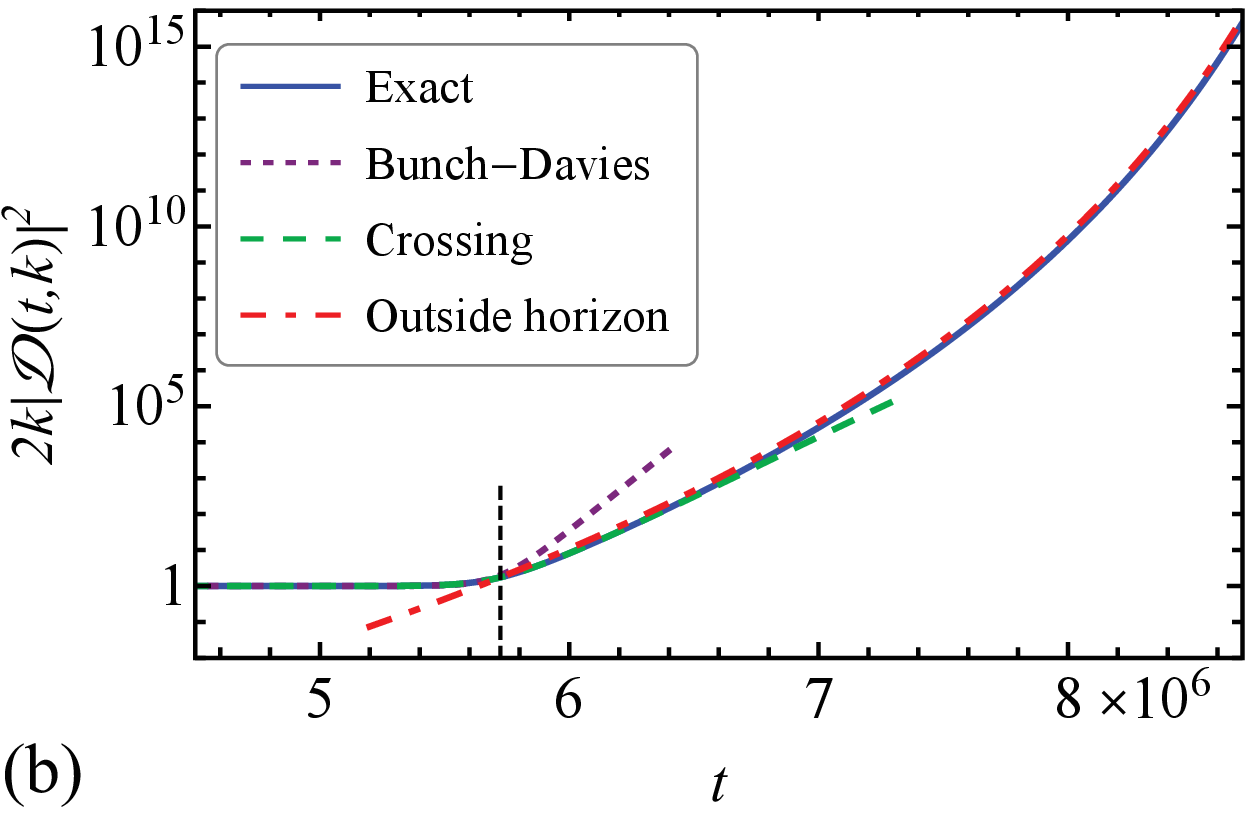}
	\caption{Time evolution of mode functions with wave number $k=10^{10}M_{p}$ in the kinetic coupling model with the Ratra coupling function $f(\phi)=\exp(\beta\phi/M_{p})$ for the coupling parameter $\beta=10$. (a)~Time dependence of the square modulus of the mode function $\mathcal{A}(t,\mathbf{k})$ multiplied by $2k$: the exact numerical solution of Eq.~(\ref{eq-mode-physical}) (blue solid line), the Bunch-Davies solution (\ref{init-BD}) (purple dotted line), the approximate solution (\ref{solution-horizon-crossing}) valid at the horizon crossing (green dashed line), and the solution (\ref{outside-horizon}) valid outside the horizon (red dash-dotted line). (b)~Time dependence of the square modulus of the auxiliary mode function $\mathcal{D}(t,\mathbf{k})$ multiplied by $2k$ [with the same notations as in panel (a)].}
	\label{fig-mode-funct}
\end{figure}

In order to illustrate our results, we consider the time evolution of the mode with $k=10^{10}M_{p}$ in the kinetic coupling model with the Ratra coupling function $f(\phi)=\exp(\beta\phi/M_{p})$ for $\beta=10$. In Fig.~\ref{fig-mode-funct}, we plot the square moduli of $\mathcal{A}(t,\mathbf{k})$ and $\mathcal{D}(t,\mathbf{k})$  multiplied by $2k$ in panels (a) and (b), respectively. We compare the exact solution of Eq.~(\ref{eq-mode-physical}) determined numerically (blue solid lines) with three different approximate solutions derived in this section. Purple dotted lines correspond to the Bunch-Davies solution [Eq.~(\ref{init-BD})]. As expected, they are in accordance with the exact solution only before the horizon crossing (shown by the vertical black dashed line). The approximate solutions (\ref{solution-horizon-crossing}) and (\ref{D-function-horizon-crossing}) are shown by green dashed lines. They nicely fit the exact solution in the vicinity of the horizon crossing from both sides (inside as well as outside the horizon). Finally, the approximate solution (\ref{outside-horizon}) is plotted by red dash-dotted lines and works well outside the horizon.

Thus, in this section we have determined the time evolution of the mode function and found relatively simple and general approximate analytic solutions valid at any stage of the mode evolution. We have shown that only the modes which cross the horizon during inflation are influenced by the kinetic coupling and contain the information about generated electromagnetic fields. On the other hand, the modes lying inside the horizon correspond to pure vacuum fluctuations and must be excluded from the analysis.

The fact that we take into account only a finite range of modes, and that this range changes during inflation, leads to an additional time dependence of the quadratic electromagnetic correlators [Eqs.~(\ref{EE})--(\ref{BB})] which is not encountered in the system of Eqs.~(\ref{eq-EE})--(\ref{eq-EB}). In order to fix this inconsistency, we would have to introduce some additional terms in the evolution equations, known as the boundary terms. They will be derived in the next section.

\section{Boundary terms}
\label{sec-boundary}

As we have shown in the previous section, the physically relevant for magnetogenesis electromagnetic modes are those which cross the horizon during inflation. Moreover, any electromagnetic observable at the moment of time $t$ is determined only by modes which crossed the horizon before this moment. This means that the number of modes contributing to physical observables changes during inflation. In order to take this into account, we need to determine spectral decompositions for all quadratic electromagnetic correlators [Eqs.~(\ref{EE})--(\ref{BB})]. 

Using the explicit expression for the field operator [Eq.~(\ref{quant-operator})] and substituting it into Eqs.~(\ref{EE})--(\ref{BB}), we obtain the following decompositions of the vacuum expectation values:
\begin{eqnarray}
\mathscr{E}^{(2n)}&=&\int\frac{dk}{k} \frac{k^{2n+3}}{\pi^{2}a^{2n+2}}f^{2}(t)\left|\frac{d}{dt}\left(\frac{\mathcal{A}(t,\mathbf{k})}{f(t)}\right)\right|^{2},\qquad \mathscr{E}^{(2n+1)}=0,\label{sp-EE}\\
\mathscr{G}^{(2n+1)}&=&\int\frac{dk}{k} \frac{k^{2n+5}}{2\pi^{2}a^{2n+4}}f^{2}(t)\frac{d}{dt}\left|\frac{\mathcal{A}(t,\mathbf{k})}{f(t)}\right|^{2},\qquad \mathscr{G}^{(2n)}=0,\label{sp-EB}\\
\mathscr{B}^{(2n)}&=&\int\frac{dk}{k} \frac{k^{2n+5}}{\pi^{2}a^{2n+4}}\left|\mathcal{A}(t,\mathbf{k})\right|^{2},\qquad \mathscr{B}^{(2n+1)}=0.\label{sp-BB}
\end{eqnarray}
Formally, the integration is performed over all possible wave numbers from 0 to infinity. Let us check that taking into account all subhorizon modes leads to divergent results. 
Deep inside the horizon, mode functions are described by the Bunch-Davies solution (\ref{init-BD}), which is valid during the whole inflation stage for modes with extremely large wave numbers $k_{h}^{\rm max}\ll k<+\infty$.
For these modes, the integrands in Eqs.~(\ref{sp-EE})--(\ref{sp-BB}) are given by
\begin{equation}
\frac{d\mathscr{E}^{(2n)}}{d\ln\,k}=\frac{k^{2n+2}}{2\pi^{2}a^{2n+2}}\left[\frac{k^{2}}{a^{2}}+\frac{\dot{f}^{2}}{f^{2}}\right],\quad
\frac{d\mathscr{B}^{(2n)}}{d\ln\,k}=\frac{k^{2n+4}}{2\pi^{2}a^{2n+4}},\quad
\frac{d\mathscr{G}^{(2n+1)}}{d\ln\,k}=-\frac{k^{2n+4}}{2\pi^{2}a^{2n+4}}\frac{\dot{f}}{f}.
\end{equation}
The UV asymptotical behavior of all three quantities is $\sim k^{2n+4}$, so that all integrals are severely UV divergent. However, as we discussed in the previous section, mode functions with large momenta $\mathcal{A}(t,k\gg k_{h})$ are not affected by the presence of the kinetic coupling function $f$ and do not undergo the quantum-to-classical transition during inflation. Therefore, we exclude such modes from the consideration.

Thus, we define the physically meaningful quadratic electromagnetic quantities as those containing only the modes crossing the horizon during inflation, i.e.,
\begin{eqnarray}
\mathscr{E}^{(2n)}&=&\int_{0}^{k_{h}(t)}\frac{dk}{k} \frac{k^{2n+5}}{\pi^{2}a^{2n+4}}\left|\mathcal{D}(t,\mathbf{k})\right|^{2},\label{sp-EE-phys}\\
\mathscr{G}^{(2n+1)}&=&\int_{0}^{k_{h}(t)}\frac{dk}{k} \frac{k^{2n+6}}{\pi^{2}a^{2n+5}}\Re e\left\{
\mathcal{A}^{*}(t,\mathbf{k})\mathcal{D}(t,\mathbf{k})\right\},\label{sp-EB-phys}\\
\mathscr{B}^{(2n)}&=&\int_{0}^{k_{h}(t)}\frac{dk}{k} \frac{k^{2n+5}}{\pi^{2}a^{2n+4}}\left|\mathcal{A}(t,\mathbf{k})\right|^{2},\label{sp-BB-phys}
\end{eqnarray}
where we express the integrands in terms of the mode function $\mathcal{A}$ and the auxiliary function $\mathcal{D}$. Despite the fact that the integration is now performed only over a finite range of modes, we will keep the same notations $\mathscr{E}^{(2n)}$, $\mathscr{B}^{(2n)}$, $\mathscr{G}^{(2n+1)}$, since in the rest of the paper only the physical quantities will be considered. Another interesting question is whether the lower integration limit in Eqs.~(\ref{sp-EE-phys})--(\ref{sp-BB-phys}) is zero, or if it should acquire some finite value. The physical meaning of modes which are outside the horizon during the whole inflation stage is still under debate in the literature \cite{Demozzi:2009,Fujita:2012,Fujita:2013,Ferreira:2013,Fujita:2016}. Moreover, the duration of the inflation itself is unknown; there is only the minimal number of $e$-foldings needed to resolve some cosmological problems, typically $N_{\rm min}\simeq 60$. That is why the lower integration limit can be a wave number somewhere between 0 and $k_{h}(t_{60})$, where $t_{60}$ is the moment of time at 60 $e$-foldings before the end of inflation. Fortunately, for the majority of coupling functions, the spectrum of generated electromagnetic fields has a blue tilt (i.e., a positive spectral index), and the exact value of the lower integration boundary is not important for numerical computations. In the rest of the work, we will take the zero value of the lower integration limit.

Thus, any electromagnetic quantity $X$ can be written in the following form:
\begin{equation}
X=\int_{0}^{k_{h}(t)}\frac{dk}{k}\frac{dX}{d\ln k},
\end{equation}
i.e., as an integral with a variable upper boundary $k_{h}(t)$ which was defined by Eq.~(\ref{horizon-crossing-mode}). The time evolution of the quadratic quantities (\ref{sp-EE-phys})--(\ref{sp-BB-phys}) cannot be described by the system of equations~(\ref{eq-EE})--(\ref{eq-EB}) because the latter does not take into account the fact that the number of relevant modes changes in time. Therefore, we have to introduce additional terms to describe the modes which cross the horizon and start to contribute to electromagnetic function $X$:
\begin{equation}
\label{boundary-general}
\left(\dot{X}\right)_{b}\equiv \left.\frac{dX}{d\ln k}\right|_{k=k_{h}(t)}\!\!\!\!\!\!\times \frac{d\ln k_{h}}{dt},
\end{equation} 
where the lower index $b$ means ``boundary term''. In order to calculate the boundary terms, we need to know the mode function in the vicinity of the horizon crossing. In Sec.~\ref{sec-mode}, we have already determined the approximate analytical expressions which work well in this region; they are given by Eqs.~(\ref{solution-horizon-crossing})--(\ref{D-function-horizon-crossing}). Then, using Eq.~(\ref{boundary-general}) and spectral decompositions [Eqs.~(\ref{sp-EE-phys})--(\ref{sp-BB-phys})], we get the boundary terms for the quadratic functions in the following form:
\begin{eqnarray}
\left[\dot{\mathscr{E}}^{(2n)}\right]_{b}&=&\frac{1}{16\pi}\left(\frac{k_{h}}{a}\right)^{2n+5}\frac{1}{H}\frac{d\ln k_{h}}{dt}\left|e^{i\frac{\alpha \pi}{2}}\right|^{2}\nonumber\\
&\times&\left|\left(1-\frac{1/2+f_{1}}{\alpha}\right)H^{(1)}_{\alpha+1}\bigg(\frac{k_{h}}{aH}\bigg)-\left(1+\frac{1/2+f_{1}}{\alpha}\right)H^{(1)}_{\alpha-1}\bigg(\frac{k_{h}}{aH}\bigg) \right|^{2},\label{bound-EE}\\
\left[\dot{\mathscr{G}}^{(2n+1)}\right]_{b}&=& \frac{1}{8\pi}\left(\frac{k_{h}}{a}\right)^{2n+6}\frac{1}{H}\frac{d\ln k_{h}}{dt}\left|e^{i\frac{\alpha \pi}{2}}\right|^{2}\Re e\Bigg\{ H^{(1)*}_{\alpha}\bigg(\frac{k_{h}}{aH}\bigg)\nonumber\\
&\times& \left[\left(1-\frac{1/2+f_{1}}{\alpha}\right)H^{(1)}_{\alpha+1}\bigg(\frac{k_{h}}{aH}\bigg)-\left(1+\frac{1/2+f_{1}}{\alpha}\right)H^{(1)}_{\alpha-1}\bigg(\frac{k_{h}}{aH}\bigg) \right] \Bigg\}, \label{bound-EB}\\
\left[\dot{\mathscr{B}}^{(2n)}\right]_{b}&=& \frac{1}{4\pi}\left(\frac{k_{h}}{a}\right)^{2n+5}\frac{1}{H}\frac{d\ln k_{h}}{dt}\left|e^{i\frac{\alpha \pi}{2}}H^{(1)}_{\alpha}\bigg(\frac{k_{h}}{aH}\bigg)\right|^{2}.\label{bound-BB}
\end{eqnarray}
In the derivation of these boundary terms, we make two approximations, which may introduce numerical errors in the result: (i)~we neglect the time dependence of functions $f_{1,2}$ near the moment of horizon crossing by an electromagnetic mode, and (ii)~we approximate the Universe expansion close to that moment of time by the de Sitter solution. Both these assumptions work well during inflation for smooth enough coupling functions. However, close to the end of inflation their impact on the result may become significant. We will analyze errors of the numerical results in the next section for different coupling functions and show that typically errors do not exceed the few-percent level.

Finally, we get the following system of equations:
\begin{eqnarray}
&&H^{2}=\frac{1}{3M_{p}^{2}}\left[\frac{1}{2}\dot{\phi}^{2}+V(\phi)+\frac{1}{2}\left(\mathscr{E}^{(0)}+\mathscr{B}^{(0)}\right)\right],\label{Friedmann-fin}\\
&&\ddot{\phi}+3H\dot{\phi}+V'_{\phi}=\frac{f'_{\phi}}{f}\left(\mathscr{E}^{(0)}-\mathscr{B}^{(0)}\right),\label{KGF-fin}\\
&&\dot{\mathscr{E}}^{(2n)}+(2n+4)H\mathscr{E}^{(2n)}+2\frac{f'_{\phi}}{f}\dot{\phi}\,\mathscr{E}^{(2n)}+2\mathscr{G}^{(2n+1)}=\left[\dot{\mathscr{E}}^{(2n)}\right]_{b},\label{eq-EE-2}\\
&&\dot{\mathscr{B}}^{(2n)}+(2n+4)H\mathscr{B}^{(2n)}-2\frac{f'_{\phi}}{f}\dot{\phi}\,\mathscr{B}^{(2n)}-2\mathscr{G}^{(2n+1)}=\left[\dot{\mathscr{B}}^{(2n)}\right]_{b},\label{eq-BB-2}\\
&&\dot{\mathscr{G}}^{(2n+1)}+(2n+5)H\mathscr{G}^{(2n+1)}+\mathscr{B}^{(2n+2)}-\mathscr{E}^{(2n+2)}=\left[\dot{\mathscr{G}}^{(2n+1)}\right]_{b}.\label{eq-EB-2}
\end{eqnarray}
This system of equations is infinite, and in the following section, we will use some approximations in order to truncate it.

\section{Numerical results}
\label{sec-numerical}

In the previous sections, we derived an infinite chain of equations [Eqs.~(\ref{eq-EE-2})--(\ref{eq-EB-2})] for the set of bilinear functions of the electromagnetic field [Eqs.~(\ref{EE})--(\ref{BB})]. However, their practical usage is impossible unless we find some physically meaningful way to truncate it. In this section, we will show that such a truncation indeed exists during the inflation stage, and then we will present some numerical results obtained from the truncated system of equations for different types of coupling functions.

\subsection{Truncation procedure}

As we discussed in Sec.~\ref{sec-boundary}, at any given moment of time $t$ the shortest mode in the electromagnetic spectra is determined by the horizon size $k^{-1}_{h}(t)$. This fact will allow us to express the higher-order electromagnetic quantities $\mathscr{E}$, $\mathscr{G}$, $\mathscr{B}$ through the lower-order ones. Let us consider, for example, the magnetic function $\mathscr{B}^{(2n+2)}$. From Eq.~(\ref{sp-BB-phys}), we conclude that for a generic momentum dependence of the mode function $\mathcal{A}(t,\mathbf{k})$, there exists high enough order $n$ for which the contribution to $\mathscr{B}^{(2n+2)}$ is dominated by the ultraviolet part of the spectrum near $k_{h}$. Let us define the spectral index $p$ in this region by the relation $|\mathcal{A}(t,k\lesssim k_{h})|^{2}\propto k^{p-1}$ (e.g., $p=0$ for the Bunch-Davies solution). Then, we can find
\begin{equation}
\mathscr{B}^{(2n)}=\frac{\tilde{C}(t)}{2n+4+p} \left[\frac{k_{h}(t)}{a}\right]^{2n+4+p},
\end{equation}
where $\tilde{C}(t)$ does not depend on $n$. Therefore, we can approximate
\begin{equation}
\mathscr{B}^{(2n+2)}=\frac{2n+4+p}{2n+6+p}\left[\frac{k_{h}(t)}{a}\right]^{2}\mathscr{B}^{(2n)}\approx \left[\frac{k_{h}(t)}{a}\right]^{2}\mathscr{B}^{(2n)}.
\end{equation}
The last approximation becomes more and more accurate for higher $n$. However, we will use it even at low orders, because the spectral index $p$ is unknown in general. Similar approximations will be used for $\mathscr{E}$ and $\mathscr{G}$ functions:
\begin{equation}
\mathscr{E}^{(2n+2)}\approx \left[\frac{k_{h}(t)}{a}\right]^{2}\mathscr{E}^{(2n)}, \qquad \mathscr{G}^{(2n+3)}\approx \left[\frac{k_{h}(t)}{a}\right]^{2}\mathscr{G}^{(2n+1)}.
\end{equation}
These relations enable us to truncate the infinite chain at a certain finite order.

Let us denote by $\mathcal{N}$ the greatest order we would like to consider. For even $\mathcal{N}$, the system of equations terminates with Eqs.~(\ref{eq-EE-2}) and (\ref{eq-BB-2}), where the above mentioned approximation for $\mathscr{G}^{(\mathcal{N}+1)}$ was used:
\begin{eqnarray}
&&\dot{\mathscr{E}}^{(\mathcal{N})}+(\mathcal{N}+4)H\mathscr{E}^{(\mathcal{N})}+2\frac{f'_{\phi}}{f}\dot{\phi}\,\mathscr{E}^{(\mathcal{N})}+2\left[\frac{k_{h}(t)}{a}\right]^{2}\mathscr{G}^{(\mathcal{N}-1)}=\left[\dot{\mathscr{E}}^{(\mathcal{N})}\right]_{b},\label{eq-EE-2-approx}\\
&&\dot{\mathscr{B}}^{(\mathcal{N})}+(\mathcal{N}+4)H\mathscr{B}^{(\mathcal{N})}-2\frac{f'_{\phi}}{f}\dot{\phi}\,\mathscr{B}^{(\mathcal{N})}-2\left[\frac{k_{h}(t)}{a}\right]^{2}\mathscr{G}^{(\mathcal{N}-1)}=\left[\dot{\mathscr{B}}^{(\mathcal{N})}\right]_{b}.\label{eq-BB-2-approx}
\end{eqnarray}
For the lowest possible order $\mathcal{N}=0$, the system can be closed by setting $\mathscr{G}^{(1)}=0$. Thus, in total, the system consists of $(3\mathcal{N}+4)/2$ equations for the electromagnetic functions and 2 equations for the scale factor [Eq.~(\ref{Friedmann-fin})] and the inflaton field [Eq.~(\ref{KGF-fin})].

For odd $\mathcal{N}$, the last equation in the system is Eq.~(\ref{eq-EB-2}), which takes the form
\begin{equation}
\dot{\mathscr{G}}^{(\mathcal{N})}+(\mathcal{N}+4)H\mathscr{G}^{(\mathcal{N})}+\left[\frac{k_{h}(t)}{a}\right]^{2}\left\{\mathscr{B}^{(\mathcal{N}-1)}-\mathscr{E}^{(\mathcal{N}-1)}\right\}=\left[\dot{\mathscr{G}}^{(\mathcal{N})}\right]_{b}.\label{eq-EB-2-approx}
\end{equation}
In this case, the total number of equations is $(3\mathcal{N}+3)/2$ for the electromagnetic functions plus 2 background equations [Eqs.~(\ref{Friedmann-fin}) and (\ref{KGF-fin})].

In the following subsections, we provide numerical solutions of the truncated system of equations for different values of $\mathcal{N}$ from 0 to 15. In order to determine the accuracy of our approximation, we need to perform a direct and independent calculation of the electromagnetic quantities. This indeed can be done in the regime when the backreaction of generated fields can be neglected. In such a situation, background Eqs.~(\ref{Friedmann-fin}) and (\ref{KGF-fin}) can be solved by setting the electromagnetic field to zero. Then, having determined the time dependences of the inflaton and scale factor, we solve Eq.~(\ref{eq-mode-physical}) numerically for all modes crossing the horizon during inflation. Finally, integrating the spectral decompositions [Eqs.~(\ref{sp-EE-phys}), (\ref{sp-EB-phys}), and (\ref{sp-BB-phys})] over the relevant range of modes, we get the exact results for electromagnetic quantities, which can be used to estimate the accuracy of the approximate solutions of the truncated system of equations.

For definiteness, we consider the Starobinsky $R^{2}$ model of inflation~\cite{Starobinsky:1980} with the effective potential (in the Einstein frame)
\begin{equation}
\label{Starobinsky-potential}
V(\phi)=V_{0}\left[1-\exp\left(-\sqrt{\frac{2}{3}}\frac{\phi}{M_{p}}\right)\right]^{2}.
\end{equation}
Predictions of this model for the spectral index of the primordial scalar perturbations and the tensor-to-scalar power ratio are in good accordance with the recent cosmic microwave background observations~\cite{Planck:2018-infl}. The amplitude of the potential, $V_{0}$, is chosen to provide the correct value of the scalar power spectrum amplitude~\cite{Planck:2018-infl}. For simplicity, we assume that the CMB pivot scale exits the Hubble horizon 60 $e$-foldings before the end of inflation.

\subsection{The Ratra coupling function}

Here we present the numerical solutions of the truncated system of equations for the Ratra coupling function 
\begin{equation}
\label{Ratra-function}
f(\phi)=\exp(\beta\phi/M_{p}),
\end{equation}
where $\beta$ is a dimensionless coupling parameter. In order to test our formalism, we consider three different values of $\beta$ which correspond to different regimes: $\beta=-7$ gives a coupling function that increases in time and does not cause the backreaction of electromagnetic fields on the background evolution; $\beta=7$ corresponds to a decreasing coupling function again without the backreaction; and finally, $\beta=15$ gives a decreasing function leading to strong backreaction.

\subsubsection{Increasing coupling function without backreaction}

\begin{figure}[ht!]
	\centering
	\includegraphics[width=0.4\textwidth]{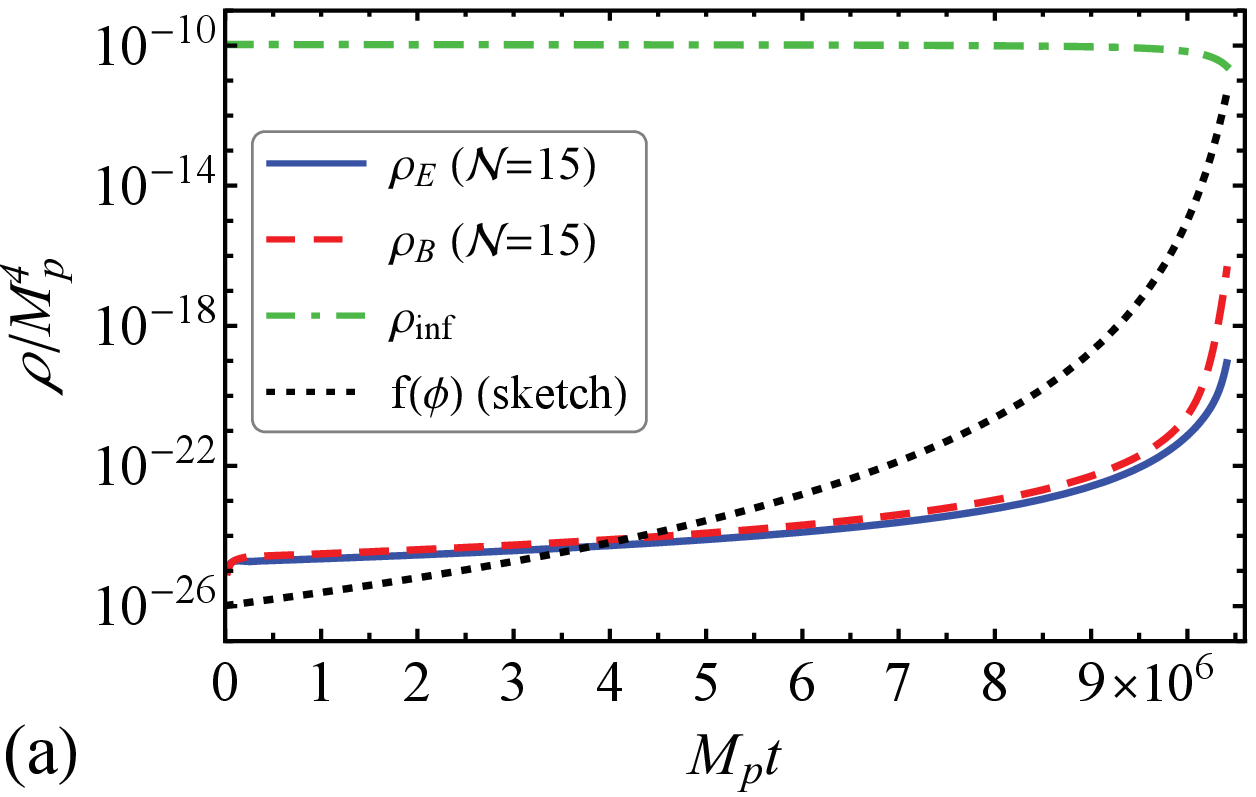}\hspace*{0.5cm}
	\includegraphics[width=0.4\textwidth]{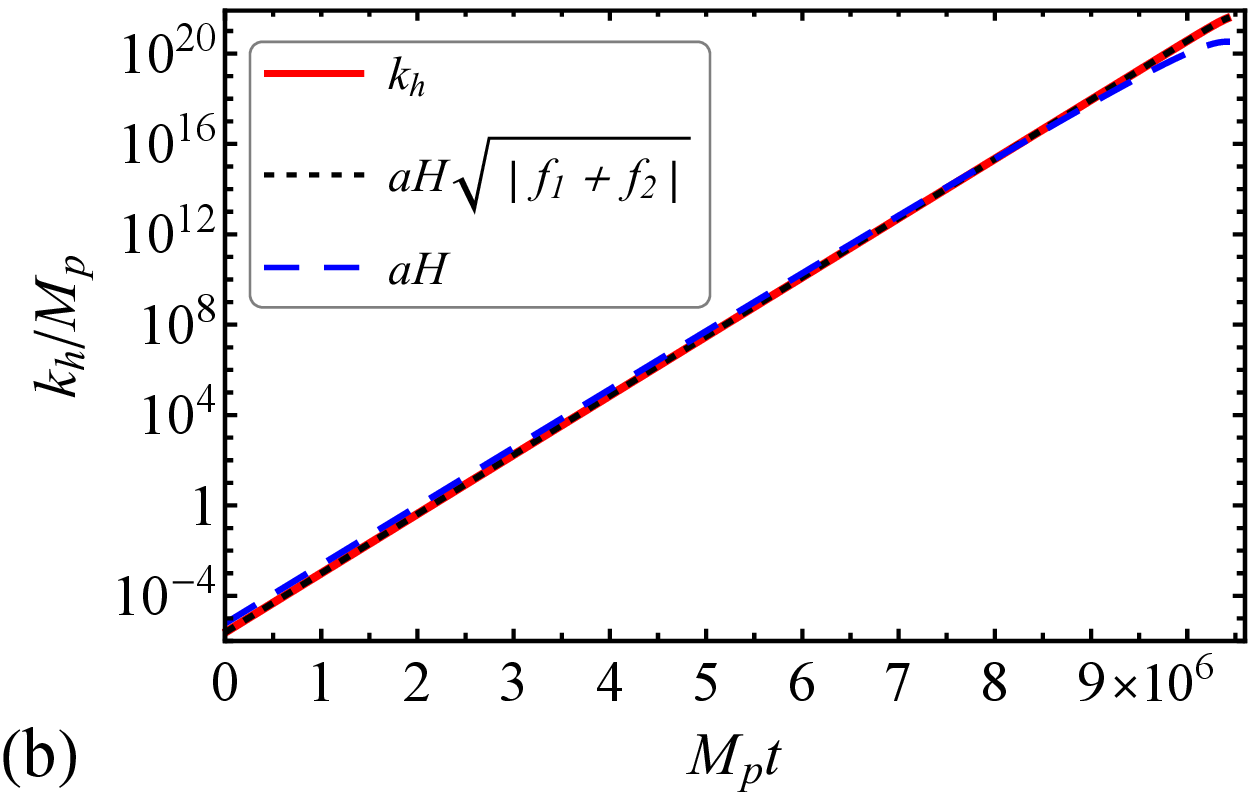}
	\caption{(a)~The time dependence of the electric (blue solid line) and magnetic (red dashed line) components of the energy density generated during inflation in the Ratra model with the coupling parameter $\beta=-7$. (The qualitative time dependence of the coupling function is sketched by the black dotted line.) The green dash-dotted line gives for comparison the time dependence of the inflaton energy density. For $\beta=-7$, the backreaction is irrelevant during inflation, and the magnetic component is always greater than the electric one. (b)~The time dependence of the horizon scale $k_{h}(t)$ (red solid line) and the Hubble horizon scale $aH$ (blue dashed line). The black dotted line shows the argument of the envelope in Eq.~(\ref{horizon-crossing-mode}). The results shown in both panels are obtained from the truncated system of equations with $\mathcal{N}=15$.}
	\label{fig-Ratra-m7-EnDens}
\end{figure}

We start with the case of a monotonically increasing coupling function with $\beta=-7$. Since this coupling function is always less than unity during inflation, the dynamics of charged fields becomes strongly coupled \cite{Demozzi:2009}. Nevertheless, we consider it in order to test the applicability of our approach to growing coupling functions. Figure~\ref{fig-Ratra-m7-EnDens} (a) shows the time dependences of the electric and magnetic energy densities obtained from the truncated system of equations with $\mathcal{N}=15$. For such a small value of the coupling parameter $\beta$, generated fields are weak; their energy density is several orders of magnitude less than that of the inflaton. The most effective generation occurs at the end of inflation, when the inflaton gets out of the slow-roll regime. Although the electric and magnetic components are of the same order of magnitude, the magnetic component is always slightly larger, because the coupling function increases with time.

The wave number of the mode crossing the horizon as a function of time $k_{h}(t)$ is shown by the red solid line in Fig.~\ref{fig-Ratra-m7-EnDens} (b). Since the argument of the envelope in Eq.~(\ref{horizon-crossing-mode}) is a monotonically increasing function (shown by the black dotted line), the envelope is identically equal to the function itself. For comparison, we also plot in the same figure the time dependence of the Hubble horizon scale $k_{H}=a(t)H(t)$ (blue dashed line). For this coupling function, the horizon for electromagnetic modes and the Hubble horizon appear to be very close. However, the biggest deviations occur close to the end of inflation, where the generation is the most effective. Therefore, it is important to use the correct definition of the horizon in order to properly extract all modes contributing to the generated electromagnetic field. 

\begin{figure}[ht!]
	\centering
	\includegraphics[width=0.32\textwidth]{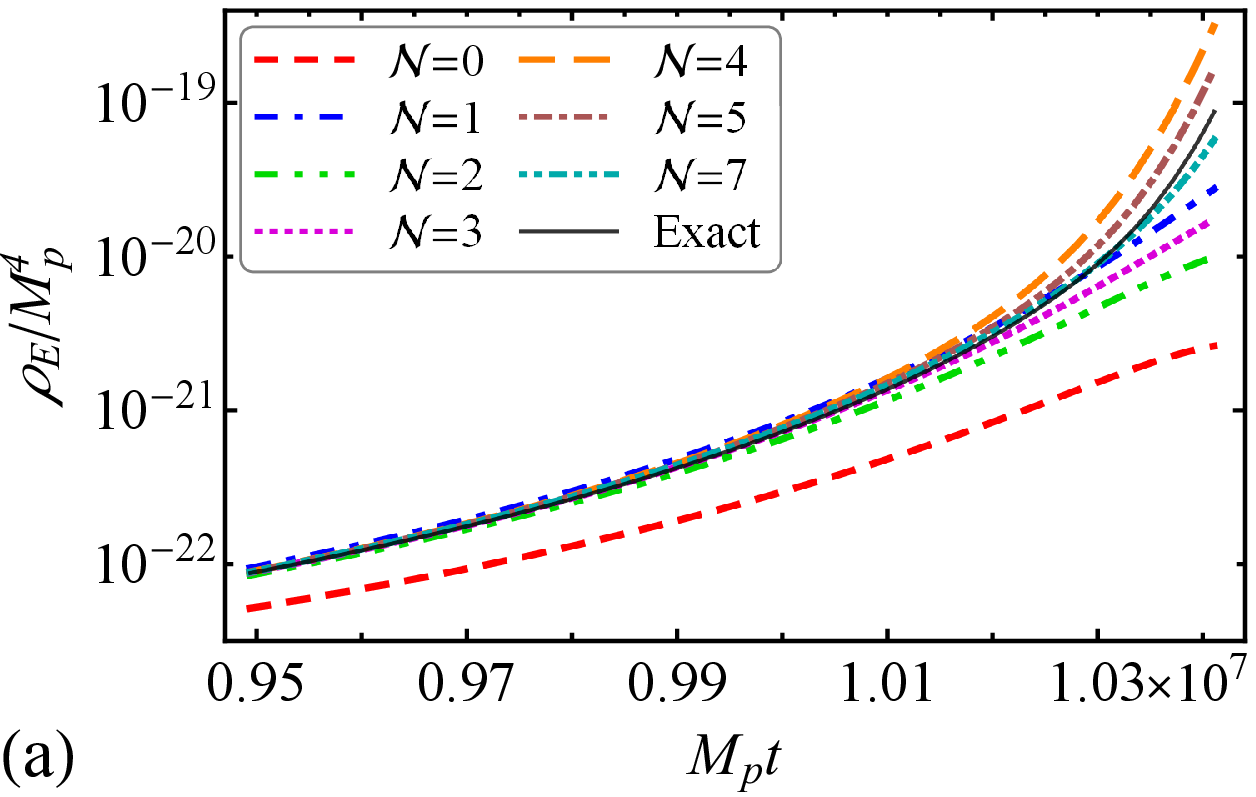}\hspace*{0.3cm}
	\includegraphics[width=0.32\textwidth]{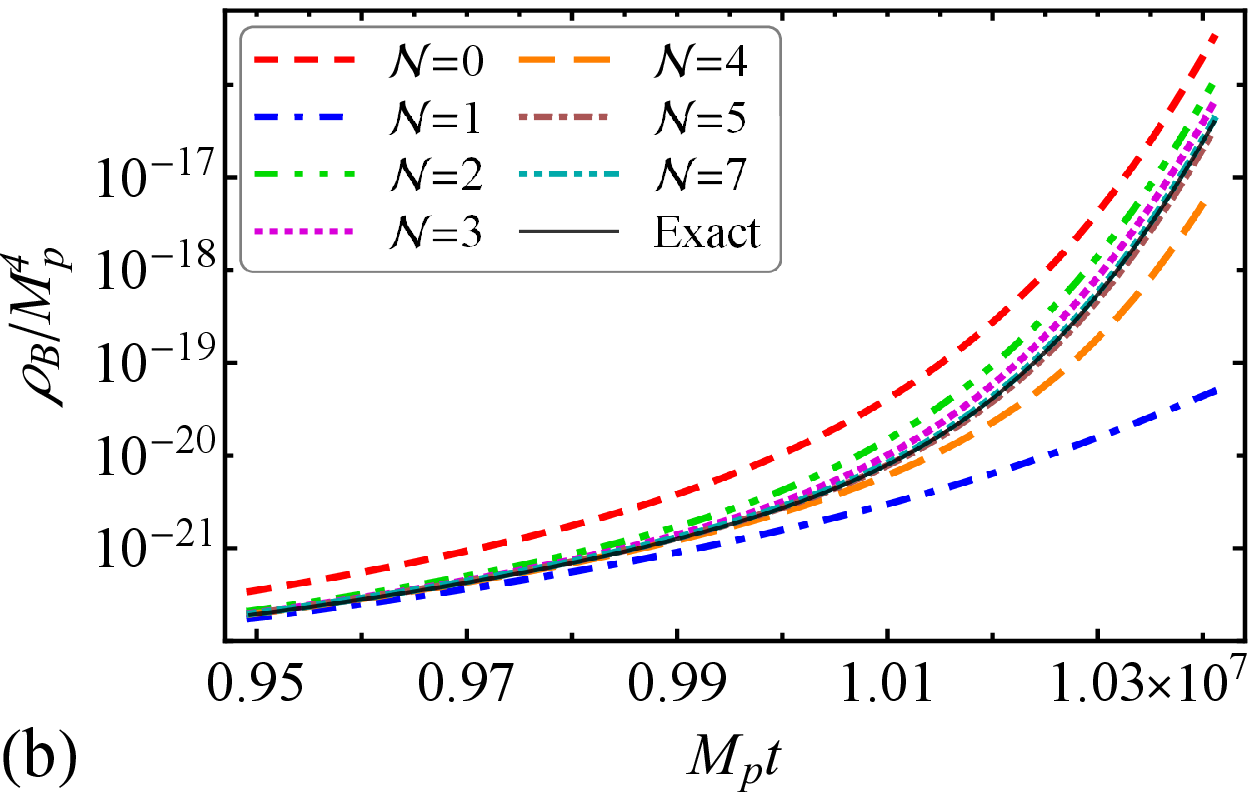}\hspace*{0.3cm}
	\includegraphics[width=0.31\textwidth]{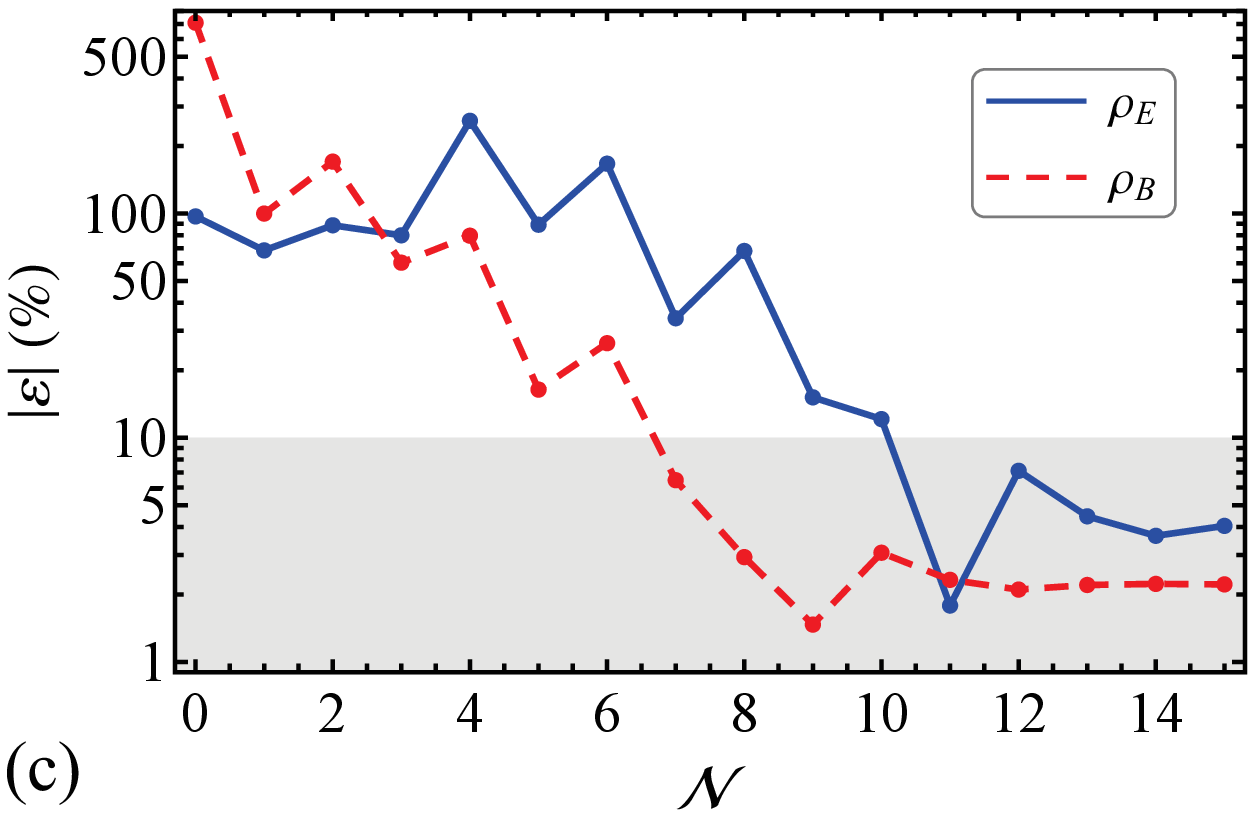}
	\caption{Comparison of the (a)~electric and (b)~magnetic energy densities calculated by using the system of equations (\ref{eq-EE-2})--(\ref{eq-EB-2}) truncated at different orders (see the legend) for the Ratra coupling with $\beta=-7$. Only the final part of the inflation stage is shown, where the difference is the most significant. The black solid lines show the time dependences for the energy densities obtained from the exact numerical solution of Eq.~(\ref{eq-mode-physical}) for all relevant modes and integration over the spectrum as in Eqs.~(\ref{sp-EE-phys}) and (\ref{sp-BB-phys}). For orders higher than $\mathcal{N}=7$, the curves become visually indistinguishable from the exact result, and therefore are not shown. (c)~The relative error of the solution of the system truncated at order $\mathcal{N}$ compared to the exact numerical solution, $\varepsilon=(\rho_{\mathcal{N}}/\rho_{\rm exact}-1)\times 100\%$, calculated at the end of inflation (when the error is maximal). The blue solid line and red dashed line correspond to the electric and magnetic energy densities, respectively. Inside the gray shaded region, the error is less than 10\%.}
	\label{fig-Ratra-m7-comparison}
\end{figure}

In Fig.~\ref{fig-Ratra-m7-comparison}, we compare the solutions of our system of equations truncated at different orders $\mathcal{N}$ with the result of the exact numerical integration of Eq.~(\ref{eq-mode-physical}) for all relevant modes. Since the backreaction is absent in this case, the latter exact solution can be used as a reference point for comparison. Panels (a) and (b) in Fig.~\ref{fig-Ratra-m7-comparison} show the time dependences of the electric and magnetic energy densities during the final part of inflation. The results show that errors introduced by the truncation of the system and by the approximate form of boundary terms are accumulated and have maximal values at the end of inflation. Nevertheless, with increasing $\mathcal{N}$, the deviations from the exact result decrease, and starting from $\mathcal{N}=7$, they become visually indistinguishable.

In order to quantify the error of the approximate result, we calculate the relative deviation from the exact solution at the end of inflation, $t=t_{e}$,
\begin{equation}
\varepsilon=\left(\frac{X_{\mathcal{N}}(t_{e})}{X_{\rm exact}(t_{e})}-1\right)\times 100\%,
\end{equation}
where $X$ denotes any of the electromagnetic quantities, in particular energy densities; and $X_{\mathcal{N}}$ corresponds to the solution of the truncated system of order $\mathcal{N}$; while $X_{\rm exact}$ is the corresponding exact solution. The errors of computation of the electric and magnetic energy densities are shown in Fig.~\ref{fig-Ratra-m7-comparison}(c) by the blue solid line and the red dashed line, respectively. There are a few features we would like to comment on. First of all, the general tendency is that increasing $\mathcal{N}$ typically leads to smaller errors. Second, the truncation at some even order gives typically worse results than at neighboring odd orders. However, for sufficiently large $\mathcal{N}$ (in this case, for $\mathcal{N}\geq 12$), the error stops decreasing and remains almost constant, of the order of a few percent. This residual error cannot be explained by the truncation of the system of equations, as it does not depend on the truncation order. Its possible origin is the approximate expressions for the boundary terms [see the discussion after Eq.~(\ref{bound-BB})]. Fortunately, this error is very small and absolutely irrelevant for practical issues.


\subsubsection{Decreasing coupling function without backreaction}

Now, let us consider the case of a Ratra coupling function that decreases in time with $\beta=7$. It does not cause the strong coupling problem during inflation and does not lead to the backreaction. This allows us to use the exact numerical solution of the mode equation (\ref{eq-mode-physical}) to test the accuracy of our approximate solutions. Before doing this, let us discuss the qualitative behavior of generated electromagnetic fields in this model.

\begin{figure}[ht!]
	\centering
	\includegraphics[width=0.4\textwidth]{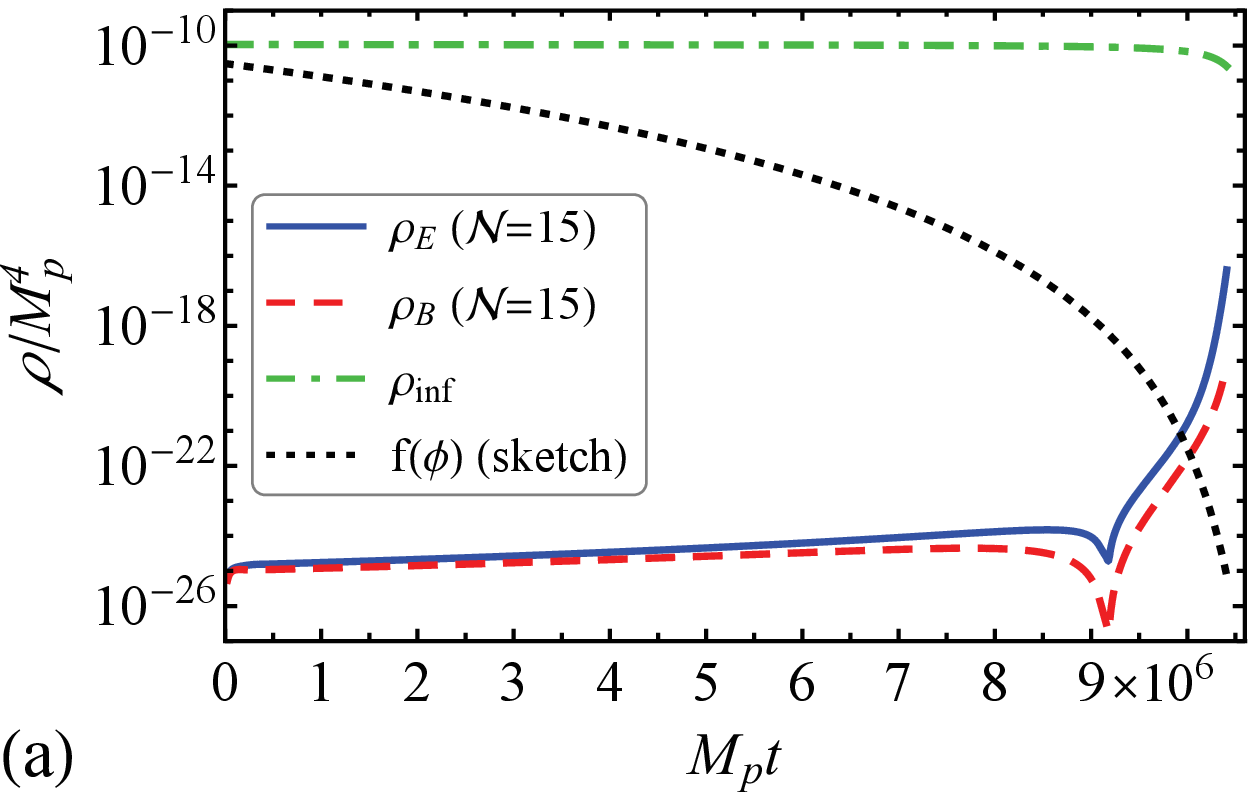}\hspace*{0.5cm}
	\includegraphics[width=0.4\textwidth]{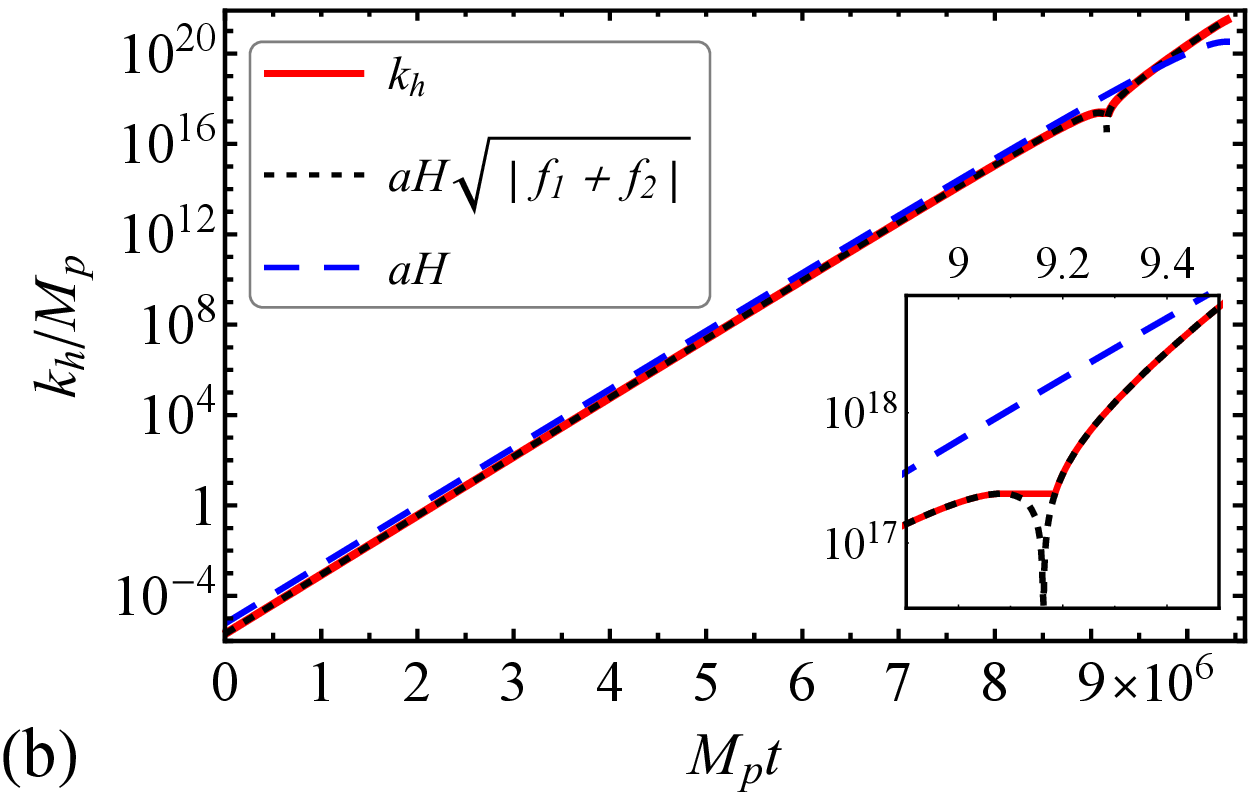}
	\caption{(a)~The time dependence of the electric (blue solid line) and magnetic (red dashed line) components of the energy density generated during inflation in the Ratra model with the coupling parameter $\beta=7$. (The time dependence of the coupling function is sketched by the black dotted line.) The green dash-dotted line gives for comparison the time dependence of the inflaton energy density. For $\beta=7$, the backreaction is irrelevant during inflation, and the electric component is always greater than the magnetic one. (b)~The time dependence of the horizon scale $k_{h}(t)$ (red solid line) and the Hubble horizon scale $aH$ (blue dashed line). The black dotted line shows the argument of the envelope in Eq.~(\ref{horizon-crossing-mode}). The inset shows the zoom-in of the region where the combination $f_{1}+f_{2}$ changes the sign during inflation. The results shown in both panels are obtained from the truncated system of equations with $\mathcal{N}=15$.}
	\label{fig-Ratra-7-EnDens}
\end{figure}

The time dependences of the electric and magnetic energy densities are depicted in Fig.~\ref{fig-Ratra-7-EnDens} (a) by the blue solid and red dashed lines, respectively. The qualitative time dependence of the coupling function is sketched (also in logarithmic scale) by the black dotted line. The energy densities are monotonically increasing functions except for a small period of time close to the end of inflation, where they exhibit nonmonotonic behavior. The origin of this minimum can be understood from Fig.~\ref{fig-Ratra-7-EnDens}(b), where the time dependence of the horizon scale is shown by the red solid line. For positive  values of $\beta$, the combination $(f_{1}+f_{2})$, which determines the horizon scale [Eq.~(\ref{horizon-crossing-mode})], changes its sign during inflation. Indeed, in the slow-roll regime, we get
\begin{equation}
f_{1}+f_{2}\approx \frac{\beta\dot{\phi}}{HM_{p}}+\left(\frac{\beta\dot{\phi}}{HM_{p}}\right)^{2}.
\end{equation}
The time derivative of the inflaton is always negative and grows in the absolute value during the Starobinsky inflation. Then, for sufficiently large positive $\beta$, the combination $|(\beta\dot{\phi})/(HM_{p})|$ may exceed unity, and the combination $f_1+f_2$ flips its sign. As a result, the argument of the envelope in Eq.~(\ref{horizon-crossing-mode}) is a nonmonotonic function; see the black dotted line in Fig.~\ref{fig-Ratra-7-EnDens}(b). In the vicinity of this moment of time, the horizon scale (the envelope) remains constant [see the inset in Fig.~\ref{fig-Ratra-7-EnDens}(b)]. Physically, this means that no new modes cross the horizon in this period of time. As a result, the Universe expansion leads to a decay of the energy densities, which is clearly seen from Fig.~\ref{fig-Ratra-7-EnDens}(a).

\begin{figure}[ht!]
	\centering
	\includegraphics[width=0.4\textwidth]{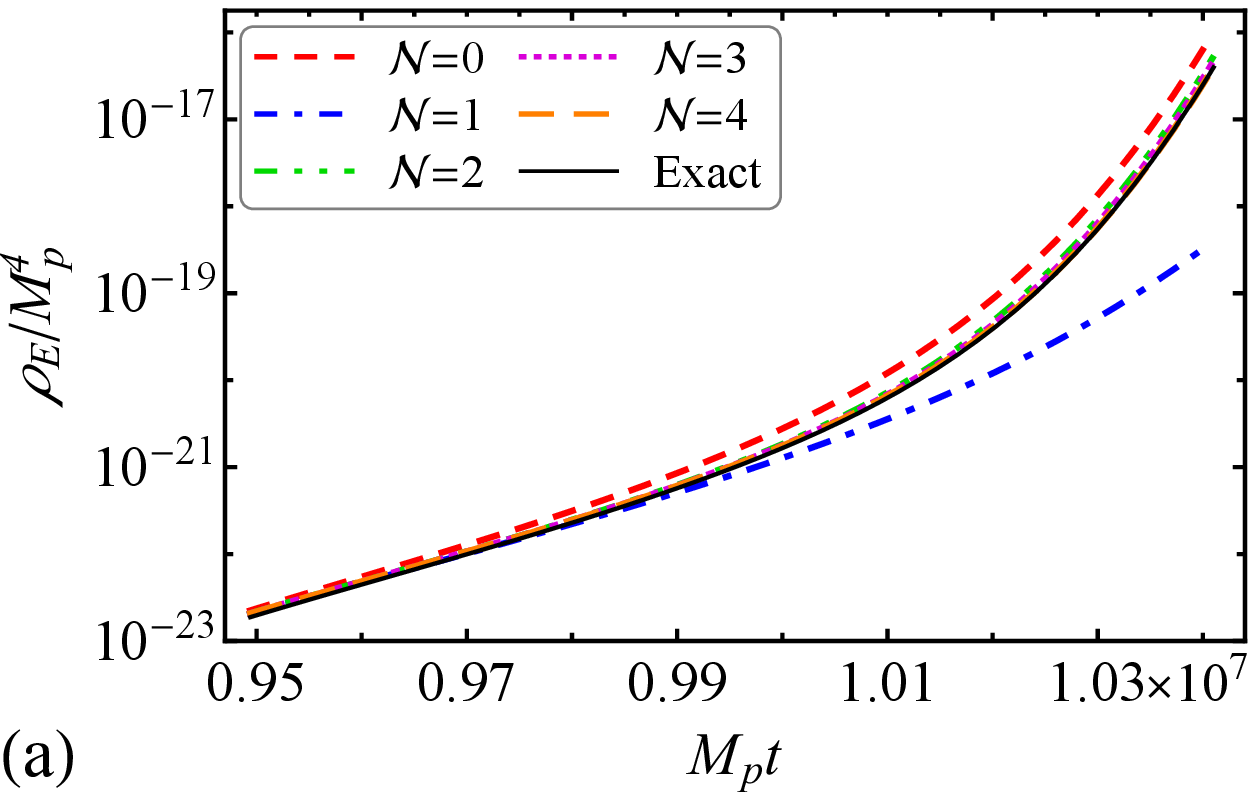}\hspace*{0.5cm}
	\includegraphics[width=0.4\textwidth]{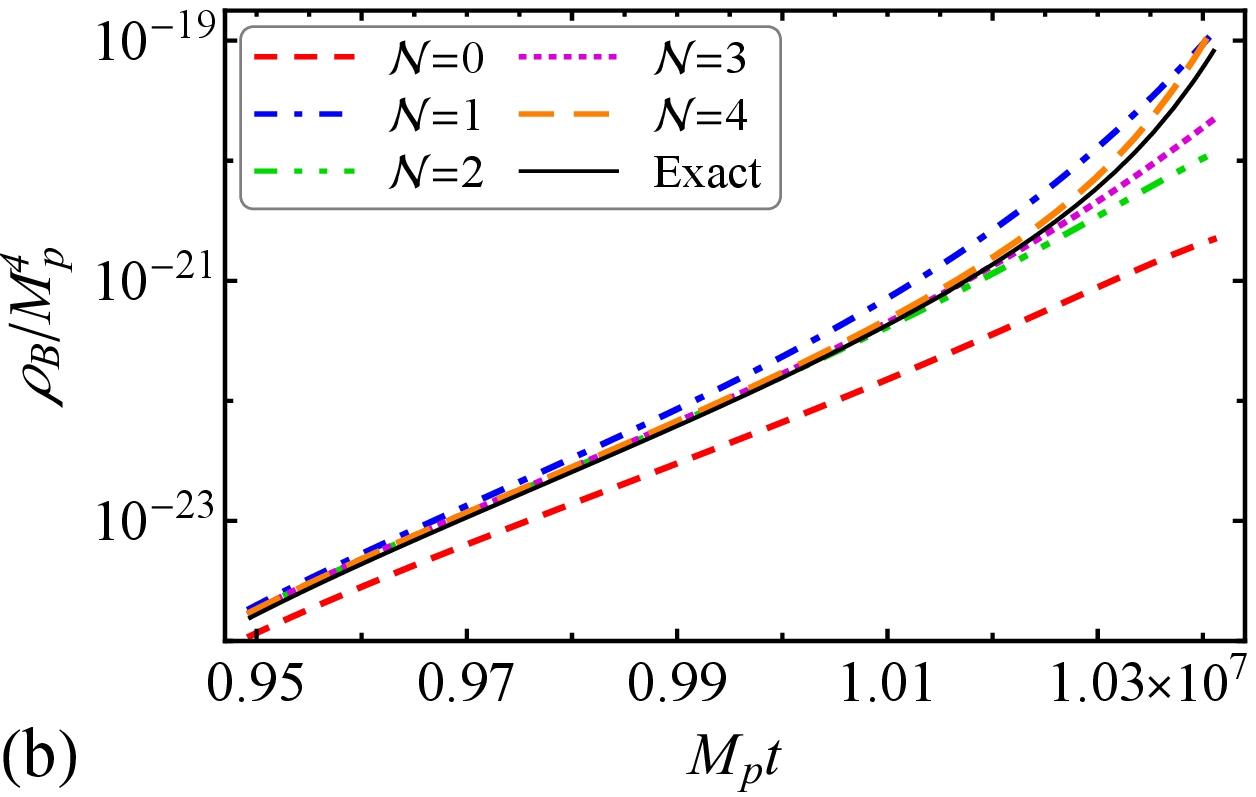}
	\caption{Comparison of the results for the (a)~electric and (b)~magnetic energy densities calculated by using the system of equations (\ref{eq-EE-2})--(\ref{eq-EB-2}) truncated at different orders (see the legend) for the Ratra coupling with $\beta=7$. Only the final part of the inflation stage is shown, where the difference is the most significant. The black solid lines show the time dependences for the energy densities obtained from the exact numerical solution of Eq.~(\ref{eq-mode-physical}) for all relevant modes and integration over the spectrum as in Eqs.~(\ref{sp-EE-phys}) and (\ref{sp-BB-phys}). The curves corresponding to the lowest $\mathcal{N}$ are shown because they can be visually distinguished from the exact result.}
	\label{fig-Ratra-7-comparison}
\end{figure}

Truncating our system at different $\mathcal{N}$, we get approximate solutions for the electric and magnetic energy densities, which are shown in Figs.~\ref{fig-Ratra-7-comparison}(a) and \ref{fig-Ratra-7-comparison}(b) by the dashed colored curves (see the legend). For comparison, the black solid curves show the exact solutions obtained from the numerical integration of the mode equation (\ref{eq-mode-physical}) for all relevant modes. For the few first values of $\mathcal{N}$, the approximate curves significantly differ from the exact result. However, this difference quickly becomes invisible for $\mathcal{N}\geq 4$. The dependence of the relative error of the approximate solution at the end of inflation (where the error is the largest) is qualitatively the same as in the case $\beta=-7$ shown in Fig.~\ref{fig-Ratra-m7-comparison}(c). Starting from $\mathcal{N}=6$, the relative errors both for the electric and magnetic energy densities become less than 10\% and asymptotically tend to residual values of the order of a few percent.


\subsubsection{Decreasing coupling function with backreaction}

\begin{figure}[ht!]
	\centering
	\includegraphics[width=0.4\textwidth]{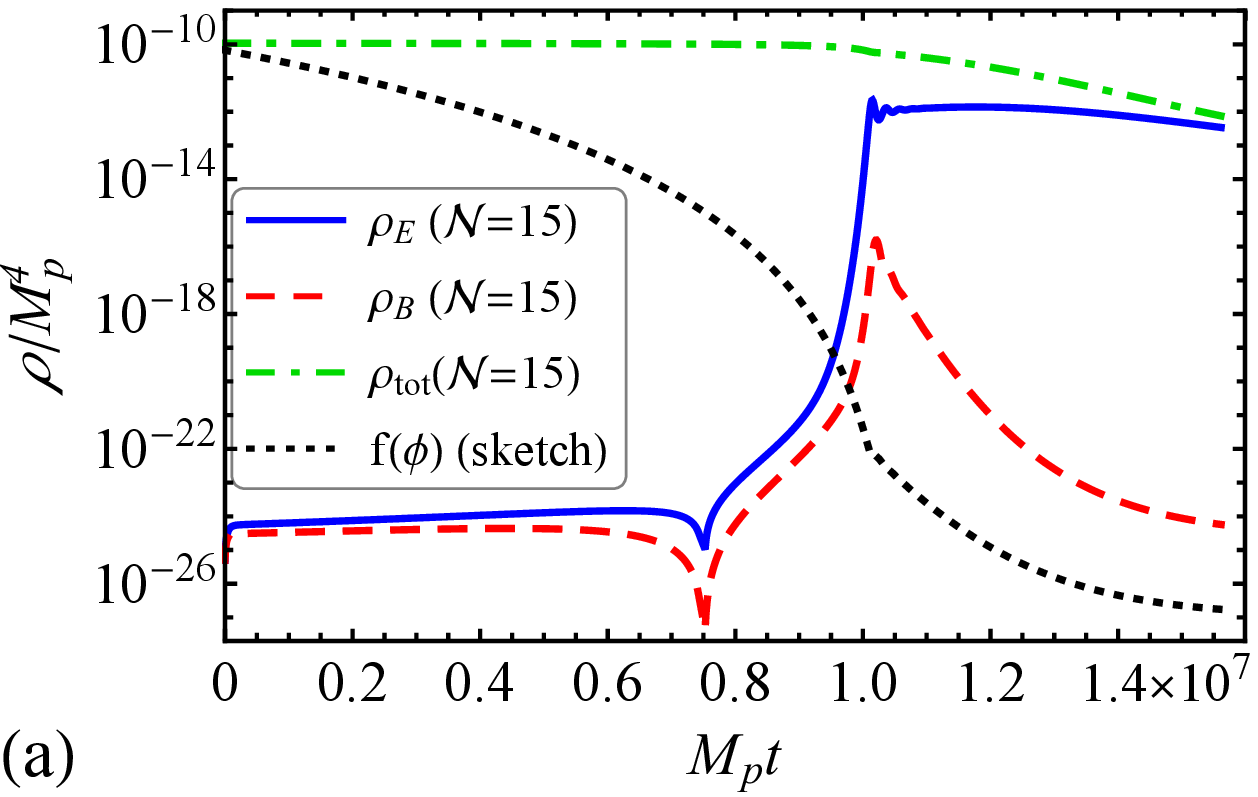}\hspace*{0.5cm}
	\includegraphics[width=0.4\textwidth]{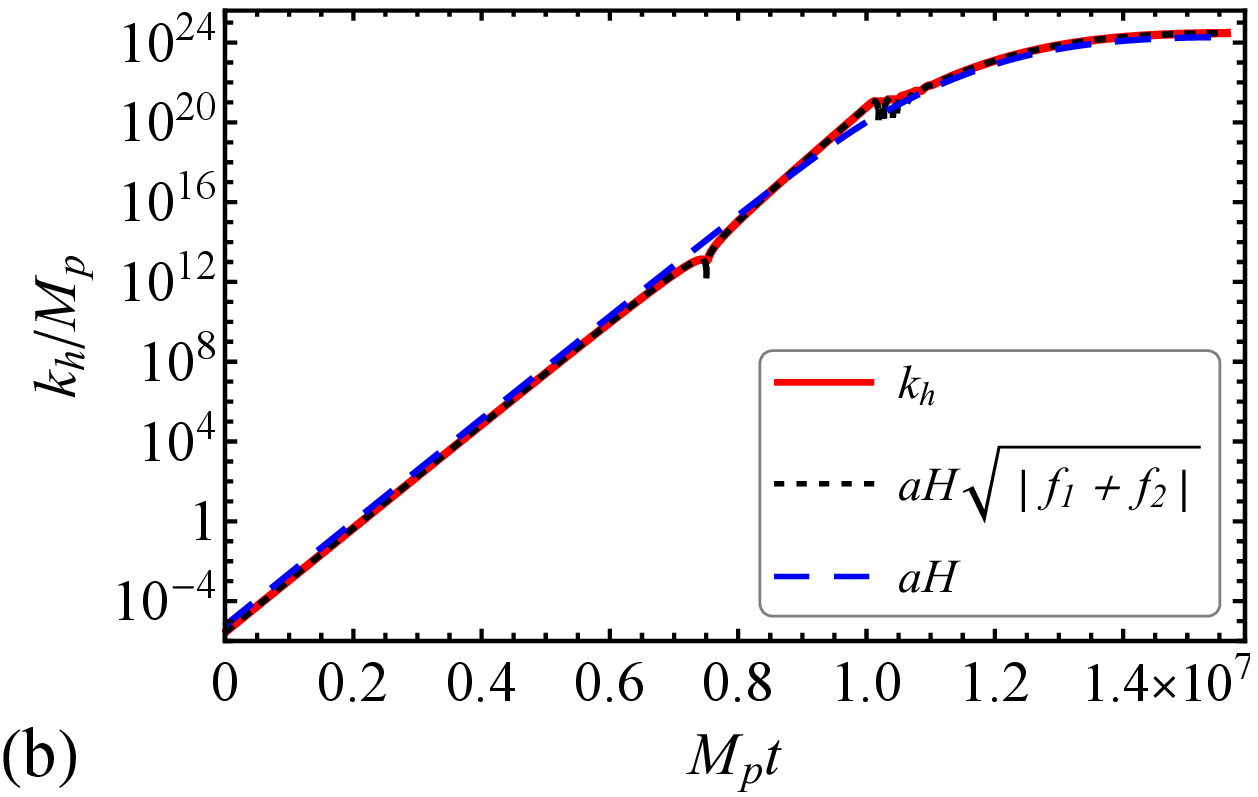}
	\caption{(a)~The time dependence of the electric (blue solid line) and magnetic (red dashed line) components of the energy density generated during inflation in the Ratra model with the coupling parameter $\beta=15$. (The time dependence of the coupling function is sketched by the black dotted line.) The green dash-dotted line gives for comparison the time dependence of the inflaton energy density. For $\beta=15$, the backreaction becomes important around $t=10^7\,M_{p}^{-1}$ and drastically changes the evolution of the Universe. The electric component is always greater than the magnetic one. (b)~The time dependence of the horizon scale $k_{h}(t)$ (red solid line) and the Hubble horizon scale $aH$ (blue dashed line). The black dotted line shows the argument of the envelope in Eq.~(\ref{horizon-crossing-mode}). The results shown in both panels are obtained from the truncated system of equations with $\mathcal{N}=15$.}
	\label{fig-Ratra-15-EnDens}
\end{figure}

Finally, let us consider the case with a large coupling parameter $\beta=15$. At first, it has the same qualitative behavior as in the previously discussed case with $\beta=7$. In fact, the coupling function is also decreasing in time---that is why the electric component slightly dominates the magnetic one; see Fig.~\ref{fig-Ratra-15-EnDens}(a). Both energy densities monotonically grow except for the short period of time when they exhibit a sharp minimum near $t\approx 0.75\times 10^{7}\,M_{p}^{-1}$. As previously, the decrease happens when the horizon scale becomes frozen for a short time near the point where $f_{1}+f_{2}$ changes the sign; see Fig.~\ref{fig-Ratra-15-EnDens}(b).

However, for $\beta=15$, the production of the electromagnetic field is much more effective. As a result, its energy density becomes comparable to that of the inflaton near $t=10^{7}\,M_{p}^{-1}$, and the backreaction becomes important (more precisely, this happens when $\rho_{\rm EM}\simeq \epsilon \rho_{\rm inf}$, where $\epsilon=(M_{p}^{2}/2)(V'/V)^{2}$ is the slow-roll parameter \cite{Sobol:2018}). This drastically changes the evolution of the system. The electric energy density (the dominant component) becomes almost constant, slowly decreasing while the magnetic one (the subdominant component) starts to decay very rapidly. This can be easily understood from the comparison of Eqs.~(\ref{eq-EE-2}) and (\ref{eq-BB-2}) for $n=0$. In fact, the backreaction slows down the inflaton rolling $\dot{\phi}$ so that the Hubble damping term with $\dot{\phi}$ in Eq.~(\ref{eq-EE-2}) is almost exactly compensated by the third term on the left-hand side with the coupling function (the last term $2\mathscr{G}^{(1)}$ as well as the boundary term are much smaller and can be neglected). On the other hand, in Eq.~(\ref{eq-BB-2}), the term with the coupling function enters with the opposite sign and thus helps the Hubble term to damp the magnetic component.

\begin{figure}[ht!]
	\centering
	\includegraphics[width=0.32\textwidth]{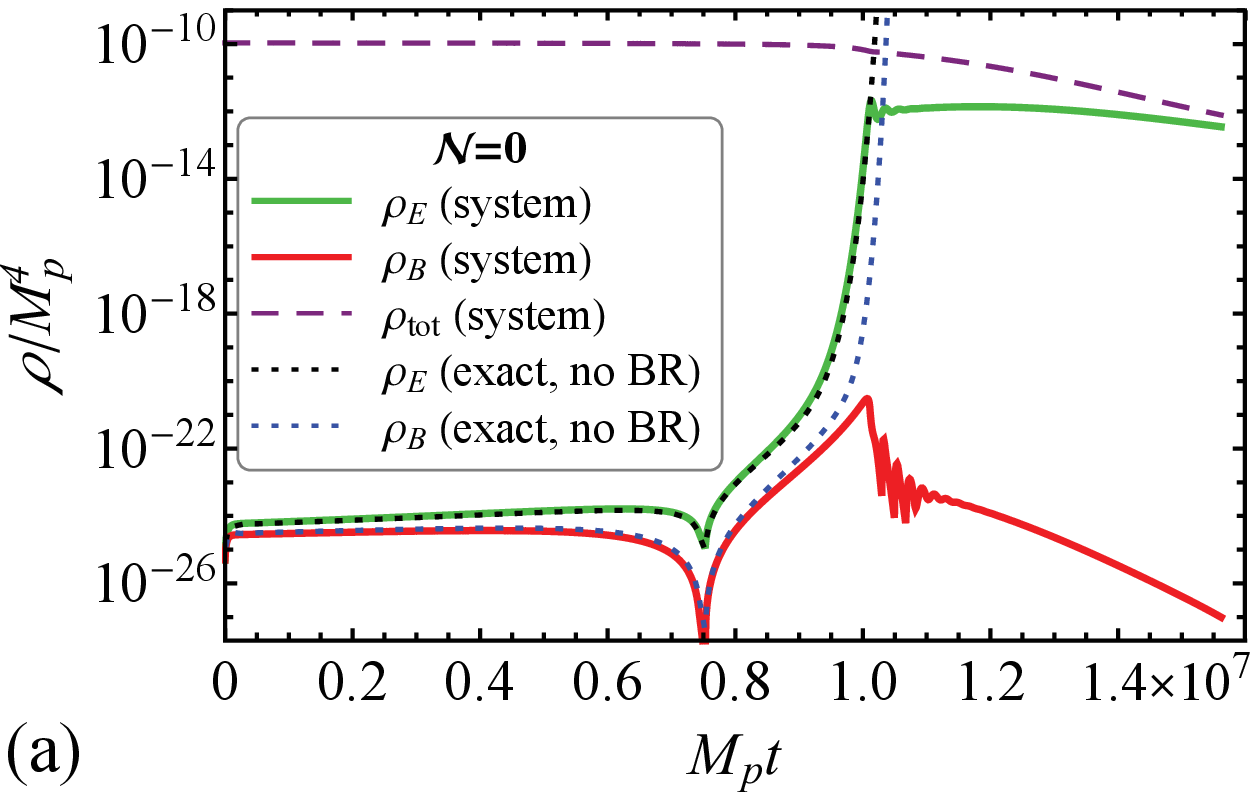}\hspace*{0.25cm}
	\includegraphics[width=0.32\textwidth]{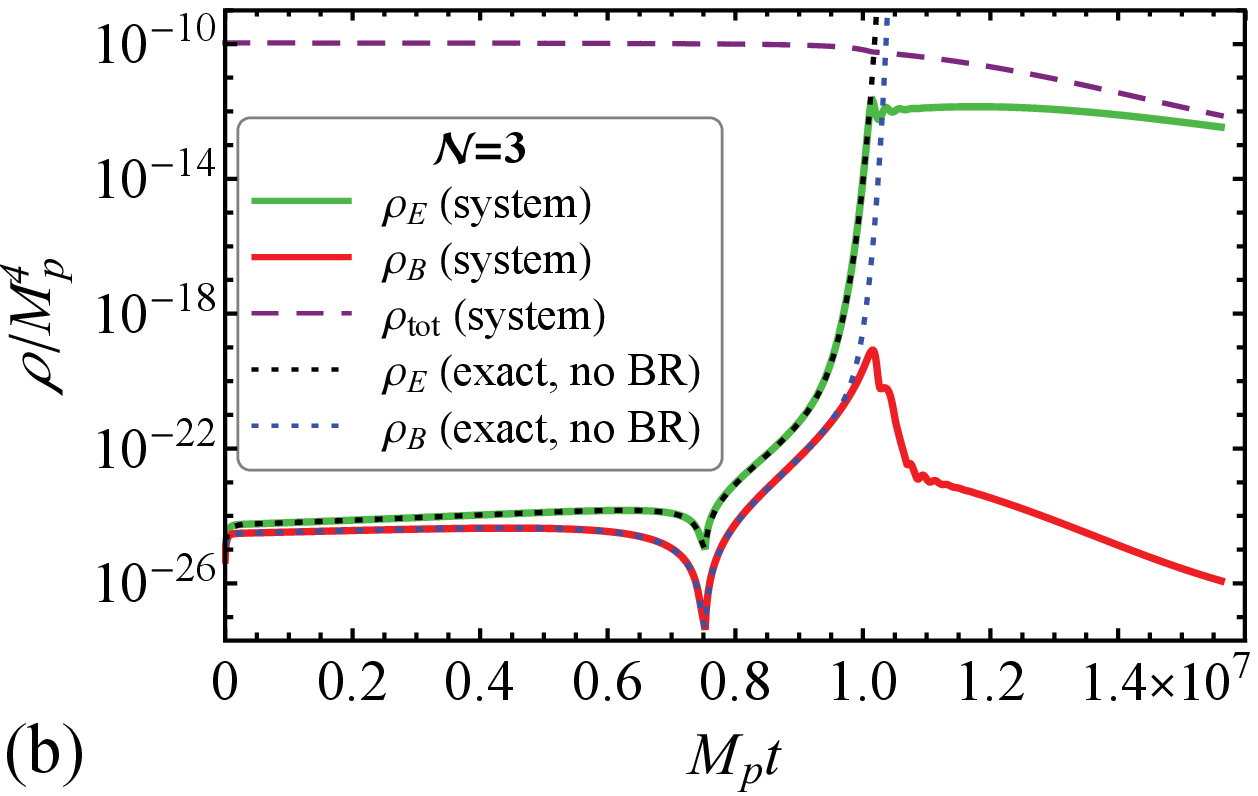}\hspace*{0.25cm}
	\includegraphics[width=0.32\textwidth]{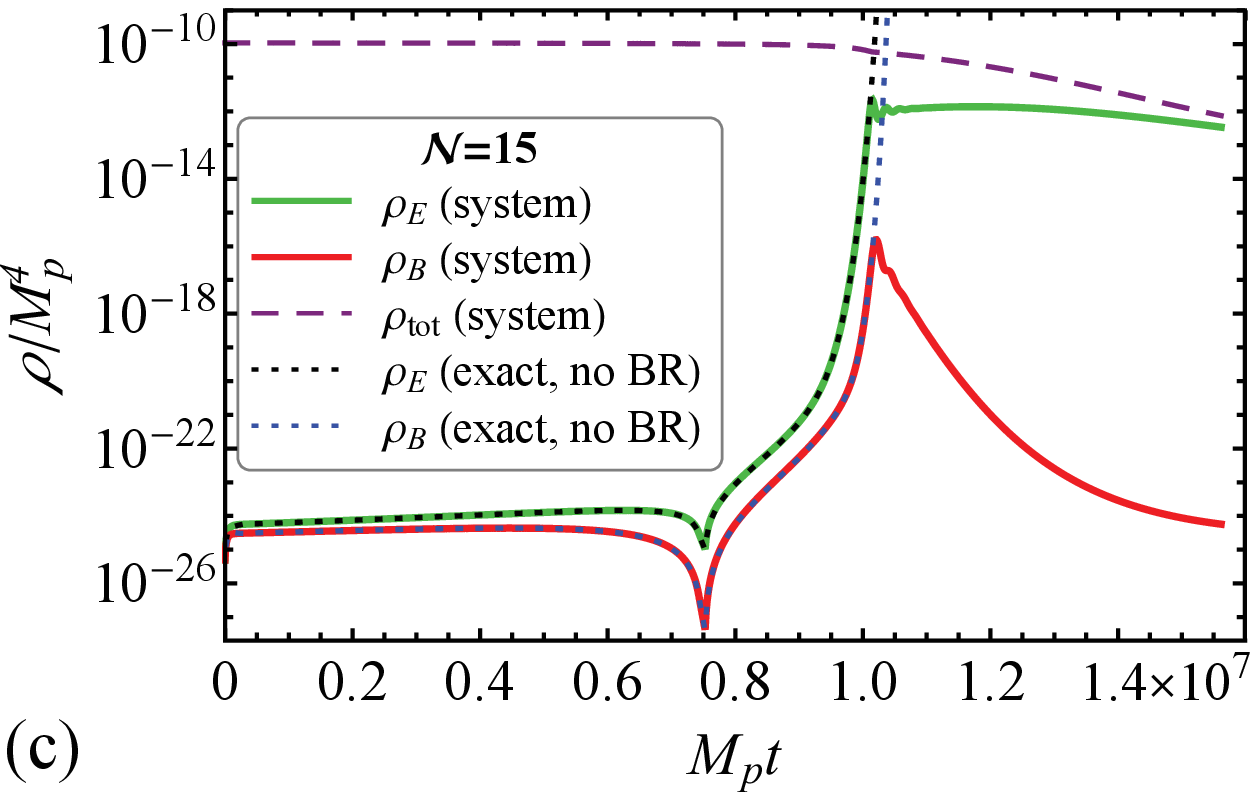}
	\caption{Solutions of the system of equations (\ref{eq-EE-2})--(\ref{eq-EB-2}) truncated at orders (a)~$\mathcal{N}=0$, (b)~$\mathcal{N}=3$, and (c)~$\mathcal{N}=15$ (solid lines), compared to the exact results obtained by the numerical integration of Eq.~(\ref{eq-mode-physical}) neglecting the backreaction (dotted lines). The electric and magnetic energy densities generated during inflation in the Ratra model with $\beta=15$ are shown. The purple dashed lines describe the time dependences of the total energy density.}
	\label{fig-Ratra-15-N}
\end{figure}

\begin{figure}[ht!]
	\centering
	\includegraphics[width=0.4\textwidth]{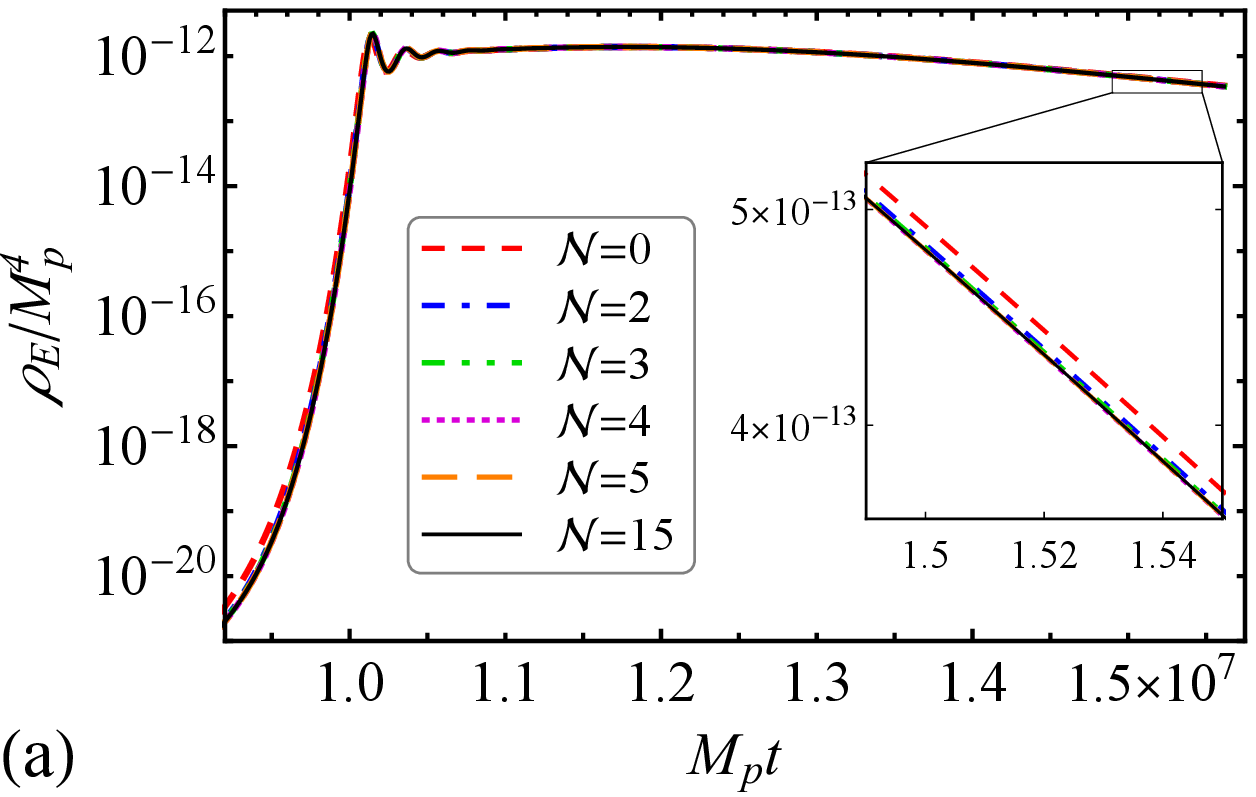}\hspace*{0.5cm}
	\includegraphics[width=0.4\textwidth]{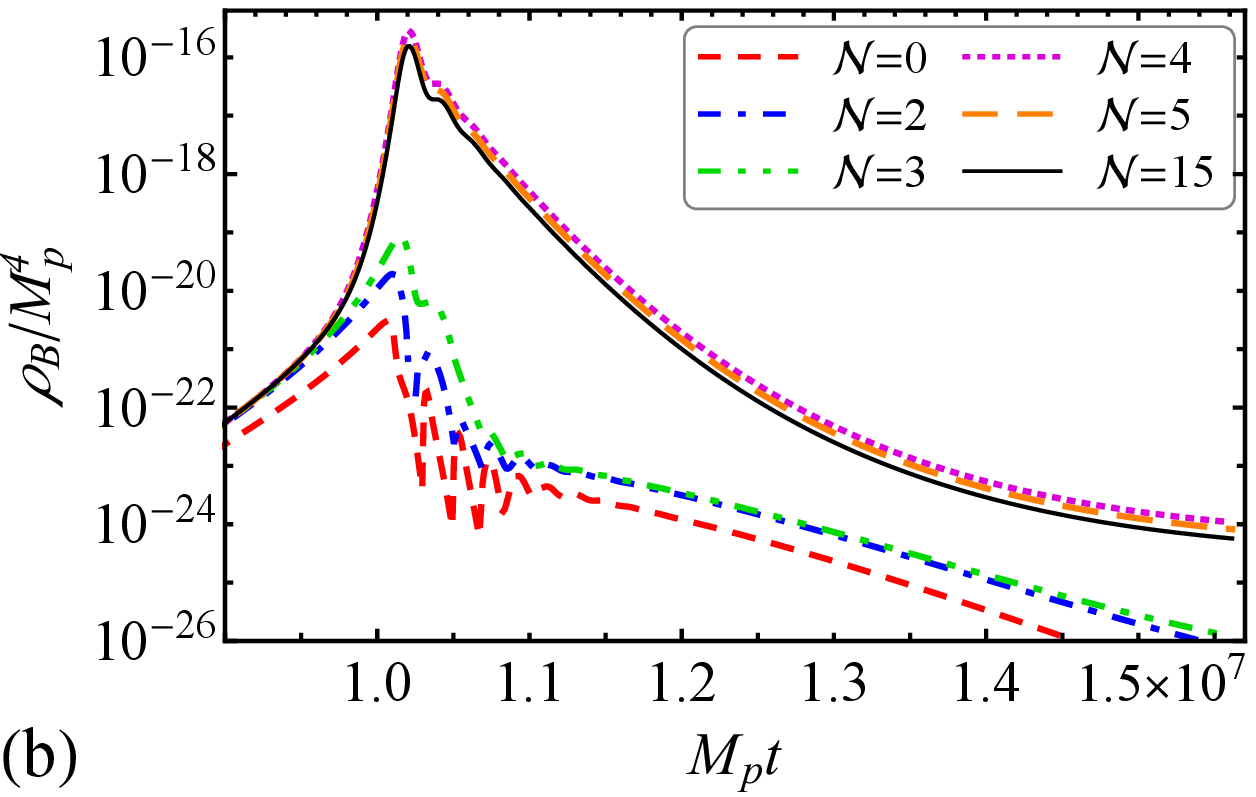}
	\caption{Comparison of the results for the (a)~electric and (b)~magnetic energy densities calculated by using the system of equations (\ref{eq-EE-2})--(\ref{eq-EB-2}) truncated at different orders (see the legend) for the Ratra coupling with $\beta=15$. Only the final part of the inflation stage is shown, where the difference is the most significant. Since the exact numerical solution of Eq.~(\ref{eq-mode-physical}) can be found only in the absence of backreaction, it is not used here as a reference point. We compare the results for a few first values of $\mathcal{N}$ with the result for $\mathcal{N}=15$ (black solid lines).}
	\label{fig-Ratra-15-comparison}
\end{figure}

For $\beta=15$, we can compare our approximate results to the exact numerical solution of the mode equation (\ref{eq-mode-physical}) only until the backreaction becomes important. Figure~\ref{fig-Ratra-15-N} compares the results of the system truncated at $\mathcal{N}=0$, 3, and 15 (solid lines) with the exact result (dotted lines). All three panels show a very good accordance when the backreaction is not important. As the latter
becomes relevant, the exact mode-by-mode solution of Eq.~(\ref{eq-mode-physical}) begins to grow rapidly and quickly becomes larger than the inflaton energy density. In this region, the comparison is no more possible. Therefore, in Fig.~\ref{fig-Ratra-15-comparison}, we compare the results of the truncated system for different $\mathcal{N}$ with the results for $\mathcal{N}=15$ (the largest of which we considered in our numerical analysis). The latter plays the role of a reference solution in the strong backreaction regime.

Figure~\ref{fig-Ratra-15-comparison}(a) shows that for the dominant component (electric energy density), all approximations even with the smallest $\mathcal{N}$ give very close results. However, the subdominant (magnetic) component is determined with a much worse accuracy; see Fig.~\ref{fig-Ratra-15-comparison}(b). The results for the magnetic component start to converge only from $\mathcal{N}=4$. The convergence is very fast, and starting from $\mathcal{N}=9$, the relative deviation from $\mathcal{N}=15$ becomes less than 10\%.


\subsection{Nonmonotonic coupling function}

In this subsection, we consider the nonmonotonic coupling function 
\begin{equation}
\label{nonmonotonic-coupling}
f(\phi)=1+\frac{C_{0}}{{\rm cosh}^{2}[\beta(\phi-\phi_{1})/M_{p}]},
\end{equation}
where $\beta$, $C_{0}$, and $\phi_{1}/M_{p}$ are three dimensionless parameter which determine the steepness, height, and location of the maximum of this function. For $C_{0}>0$, it is always greater than unity, which ensures that the strong coupling problem never occurs during inflation. The values of coupling parameters should be chosen in such a way that (i)~the maximum of the coupling function is reached during the simulation time (so that we exploit its nonmonotonic behavior); (ii)~the value $f\simeq 1$ is reached at the end of inflation (electric charges of all fields are not rescaled); and (iii)~$C_{0}$ is large enough to provide a significant generation of electromagnetic fields. We do not claim that such a function has some physical origin, but only use it to test the applicability of our formalism to more complicated (nonmonotonic) functions. We will see that the evolution of the electromagnetic energy density has much more interesting features, and our truncated system already can reproduce all of them even for small $\mathcal{N}$.

In the numerical analysis, we use two different sets of parameters, which represent the regimes without backreaction and those with it.

\subsubsection{Nonmonotonic function without backreaction}

In order to be able to compare our approximate results with the exact solution of Eq.~(\ref{eq-mode-physical}), we first consider the coupling function which does not lead to the backreaction on the inflaton evolution. This is realized, e.g., for the following set of coupling parameters: $\beta=10$, $C_{0}=10^{19}$, $\phi_{1}=4.5\,M_{p}$. 

\begin{figure}[ht!]
	\centering
	\includegraphics[width=0.4\textwidth]{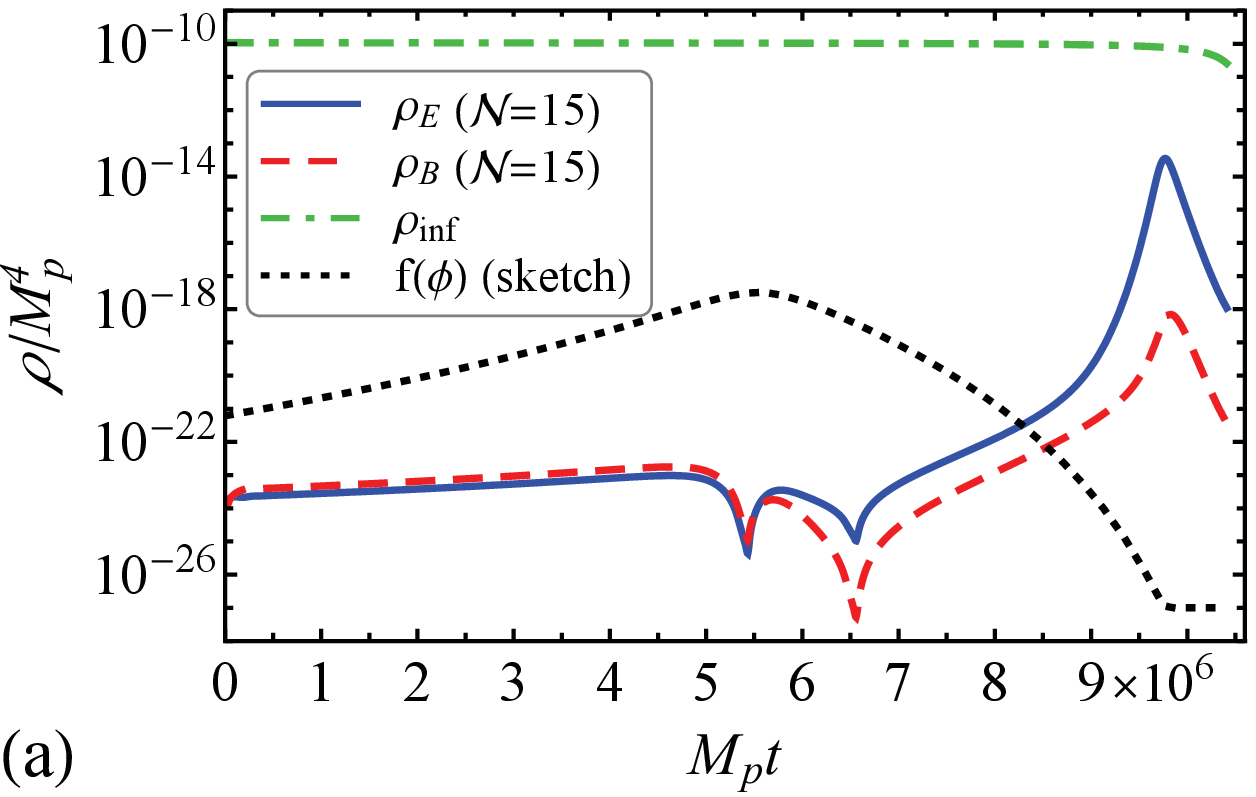}\hspace*{0.5cm}
	\includegraphics[width=0.4\textwidth]{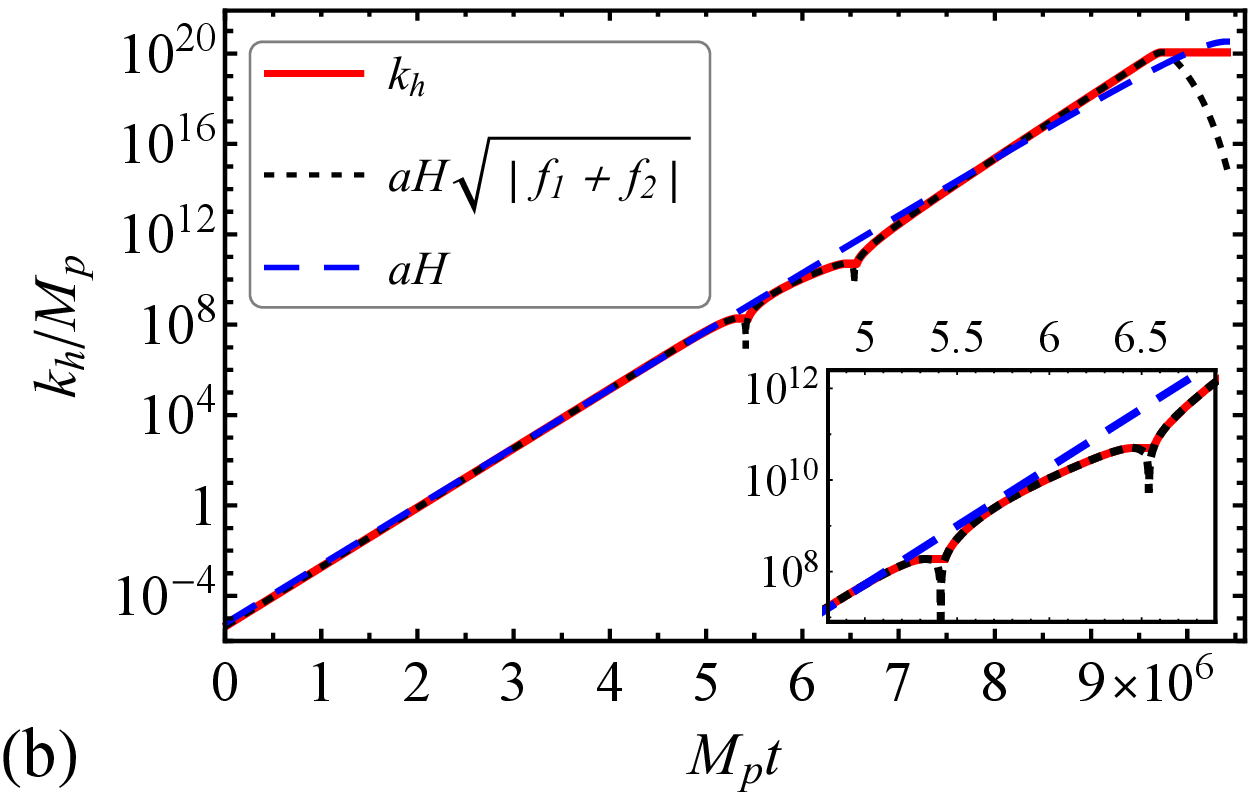}
	\caption{(a)~The time dependence of the electric (blue solid line) and magnetic (red dashed line) components of the energy density generated during inflation in the kinetic coupling model with the nonmonotonic function (\ref{nonmonotonic-coupling}) for parameters $\beta=10$, $C_{0}=10^{19}$, $\phi_{1}=4.5\,M_{p}$. (The time dependence of the coupling function is sketched by the black dotted line.) The green dash-dotted line gives for comparison the time dependence of the inflaton energy density. The backreaction is irrelevant during inflation; the magnetic component is greater than the electric one when the coupling function grows, while the situation changes to the opposite when the coupling function starts to decrease. (b)~The time dependence of the horizon scale $k_{h}(t)$ (red solid line) and the Hubble horizon scale $aH$ (blue dashed line). The black dotted line shows the argument of the envelope in Eq.~(\ref{horizon-crossing-mode}). The inset shows the zoom-in of the region where the combination $f_{1}+f_{2}$ changes sign during inflation. The results shown in both panels are obtained from the truncated system of equations with $\mathcal{N}=15$.}
	\label{fig-NM1-EnDens}
\end{figure}

The time dependences of the energy densities in this case are shown in Fig.~\ref{fig-NM1-EnDens}(a). As long as the function increases (it is sketched by the black dotted line), the magnetic component is slightly greater than the electric one. As the coupling function starts to decrease, their roles interchange. At first, the curves grow monotonically, but then two sharp minima appear. They happen when the combination $f_{1}+f_{2}$ changes sign. For the nonmonotonic function [Eq.~(\ref{nonmonotonic-coupling})], this happens twice: once close to its maximum, and once during the decreasing stage. At these moments of time, the horizon scale $k_{h}$, which is the envelope of $aH |f_{1}+f_{2}|$, is frozen; see Fig.~\ref{fig-NM1-EnDens}(b). No new modes exit the horizon, and the boundary terms vanish at that time. Therefore, the electromagnetic field undergoes short-term damping due to the Universe expansion.

The generation becomes more efficient when the inflaton starts rolling faster close to the end of inflation. However, this lasts until the hyperbolic cosine in Eq.~(\ref{nonmonotonic-coupling}) becomes greater than $\sqrt{C_{0}}$. After that, the coupling function remains almost constant with $f\approx 1$; see the final horizontal part of the black dotted line in Fig.~\ref{fig-NM1-EnDens}(a). At this stage, the horizon scale $k_{h}$ remains constant; see Fig.~\ref{fig-NM1-EnDens}(b). As a result, both the generation terms $\propto (\dot{f}/f)$ and the boundary terms in Eqs.~(\ref{eq-EE-2})--(\ref{eq-EB-2}) vanish. Then, the Hubble expansion leads to a fast damping of generated fields.

\begin{figure}[ht!]
	\centering
	\includegraphics[width=0.32\textwidth]{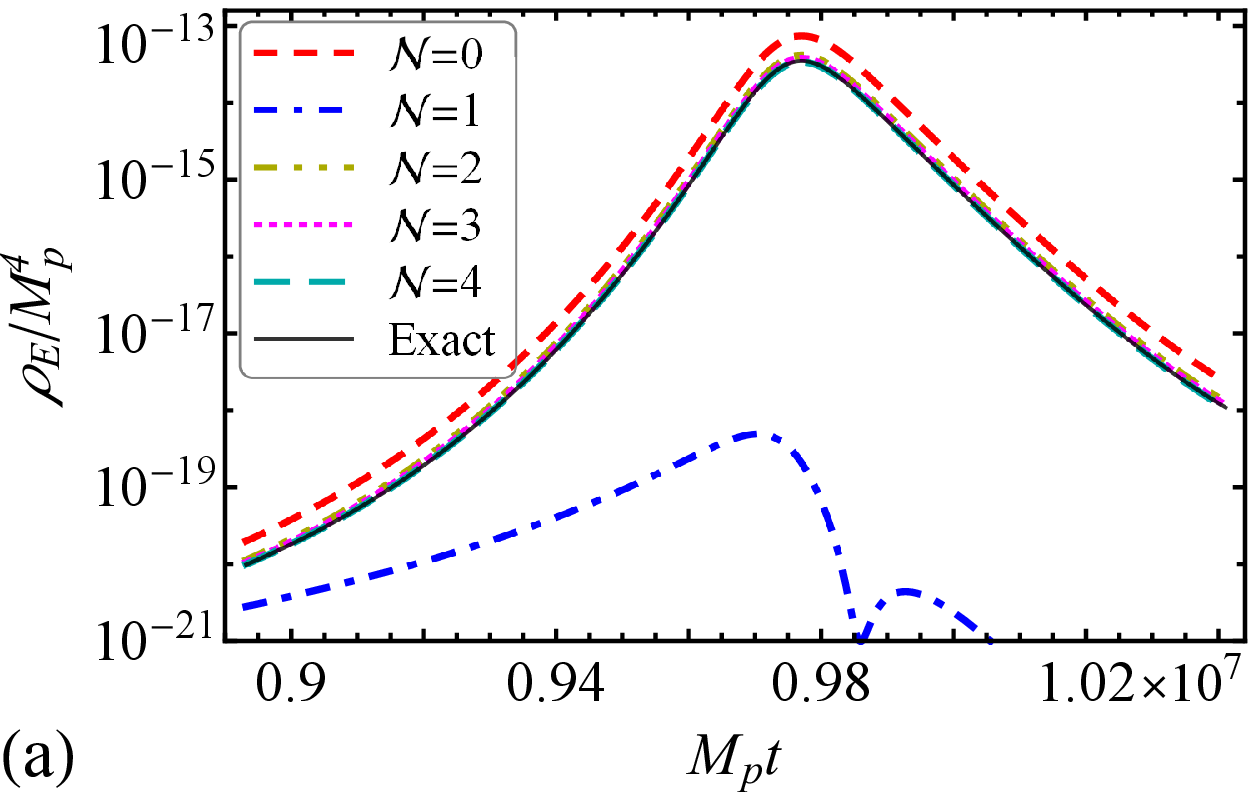}\hspace*{0.3cm}
	\includegraphics[width=0.32\textwidth]{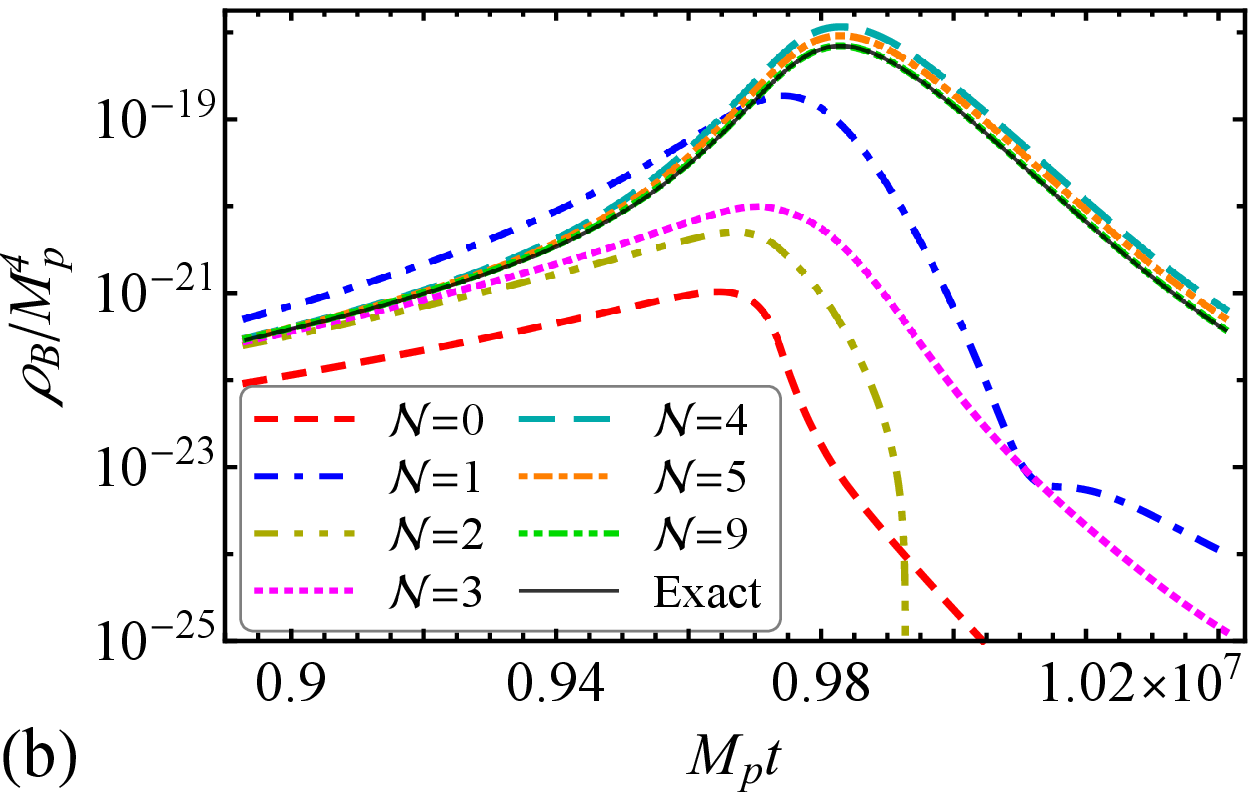}\hspace*{0.3cm}
	\includegraphics[width=0.31\textwidth]{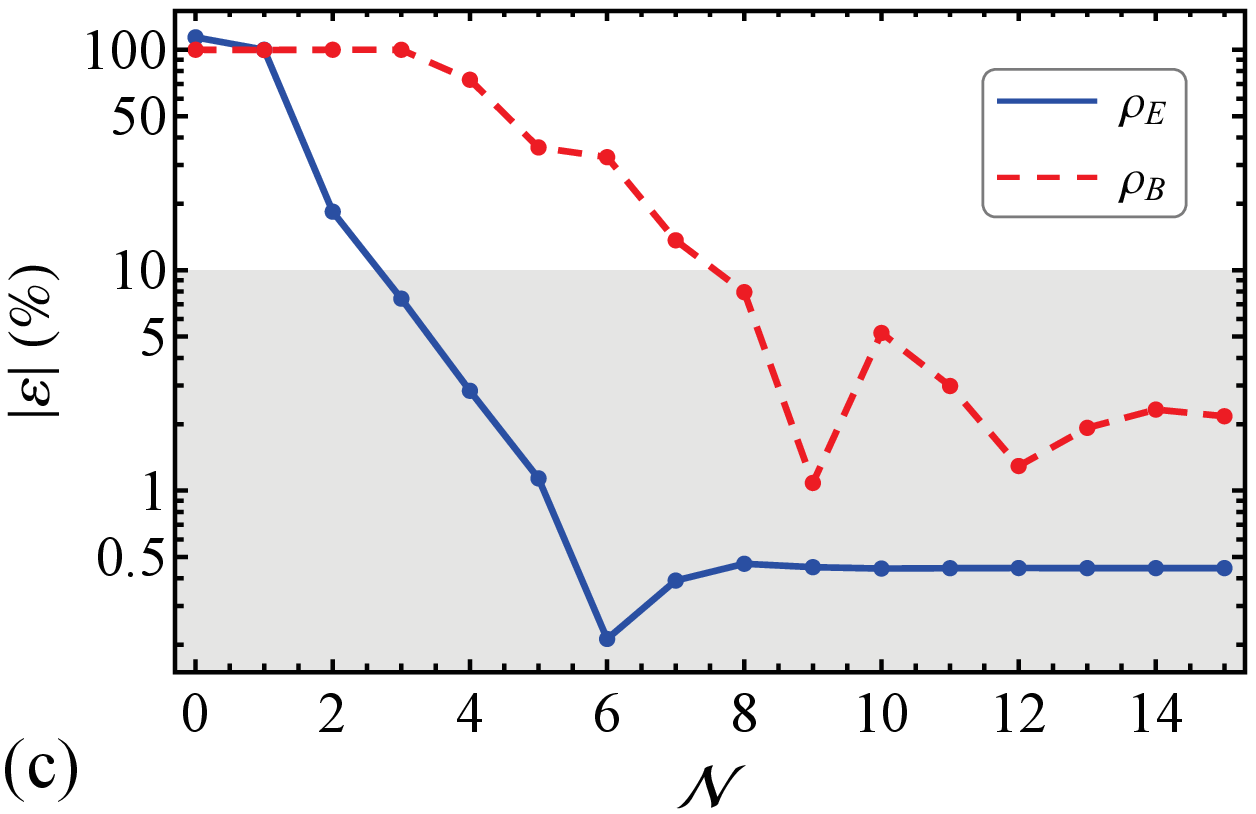}
	\caption{Comparison of the results for the (a)~electric and (b)~magnetic energy densities calculated by using the system of equations (\ref{eq-EE-2})--(\ref{eq-EB-2}) truncated at different orders (see the legend) for the nonmonotonic coupling function (\ref{nonmonotonic-coupling}) with parameters $\beta=10$, $C_{0}=10^{19}$, and $\phi_{1}=4.5\,M_{p}$. Only the final part of the inflation stage is shown, where the difference is the most significant. The black solid lines show the time dependences for the energy densities obtained from the exact numerical solution of Eq.~(\ref{eq-mode-physical}) for all relevant modes and integration over the spectrum as in Eqs.~(\ref{sp-EE-phys}) and (\ref{sp-BB-phys}). Starting from $\mathcal{N}=4$, the curves become visually very close to the exact result. (c)~The relative error of the solution of the system truncated at order $\mathcal{N}$ compared to the exact numerical solution, $\varepsilon=(\rho_{\mathcal{N}}/\rho_{\rm exact}-1)\times 100\%$, calculated at the end of inflation (when the error is maximal). The blue solid line and red dashed line correspond to the electric and magnetic energy densities, respectively. Inside the gray shaded region, the error is less than 10\%.}
	\label{fig-NM1-comparison}
\end{figure}

Let us now compare the results of the approximate system of equations truncated at different orders. The electric and magnetic energy densities are shown in Figs.~\ref{fig-NM1-comparison}(a) and \ref{fig-NM1-comparison}(b), respectively. The dashed colored curves corresponding to different $\mathcal{N}$ are compared to the exact numerical result obtained from the mode-by-mode solution of Eq.~(\ref{eq-mode-physical}), which is shown by the black solid line. As in the previously discussed cases with monotonic coupling functions, the dominant (electric) component is satisfactorily described already by the approximation with $\mathcal{N}=2$. However, the subdominant (magnetic) component starts to converge to the exact solution only after $\mathcal{N}=4$. The maximal error is, as usually, reached at the end of inflation. We show its dependence on the truncation order $\mathcal{N}$ in Fig.~\ref{fig-NM1-comparison}(c). Starting from $\mathcal{N}=8$, the errors for both the electric and magnetic components are less than 10\%. With further increase of $\mathcal{N}$, the errors stabilize and remain at the level of order 1\%. This residual error is caused not by the truncation but by other approximations we made in the derivation of our system.


\subsubsection{Nonmonotonic function with backreaction}

Finally, let us discuss the situation when the nonmonotonic coupling function [Eq.~(\ref{nonmonotonic-coupling})] leads to a strong backreaction of generated electromagnetic fields. For example, this happens for the coupling  parameters $\beta=12$, $C_{0}=10^{21}$, $\phi_{1}=2.8\,M_{p}$. 

\begin{figure}[ht!]
	\centering
	\includegraphics[width=0.4\textwidth]{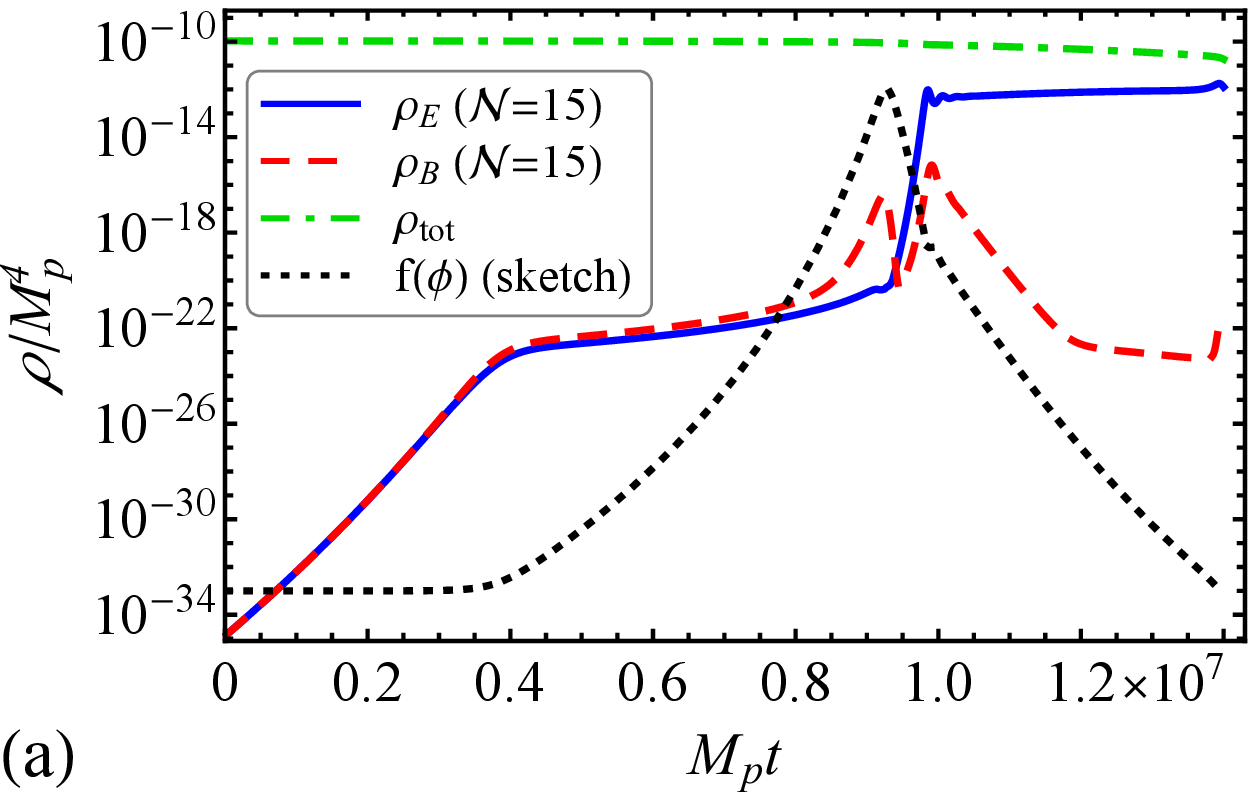}\hspace*{0.5cm}
	\includegraphics[width=0.4\textwidth]{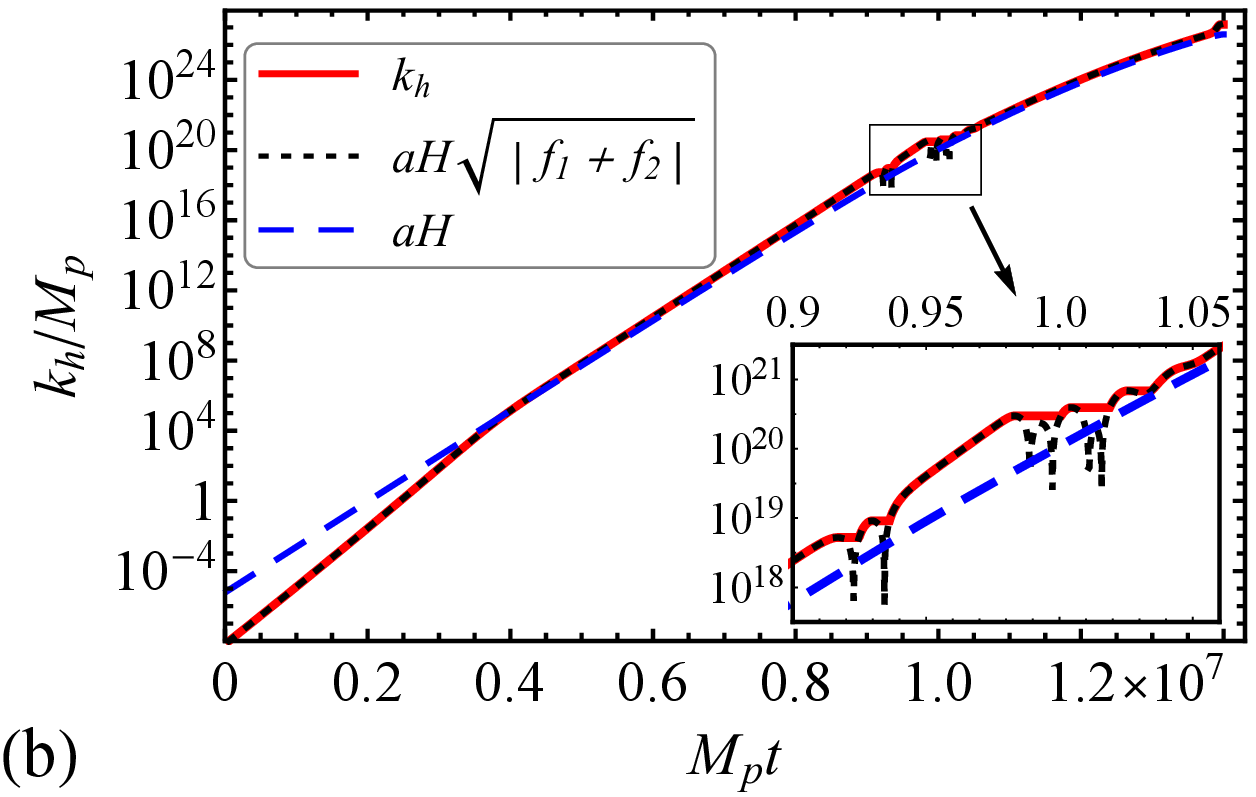}
	\caption{(a)~The time dependence of the electric (blue solid line) and magnetic (red dashed line) components of the energy density generated during inflation in the kinetic coupling model with the nonmonotonic function (\ref{nonmonotonic-coupling}) for parameters $\beta=12$, $C_{0}=10^{21}$, $\phi_{1}=2.8\,M_{p}$. (The time dependence of the coupling function is sketched by the black dotted line.) The green dash-dotted line gives for comparison the time dependence of the inflaton energy density. The backreaction becomes important around $t=10^7\,M_{p}^{-1}$ and drastically changes the evolution of the Universe. The magnetic component is greater than the electric one when the coupling function grows, while the opposite relation occurs when the function starts to decrease. (b)~The time dependence of the horizon scale $k_{h}(t)$ (red solid line) and the Hubble horizon scale $aH$ (blue dashed line). The black dotted line shows the argument of the envelope in Eq.~(\ref{horizon-crossing-mode}). The inset shows the zoom-in of the region when the latter changes nonmonotonically. The results shown in both panels are obtained from the truncated system of equations with $\mathcal{N}=15$.}
	\label{fig-NM2-EnDens}
\end{figure}

Energy densities of the electric and magnetic fields generated in this case are shown in Fig.~\ref{fig-NM2-EnDens}(a). Their time dependences have several interesting features. At first, when the function is almost constant and close to unity (see the horizontal part of the black dotted curve), generated fields are very weak. Further, the magnetic field slightly dominates over the electric one as the function grows. Then, when the coupling function starts to decrease, their roles change. A rapid decay of the magnetic field is caused by two factors: (i)~the horizon scale becomes frozen twice for a short time (again, this happens when $f_{1}+f_{2}$ changes sign)---see Fig.~\ref{fig-NM2-EnDens}(b)---and the electromagnetic field is no longer enhanced by new modes crossing the horizon; and (ii)~the third term on the left-hand side in Eq.~(\ref{eq-BB-2}) leads to a decay of the magnetic field. 

A rapid decrease of the coupling function generates very strong electric fields, which then induce the growth of the magnetic field. This is due to the Faraday law, which is incorporated in Eq.~(\ref{eq-BB-2}) by the last term on the left-hand side. This explains the second increase of the magnetic energy density. Finally, the total electromagnetic energy density reaches the value $\rho_{\rm EM}=\epsilon \rho_{\rm inf}$, and the backreaction occurs. The electric energy density, which is the dominant component, stops growing and remains almost constant (as a result of the dynamical equilibrium between the inflaton and the generated field). The magnetic field at this time decays due to the Universe expansion. This analysis explains the ``zig-zag'' form of the red dashed curve in Fig.~\ref{fig-NM2-EnDens}.

\begin{figure}[ht!]
	\centering
	\includegraphics[width=0.32\textwidth]{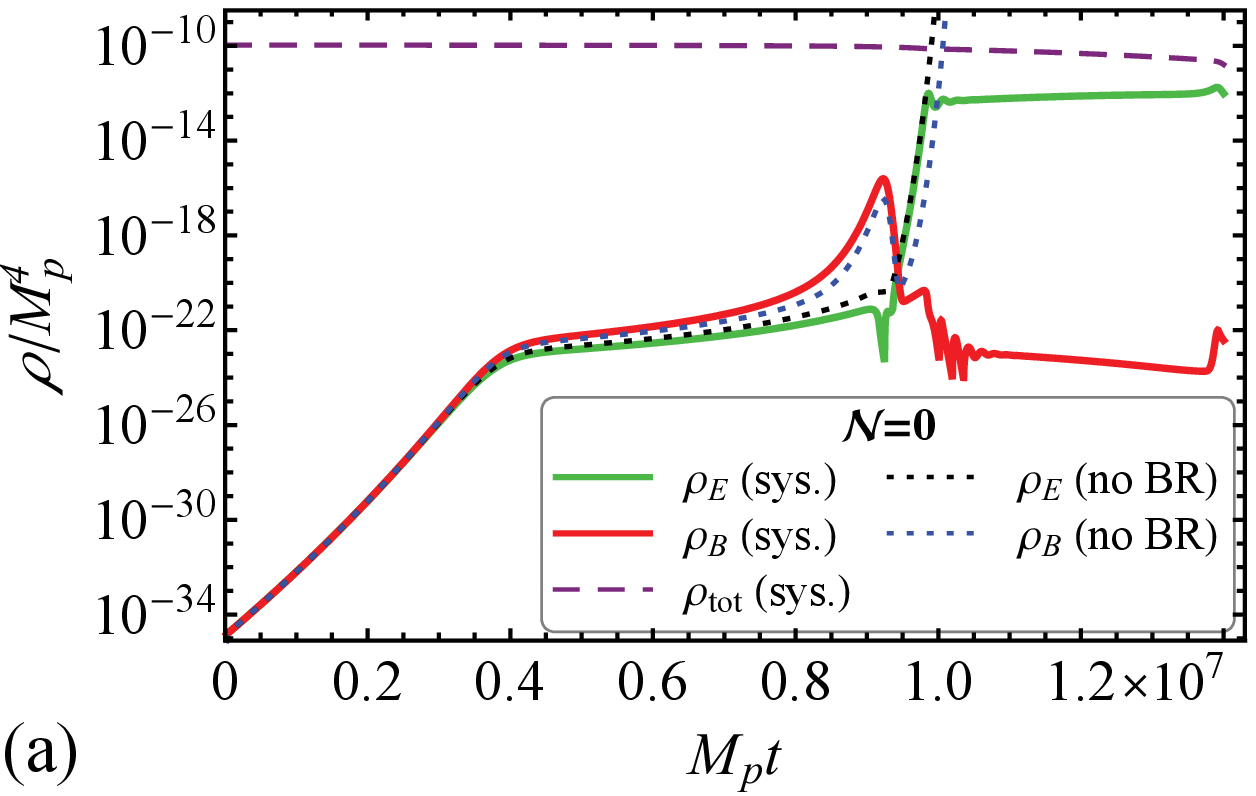}\hspace*{0.25cm}
	\includegraphics[width=0.32\textwidth]{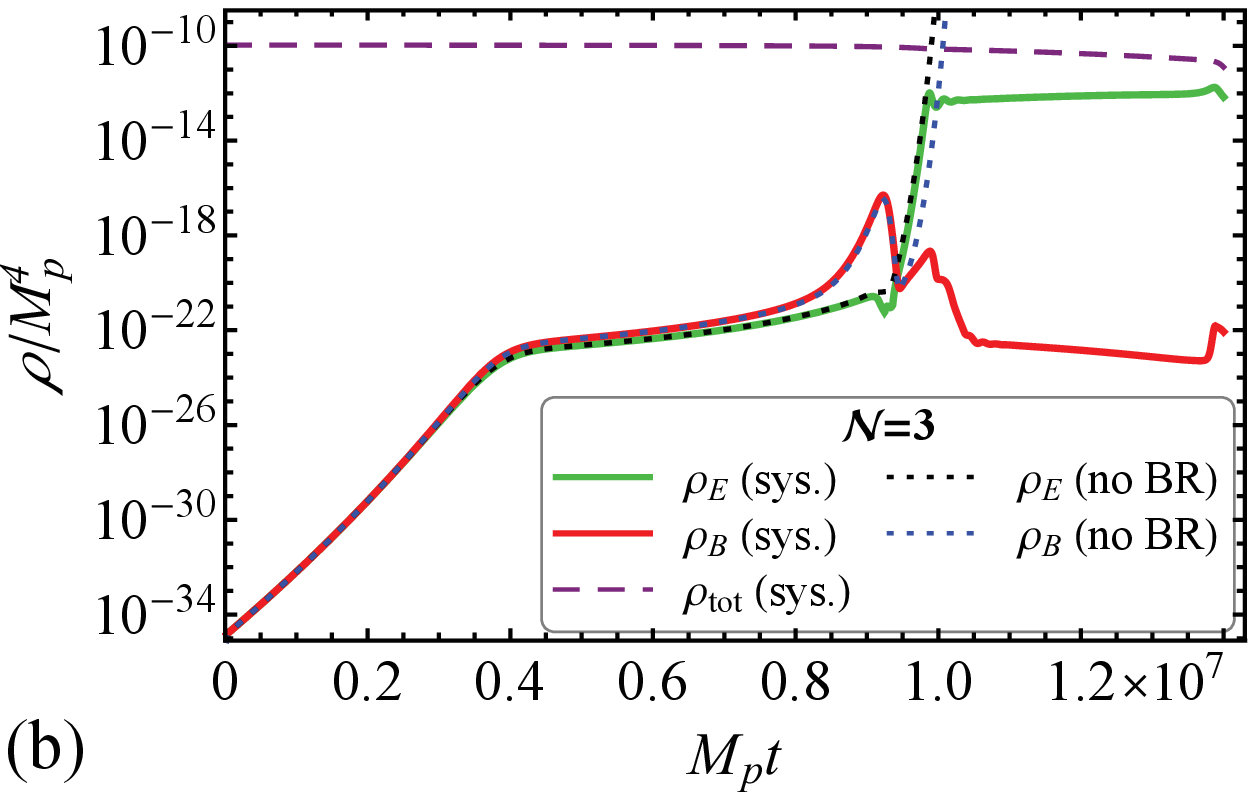}\hspace*{0.25cm}
	\includegraphics[width=0.32\textwidth]{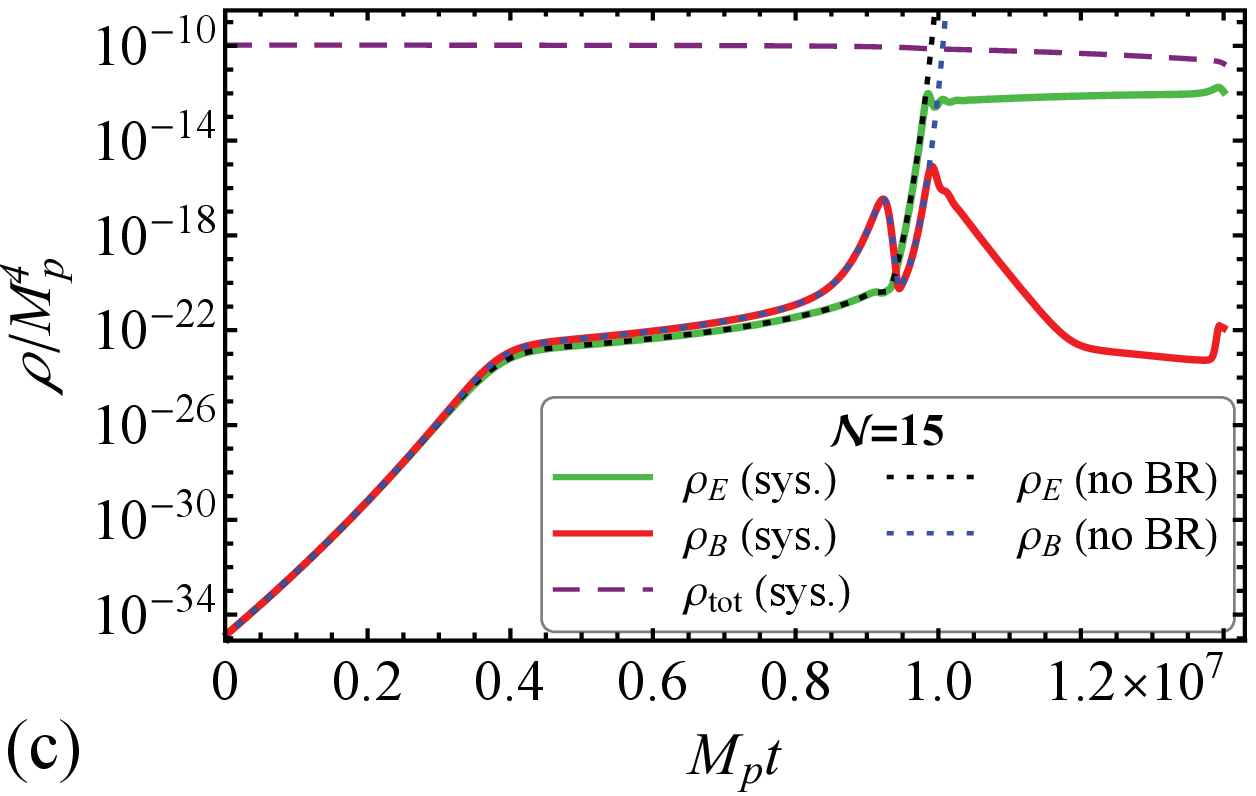}
	\caption{Solutions of the system of equations (\ref{eq-EE-2})--(\ref{eq-EB-2}) truncated at orders (a)~$\mathcal{N}=0$, (b)~$\mathcal{N}=3$, and (c)~$\mathcal{N}=15$ (solid lines), compared to the exact results obtained by the numerical integration of Eq.~(\ref{eq-mode-physical}) neglecting the backreaction (dotted lines). The electric and magnetic energy densities generated during inflation in the kinetic coupling model with the nonmonotonic function (\ref{nonmonotonic-coupling}) with parameters $\beta=12$, $C_{0}=10^{21}$, and $\phi_{1}=2.8\,M_{p}$ are shown. The purple dashed lines describe the time dependences of the total energy density.}
	\label{fig-NM2-N}
\end{figure}

Numerical results for the energy densities obtained from the exact solution of Eq.~(\ref{eq-mode-physical}) for all relevant modes naturally do not take into account the backreaction. Therefore, we can use it for comparison with our approximate results only until the backreaction becomes important. Such a comparison is shown in Fig.~\ref{fig-NM2-N}, where the solid lines correspond to the results for the truncated system of equations, while the dotted lines show the mode-by-mode solutions. As can be seen from this figure, the truncated system reproduces very well the exact solution until the backreaction turns on.

\begin{figure}[ht!]
	\centering
	\includegraphics[width=0.4\textwidth]{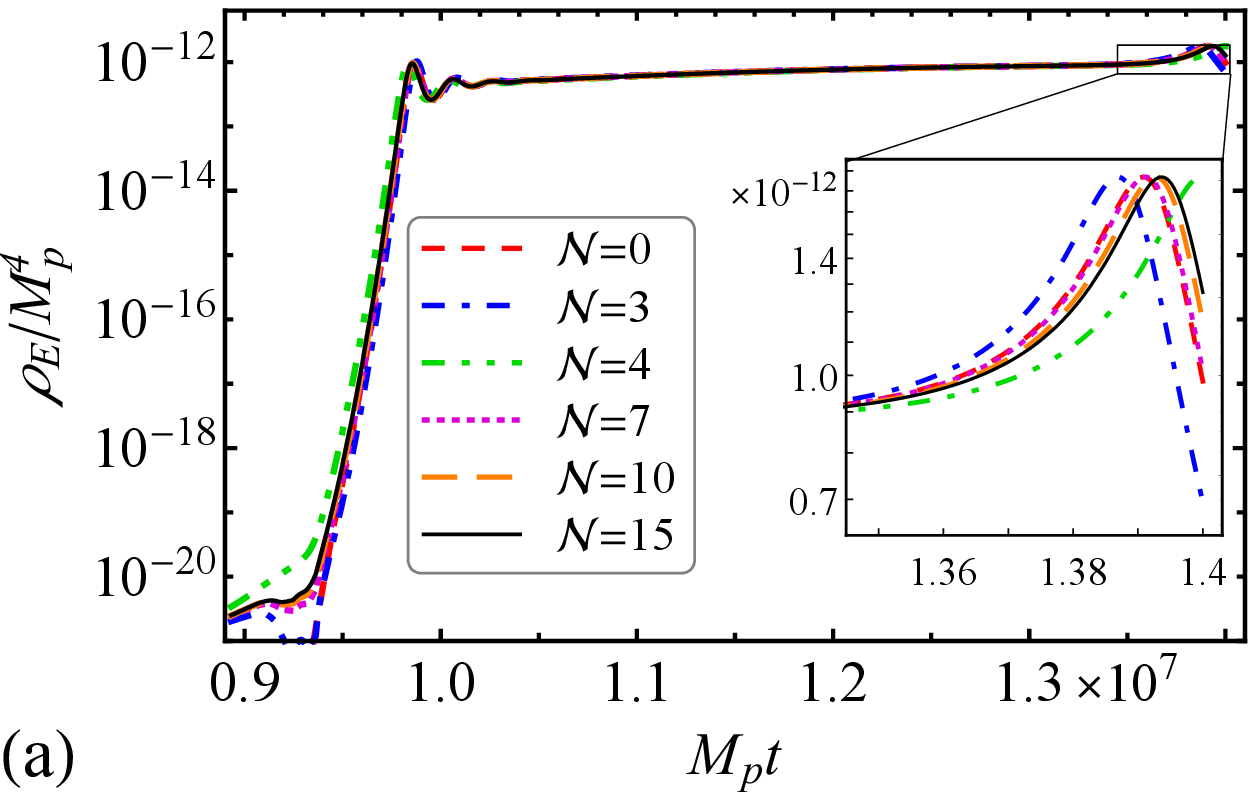}\hspace*{0.5cm}
	\includegraphics[width=0.4\textwidth]{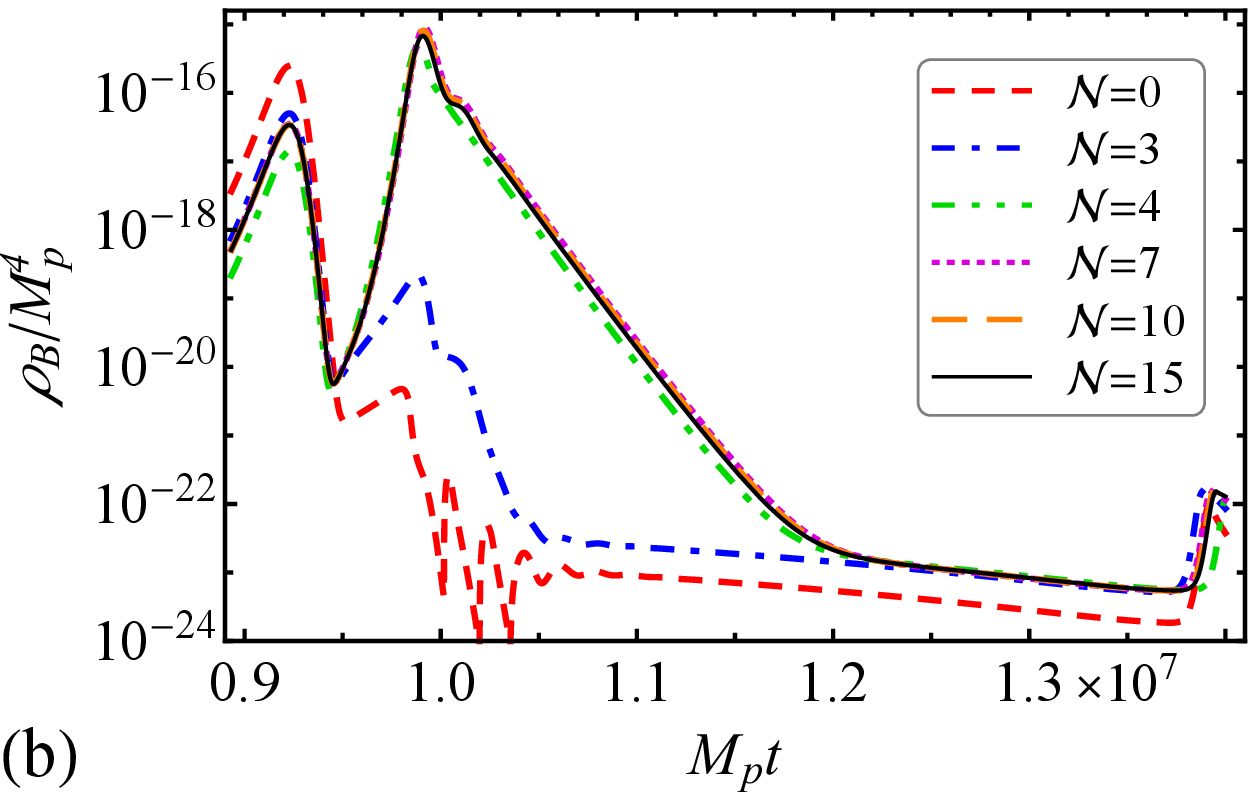}
	\caption{Comparison of the results for the (a)~electric and (b)~magnetic energy densities calculated by using the system of equations (\ref{eq-EE-2})--(\ref{eq-EB-2}) truncated at different orders (see the legend) for the nonmonotonic coupling function (\ref{nonmonotonic-coupling}) with parameters $\beta=12$, $C_{0}=10^{21}$, and $\phi_{1}=2.8\,M_{p}$. Only the final part of the inflation stage is shown, where the difference is the most significant. Since the exact numerical solution of Eq.~(\ref{eq-mode-physical}) can be found only in the absence of backreaction, it is not used here as a reference point. We compare the results for the few first values of $\mathcal{N}$ with the result for $\mathcal{N}=15$ (black solid lines).}
	\label{fig-NM2-comparison}
\end{figure}

Because of the backreaction, we cannot use the exact mode-by-mode solution as a reference for quantifying the error of our approximate results. That is why we use the result of the truncated system with $\mathcal{N}=15$ for this purpose. The latter is shown by black solid lines in Fig.~\ref{fig-NM2-comparison}, while the solutions of the system with smaller $\mathcal{N}$ are shown by the colored dashed lines (see the legend). Again, the dominant component of the generated energy density (the electric one) is reproduced with high accuracy even by the low-$\mathcal{N}$ approximations, while for the magnetic component the convergence starts from $\mathcal{N}\geq 4$.

\FloatBarrier

\section{Conclusion}
\label{sec-concl}

In this work, we extended the gradient expansion formalism (proposed in Ref.~\cite{Sobol:2019} for the axion coupling model) to the kinetic coupling model of the generation of electromagnetic fields during inflation. In contrast to the standard approach operating with separate Fourier modes of the electromagnetic field, we worked with physical quantities in the coordinate space which include all physically relevant modes at once. Although, in principle, there are infinitely many of such quantities, at a practical level it appears to be sufficient to use a finite number of them to satisfactorily describe the magnetogenesis process during inflation. We found that such a formalism works very efficiently and has an excellent convergence.

An important feature of inflationary magnetogenesis is the fact that the number of relevant modes grows in time as more and more new modes cross the horizon and undergo the quantum-to-classical transition. We take this into account by introducing the boundary terms in the equations of motion for the electromagnetic quantities. Their origin can be described as follows: Any of the electromagnetic quantities [Eqs.~(\ref{EE})--(\ref{BB})] is a result of integration of the corresponding spectral densities over the range of physically relevant modes. Since new modes cross the horizon, the upper integration limit continuously grows in time during inflation, and the electromagnetic quantities acquire additional time dependence. The latter is described by an additional term in the corresponding equation of motion.  Thus, we realize the idea of the adaptive UV cutoff of the electromagnetic power spectrum.

The central result of our paper is the system of coupled equations (\ref{eq-EE-2})--(\ref{eq-EB-2}) for the set of bilinear electromagnetic quantities $\mathscr{E}^{(n)}$, $\mathscr{G}^{(n)}$, $\mathscr{B}^{(n)}$, which contain $n$ spatial curls acting on the electric or magnetic field. The equation of motion for the $n$th-order quantity always contains at least one quantity with $n+1$ curl. As a result, all equations are coupled into an infinite chain. There is, however, a physical argument allowing us to truncate this chain at some finite order. Indeed, for the quantities with a large number of curls, the dominating contribution is made by the UV edge of the spectrum. The shortest relevant mode is determined by the horizon scale [Eq.~(\ref{horizon-crossing-mode})] and can be easily determined. This provides an opportunity to express the higher-order quantity in terms of the lower-order one, and thus, to truncate the chain.

As a test bed for our formalism we used the Starobinsky model of inflation, which is one of the most favored by the cosmic microwave background observations~\cite{Planck:2018-infl}. We considered two types of coupling functions: the Ratra function [Eq.~(\ref{Ratra-function})] and the nonmonotonic inverse cosine function [Eq.~(\ref{nonmonotonic-coupling})]. Choosing different sets of coupling parameters, we studied qualitatively different cases of the increasing, decreasing, and nonmonotonic coupling functions, both in the absence of the backreaction and taking it into account. We compared the numerical results for the electric and magnetic energy densities obtained from the approximate system of equations truncated at different orders $\mathcal{N}=0$--15. For all types of coupling function, we found a convergence of the numerical results to some limiting solution which is independent of $\mathcal{N}$. The residual error of this solution [compared to the exact result obtained from the numerical integration of the mode equation (\ref{eq-mode-physical}) for all physically relevant modes] has the maximal value of the order of a few percent at the end of inflation. This error is caused not by the truncation of the system but by other approximations which were adopted in its derivation (an approximate solution of the mode equation near the moment of horizon crossing, the slow-roll approximation, etc.). We observed that the approximate solution has a less than 10\% error already for the truncation order $\mathcal{N}\simeq 10$. At this order, one has to solve $\sim 20$ coupled ordinary differential equations, which is a simple task even for an ordinary personal computer.

The nonmonotonic coupling function [Eq.~(\ref{nonmonotonic-coupling})] leads to much more complicated time behavior of the energy densities than in the case of a monotonic Ratra function. At first, when the function grows in time, the magnetic component is slightly greater than the electric one. Further, when the coupling function starts to decrease, the roles of these components interchange. However, this does not have any important effect on the final value of the magnetic field. The reason for this is pretty clear: the most intensive generation occurs close to the end of inflation when the inflaton rolls faster. At that time, the coupling function is already a decreasing function, and therefore, the generation occurs qualitatively in the same way as for a function monotonically decreasing from the very beginning. Nonmonotonic coupling functions were studied previously in Refs.~\cite{Ferreira:2013,Ferreira:2014}. The authors considered the piecewise smooth functions composed of two or three pieces with a power-like dependence on the conformal time $f\propto \eta^{\alpha}$ with different $\alpha$'s. For a nonmonotonic function with one maximum, their conclusions are similar to ours: they do not observe a significant enhancement compared to the case of a monotonic function.

Our method fails to work after the end of inflation. First of all, this happens because the horizon scale stops increasing and no new modes exit the horizon. As a result, the enhancement occurs only for modes which crossed the horizon during inflation, and the spectrum of such modes strongly depends on time. In such a case, there is no simple relation between the quantities of different order $n$ anymore, and we cannot truncate the chain of equations. There are, however, even more fundamental problems in the description of preheating. First, at this stage, the inflaton cannot be treated as spatially homogeneous~\cite{Cuissa:2018}. Second, one has to take into account the inflaton damping due to the perturbative and nonperturbative (e.g., parametric resonance) particle production during preheating~\cite{Kofman:1997,Amin:2015}. Last but not the least, the Schwinger effect may become extremely important after the end of inflation. All these phenomena can be taken into account only in a complicated lattice simulation. Fortunately, during preheating the scale factor grows in time slowly (powerlike time dependence in contrast to the exponential one during inflation), and the lattice spacing remains of the same order of magnitude during the simulation. Indeed, this method has been already applied to the description of preheating; see, e.g., Ref.~\cite{Cuissa:2018}. The lattice simulations require the initial conditions for all physical quantities imposed at the end of inflation (or a few $e$-foldings before its end). Our approach can be used to generate such initial conditions.

Another interesting question is about the role of the Schwinger effect during inflation. In the present work, we completely neglected this phenomenon. However, it is an important problem to take it into consideration. For instance, it may be particularly important in the backreaction regime when the Schwinger process decreases the electric energy density and helps to end the inflation stage~\cite{Sobol:2018}. This issue deserves a separate investigation and will be addressed elsewhere.

\begin{acknowledgments}
	
	The work of O.~O.~S. was supported by Swiss National Science Foundation Grant No.~200020B\_182864.
	The work of E.~V.~G. was supported by National Research Foundation of Ukraine Project No.~2020.02/0062.	
	The work of S.~I.~V. was supported by Swiss National Science Foundation Grant No.~SCOPE IZSEZ0-186551 and by German Academic Exchange Service (DAAD) Grant No.~57387479.
		
\end{acknowledgments}

\appendix

\section{Upper and lower monotonic envelopes of a function}
\label{app-envelope}

In this appendix, we give the definition, properties, and examples of the upper and lower monotonic envelopes of an arbitrary function.

\textbf{Definition:} The upper (lower) monotonic envelope of a function $F(t)$ in the interval $[a,\ b]$ is a \textit{nondecreasing} (\textit{nonincreasing}) function which is the closest to $F(t)$ at any point of the interval. We will denote it as ${\rm Env}^{\uparrow (\downarrow)}_{[a,b]}\{F(t)\}$.
For any given $t$, it equals the maximal (minimal) value of the function $F(t')$ in the interval $t'\in[a,\,t]$.

In simple words, if we start from the left boundary of the interval, $a$, the upper envelope follows the function $F(t)$ until it reaches its local maximum and starts to decrease. From that point, the upper envelope remains constant until the moment when the function $F(t)$ again reaches the same value (as in its first local maximum). Starting from then, the envelope again follows $F(t)$ until it reaches the local maximum, etc. It can be defined also in a recursive way:
\begin{equation}
\label{upper-envelope-recursive}
{\rm Env}^{\uparrow}_{[a,b]}\{F(t)\}=\left\{
\begin{array}{ll}
F(a), & \text{if  } t=a,\\
F(t), & \text{if  } F(t)\geq {\rm Env}^{\uparrow}_{[a,b]}\{F(t-\varepsilon)\},\\
{\rm Env}^{\uparrow}_{[a,b]}\{F(t-\varepsilon)\}, & \text{if  } F(t)< {\rm Env}^{\uparrow}_{[a,b]}\{F(t-\varepsilon)\},
\end{array}
\right.
\end{equation}
where $\varepsilon \to +0$. For the lower monotonic envelope, one should replace ``maximum'' with ``minimum'' and change the signs of inequalities in Eq.~(\ref{upper-envelope-recursive}).

\begin{figure}[ht!]
	\centering
	\includegraphics[width=0.4\textwidth]{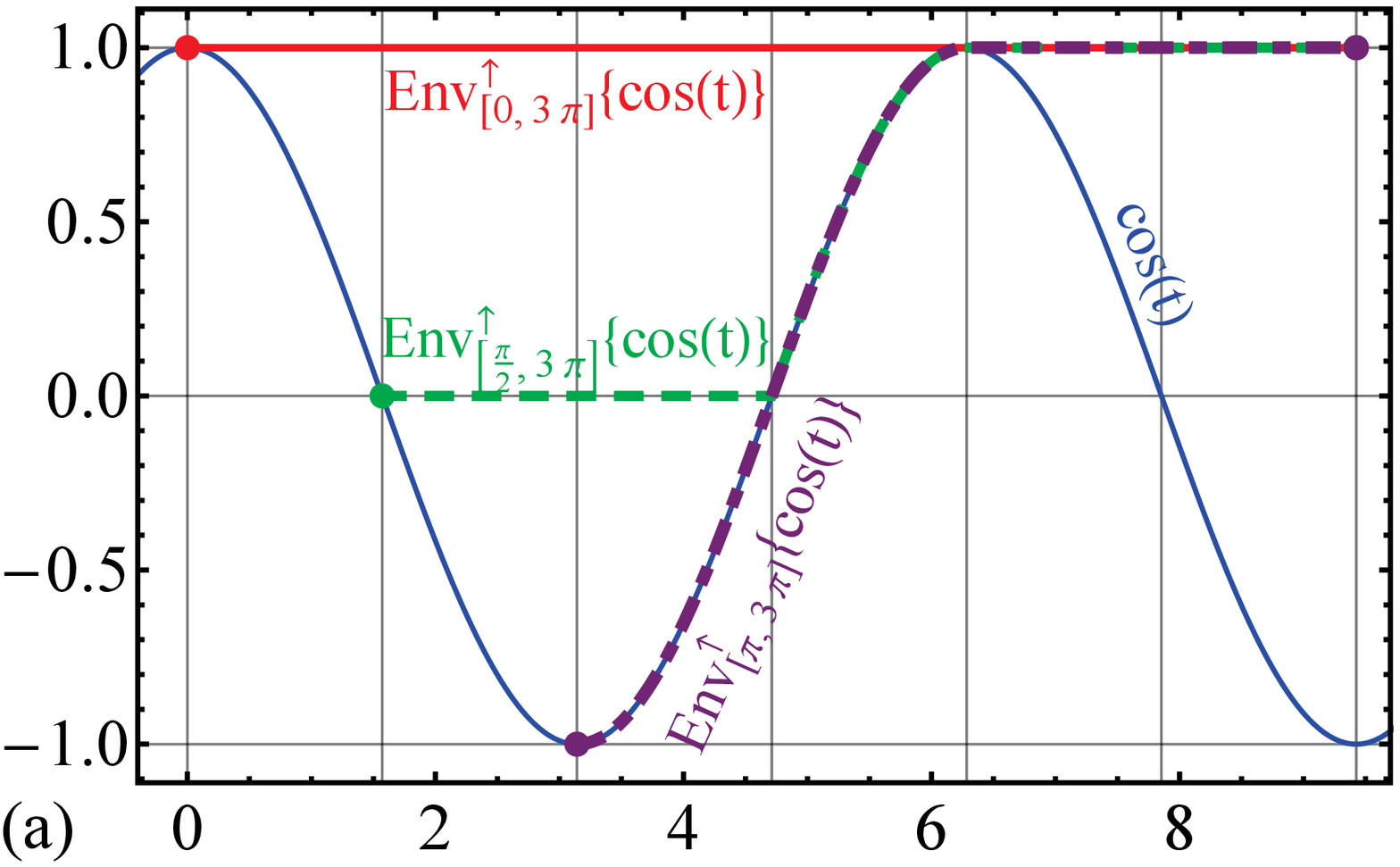}\hspace*{0.5cm}
	\includegraphics[width=0.4\textwidth]{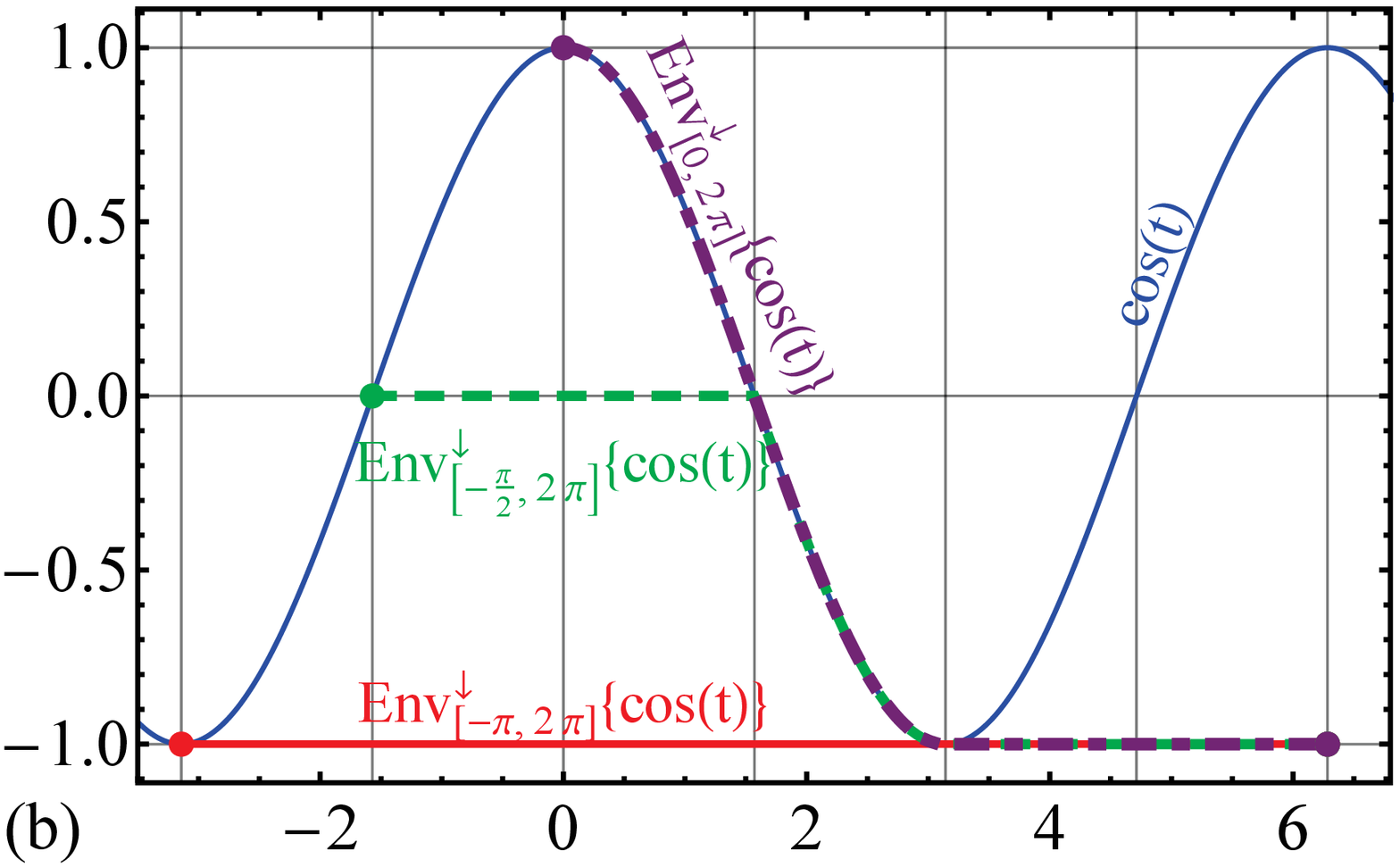}
	\caption{The cosine function (blue solid line) and its (a)~upper and (b)~lower monotonic envelopes in three different intervals (red solid lines, green dashed lines, and purple dash-dotted lines).}
	\label{fig-app-env}
\end{figure}

\textbf{Properties:}
\vspace*{-0.3cm}
\begin{enumerate}
	\item The upper monotonic envelope is a solution of the following Cauchy problem:
	\begin{equation}
	\left\{
	\begin{array}{l}
	\dfrac{d}{dt}{\rm Env}^{\uparrow}_{[a,b]}\{F(t)\}=\dot{F}(t) \theta\big(\dot{F}(t)\big)\theta\big(F(t)-{\rm Env}^{\uparrow}_{[a,b]}\{F(t)\}\big),\\
	{\rm Env}^{\uparrow}_{[a,b]}\{F(a)\}=F(a).
	\end{array}
	\right.
	\end{equation}
	Here $\theta(x)=1$ for $x\geq 0$ and $\theta(x)=0$ for $x< 0$ is the Heaviside $\theta$ function or the unit step function. For the lower monotonic envelope, one should change the signs in the arguments of both $\theta$ functions. This definition is more convenient for practical usage.
	
	\item For nondecreasing (nonincreasing) functions, the upper (lower) monotonic envelope coincides with the function itself.
	
	\item If $F(a)$ is the maximal (minimal) value in the interval $[a,b]$, the upper (lower) envelope ${\rm Env}^{\uparrow (\downarrow)}_{[a,b]}\{F(t)\}=F(a)={\rm const}$. This is shown by the red solid lines in Fig.~\ref{fig-app-env}.
	
	\item The upper (lower) monotonic envelopes are not uniquely defined for a nonmonotonic function if the interval is not specified. That is why it is important to indicate at least the left boundary of the desired interval (the starting point). This is illustrated by Fig.~\ref{fig-app-env}, where three different envelopes of the cosine are shown in red solid, green dashed, and purple dash-dotted lines.
	
	\item ${\rm Env}^{\uparrow}_{[a,b]}\{-F(t)\}=-{\rm Env}^{\downarrow}_{[a,b]}\{F(t)\}$ and vice versa.
\end{enumerate}

In the main text, we use only the upper monotonic envelope of the function during inflation---i.e., for $0\leq t\leq t_{e}$. In order to simplify the notation, we write ${\rm Env}^{\uparrow}_{[0,t_{e}]}\{F(t)\}\equiv {\rm Env}\{F(t)\}$.

\section{Properties of cylindric functions}
\label{app-cylindric}

By changing the variable $x=-k\eta$ and introducing the function $\mathcal{A}(\eta,\mathbf{k})=\sqrt{x}\psi(x)$, we rewrite Eq.~(\ref{eq-mode-conformal-horizon-crossing}) in the form
\begin{equation}
\psi''(x)+\frac{1}{x}\psi'(x)+\left(1-\frac{1/4+f_{1}+f_{2}}{x^{2}}\right)\psi(x)=0,
\end{equation}
which is Bessel's differential equation with the index $\alpha=\sqrt{1/4+f_{1}+f_{2}}$~\cite{Luke:book}. Its general solution is a linear combination of the Bessel $J_{\alpha}$ and Neumann $Y_{\alpha}$ functions which leads to Eq.~(\ref{solution-general-cylindric}). It is convenient to introduce also the Hankel functions of the first and second kind:
\begin{equation}
H^{(1,2)}_{\alpha}(x)=J_{\alpha}(x)\pm i Y_{\alpha}(x).
\end{equation}
Their asymptotical behavior for large values of the argument is the following:
\begin{equation}
\label{Hankel-asymptotics}
H^{(1,2)}_{\alpha}(x)\sim\sqrt{\frac{2}{\pi x}}\exp\big[\pm i\big(x-\frac{\pi\alpha}{2}-\frac{\pi}{4}\big)\big], \qquad x\to +\infty.
\end{equation}
From this expression, we conclude that $H^{(1)}(x)$ matches the correct asymptotics of the Bunch-Davies solution (\ref{init-BD}) for $-k\eta\to +\infty$. Comparing the prefactors in Eqs.~(\ref{init-BD}), (\ref{solution-general-cylindric}), and (\ref{Hankel-asymptotics}), we finally get solution (\ref{solution-horizon-crossing}).

Below, we list two properties of cylindric functions which were used in algebraic calculations in the main text. They are valid for any kind of cylindric functions $Z_{\alpha}=\{J_{\alpha},\,Y_{\alpha},\,H^{(1,2)}_{\alpha}\}$~\cite{Luke:book}:
\begin{eqnarray}
2\frac{dZ_{\alpha}(x)}{dx}&=&Z_{\alpha-1}(x)-Z_{\alpha+1}(x),\label{property-1}\\
\frac{2\alpha}{x}Z_{\alpha}(x)&=&Z_{\alpha-1}(x)+Z_{\alpha+1}(x).\label{property-2}
\end{eqnarray}


\vspace*{-0.5cm}

\begin{thebibliography}{99}
	
	
	\bibitem{Tavecchio:2010} F.~Tavecchio, G.~Ghisellini, L.~Foschini, G.~Bonnoli, G.~Ghirlanda, and P.~Coppi,  {The intergalactic magnetic field constrained by Fermi/Large Area Telescope observations of the TeV blazar 1ES 0229+200}, \href{https://doi.org/10.1111/j.1745-3933.2010.00884.x}{Mon. Not. R. Astron. Soc. \textbf{406}, L70 (2010)} [\href{https://arxiv.org/abs/1004.1329}{arXiv: 1004.1329 [astro-ph.CO]}]. 
	
	\bibitem{Ando:2010}	S.~Ando and A.~Kusenko, Evidence for gamma-ray halos around active galactic nuclei and the first measurement of intergalactic magnetic fields,
	\href{https://doi.org/10.1088/2041-8205/722/1/L39}{Astrophys. J. Lett. \textbf{722}, L39 (2010)}	[\href{https://arxiv.org/abs/1005.1924}{arXiv: 1005.1924 [astro-ph.HE]}]. 
	
	\bibitem{Neronov:2010} A.~Neronov and I.~Vovk,  {Evidence for strong extragalactic magnetic fields from Fermi observations of TeV blazars}, \href{https://doi.org/10.1126/science.1184192}{Science \textbf{328}, 73 (2010) [\href{https://arxiv.org/abs/1006.3504}{arXiv: 1006.3504 [astro-ph.HE]}].} 
	
	\bibitem{Dolag:2010} K.~Dolag, M.~Kachelriess, S.~Ostapchenko, and R.~Tomas,
	Lower limit on the strength and filling factor of extragalactic magnetic fields,
	\href{https://doi.org/10.1088/2041-8205/727/1/L4}{Astrophys. J. Lett. \textbf{727}, L4 (2011)} [\href{https://arxiv.org/abs/1009.1782}{arXiv: 1009.1782 [astro-ph.HE]}]. 
	
	\bibitem{Dermer:2011} C.D.~Dermer, M.~Cavadini, S.~Razzaque, J.D.~Finke, J.~Chiang, and B.~Lott,  {Time delay of cascade radiation for TeV blazars and the measurement of the intergalactic magnetic field}, \href{https://doi.org/10.1088/2041-8205/733/2/L21}{Astrophys. J. Lett. \textbf{733}, L21 (2011)} [\href{https://arxiv.org/abs/1011.6660}{arXiv: 1011.6660 [astro-ph.HE]}]. 
	
	\bibitem{Taylor:2011} A.M.~Taylor, I.~Vovk, and A.~Neronov,  {Extragalactic magnetic fields constraints from simultaneous GeV-TeV observations of blazars}, \href{https://doi.org/10.1051/0004-6361/201116441}{Astron. Astrophys. \textbf{529}, A144 (2011)} [\href{https://arxiv.org/abs/1101.0932}{arXiv: 1101.0932 [astro-ph.HE]}]. 
	
	\bibitem{Caprini:2015} C.~Caprini and S.~Gabici,  {Gamma-ray observations of blazars and the intergalactic magnetic field spectrum}, \href{https://doi.org/10.1103/PhysRevD.91.123514}{Phys. Rev. D \textbf{91}, 123514 (2015)} [\href{https://arxiv.org/abs/1504.00383}{arXiv: 1504.00383 [astro-ph.CO]}]. 
	
	
	
	
	\bibitem{Kronberg:1994} P.P.~Kronberg, {Extragalactic magnetic fields}, \href{https://doi.org/10.1088/0034-4885/57/4/001}{Rep. Prog. Phys. \textbf{57}, 325 (1994).} 
	
	\bibitem{Grasso:2001} D.~Grasso and H.R.~Rubinstein, {Magnetic fields in the early Universe}, \href{https://doi.org/10.1016/S0370-1573(00)00110-1}{Phys. Rep. \textbf{348}, 163 (2001)} [\href{https://arxiv.org/abs/astro-ph/0009061}{arXiv: astro-ph/0009061}]. 
	
	\bibitem{Widrow:2002} L.M.~Widrow, {Origin of galactic and extragalactic magnetic fields}, \href{https://doi.org/10.1103/RevModPhys.74.775}{Rev. Mod. Phys. \textbf{74}, 775 (2002)} [\href{https://arxiv.org/abs/astro-ph/0207240}{arXiv: astro-ph/0207240}]. 
	
	\bibitem{Giovannini:2004} M.~Giovannini, {The magnetized Universe}, \href{https://doi.org/10.1142/S0218271804004530}{Int. J. Mod. Phys. D \textbf{13}, 391 (2004)} [\href{https://arxiv.org/abs/astro-ph/0312614}{arXiv: astro-ph/0312614}]. 
	
	\bibitem{Kandus:2011} A.~Kandus, K.E.~Kunze, and C.~G.~Tsagas, {Primordial magnetogenesis}, \href{https://doi.org/10.1016/j.physrep.2011.03.001}{Phys. Rep. \textbf{505}, 1 (2011)} [\href{https://arxiv.org/abs/1007.3891}{arXiv: 1007.3891 [astro-ph.CO]}]. 
	
	\bibitem{Durrer:2013} R.~Durrer and A.~Neronov, {Cosmological magnetic fields: Their generation, evolution and observation}, \href{https://doi.org/10.1007/s00159-013-0062-7}{Astron. Astrophys. Rev. \textbf{21}, 62 (2013)} [\href{https://arxiv.org/abs/1303.7121}{arXiv: 1303.7121 [astro-ph.CO]}]. 
	
	\bibitem{Subramanian:2016} K.~Subramanian, {The origin, evolution and signatures of primordial magnetic fields}, \href{https://doi.org/10.1088/0034-4885/79/7/076901}{Rep. Prog. Phys. \textbf{79}, 076901 (2016)} [\href{https://arxiv.org/abs/1504.02311}{arXiv: 1504.02311 [astro-ph.CO]}]. 
	
	
	\bibitem{Harrison:1970} E.R.~Harrison, Fluctuations at the threshold of classical cosmology,
	\href{https://doi.org/10.1103/PhysRevD.1.2726}{Phys. Rev. D \textbf{1}, 2726--2730 (1970)}.
	
	\bibitem{Zeldovich:1972} Ya.B.~Zeldovich, A Hypothesis, Unifying the Structure and the Entropy of the Universe, \href{https://doi.org/10.1093/mnras/160.1.1P}{Mon. Not. R. Astron. Soc. \textbf{160}, 1P--3P (1972)}. 
	
	\bibitem{Chibisov:1982} G.V.~Chibisov and V.F.~Mukhanov, Galaxy formation and phonons,
	\href{https://doi.org/10.1093/mnras/200.3.535}{Mon. Not. Roy. Astron. Soc. \textbf{200}, 535--550 (1982)}. 
	
	\bibitem{Mukhanov:1992} V.F.~Mukhanov, H.A.~Feldman, and R.H.~Brandenberger,
	Theory of cosmological perturbations. Part 1. Classical perturbations. Part 2. Quantum theory of perturbations. Part 3. Extensions,
	\href{https://doi.org/10.1016/0370-1573(92)90044-Z}{Phys. Rept. \textbf{215}, 203--333 (1992)}.
	
	\bibitem{Durrer:book} R.~Durrer, \textit{The Cosmic Microwave Background} (Cambridge University Press, New York, 2008).
	
	\bibitem{Parker:1968} L.~Parker, \textit{Particle Creation in Expanding Universes}, \href{https://doi.org/10.1103/PhysRevLett.21.562}{Phys. Rev. Lett. \textbf{21}, 562 (1968).} 
	
	
	
	\bibitem{Turner:1988} M.S.~Turner and L.M.~Widrow,  {Inflation-produced, large-scale magnetic fields}, \href{https://doi.org/10.1103/PhysRevD.37.2743}{Phys. Rev. D \textbf{37}, 2743 (1988).} 
	
	\bibitem{Ratra:1992} B.~Ratra,  {Cosmological `seed' magnetic field from inflation}, \href{https://doi.org/10.1086/186384}{Astrophys. J. \textbf{391}, L1 (1992)}. 
	
	\bibitem{Garretson:1992} W.~D.~Garretson, G.~B.~Field, and S.~M.~Carroll,  {Primordial magnetic fields from pseudo-Goldstone bosons}, \href{https://doi.org/10.1103/PhysRevD.46.5346}{Phys. Rev. D {\bf 46}, 5346 (1992)} [\href{https://arxiv.org/abs/hep-ph/9209238}{arXiv: hep-ph/9209238}]. 
	
	\bibitem{Dolgov:1993} A.D.~Dolgov,  {Breaking of conformal invariance and electromagnetic field generation in the Universe}, \href{https://doi.org/10.1103/PhysRevD.48.2499}{Phys. Rev. D \textbf{48}, 2499 (1993)} [\href{https://arxiv.org/abs/hep-ph/9301280}{arXiv: hep-ph/9301280}]. 
	


	\bibitem{Giovannini:2001} M.~Giovannini,  {Variation of the gauge couplings during inflation}, \href{https://doi.org/10.1103/PhysRevD.64.061301}{Phys. Rev. D \textbf{64}, 061301(R) (2001)}. 
	
	\bibitem{Bamba:2004} K.~Bamba and J.~Yokoyama,  {Large-scale magnetic fields from inflation in dilaton electromagnetism}, \href{https://doi.org/10.1103/PhysRevD.69.043507}{Phys. Rev. D \textbf{69}, 043507 (2004)} [\href{https://arxiv.org/abs/astro-ph/0310824}{arXiv: astro-ph/0310824}]. 
	
	\bibitem{Martin:2008} J.~Martin and J.~Yokoyama,  {Generation of large scale magnetic fields in single-field inflation}, \href{https://doi.org/10.1088/1475-7516/2008/01/025}{J. Cosmol. Astropart. Phys. 01 (2008) 025} [\href{https://arxiv.org/abs/0711.4307}{arXiv: 0711.4307 [astro-ph]}]. 
	
	\bibitem{Kanno:2009} S.~Kanno, J.~Soda, and M.~Watanabe,  {Cosmological magnetic fields from inflation and backreaction}, \href{https://doi.org/10.1088/1475-7516/2009/12/009}{J. Cosmol. Astropart. Phys. 12 (2009) 009} [\href{https://arxiv.org/abs/0908.3509}{arXiv: 0908.3509 [astro-ph.CO]}]. 
	
	\bibitem{Demozzi:2009} V.~Demozzi, V.M.~Mukhanov, and H.~Rubinstein,  {Magnetic fields from inflation?}, \href{https://doi.org/10.1088/1475-7516/2009/08/025}{J. Cosmol. Astropart. Phys. 08 (2009) 025} [\href{https://arxiv.org/abs/0907.1030}{arXiv: 0907.1030 [astro-ph.CO]}]. 
	
	\bibitem{Fujita:2012} T.~Fujita and S.~Mukohyama, Universal upper limit on inflation energy scale from cosmic magnetic field,
	\href{https://doi.org/10.1088/1475-7516/2012/10/034}{J. Cosmol. Astropart. Phys. 10 (2012) 034} [\href{https://arxiv.org/abs/1205.5031}{arXiv: 1205.5031 [astro-ph.CO]}]. 
	
	\bibitem{Fujita:2013} T.~Fujita and S.~Yokoyama, Higher order statistics of curvature perturbations in IFF model and its Planck constraints,
	\href{https://doi.org/10.1088/1475-7516/2013/09/009}{J. Cosmol. Astropart. Phys. 09 (2013) 009} [\href{https://arxiv.org/abs/1306.2992}{arXiv: 1306.2992 [astro-ph.CO]}]. 
	
	\bibitem{Ferreira:2013} R.J.Z.~Ferreira, R.K.~Jain, and M.S.~Sloth,  {Inflationary magnetogenesis without the strong coupling problem}, \href{https://doi.org/10.1088/1475-7516/2013/10/004}{J. Cosmol. Astropart. Phys. 10 (2013) 004} [\href{https://arxiv.org/abs/1305.7151}{arXiv: 1305.7151 [astro-ph.CO]}]. 
	
	\bibitem{Ferreira:2014} R.J.Z.~Ferreira, R.K.~Jain, and M.S.~Sloth,  {Inflationary magnetogenesis without the strong coupling problem. II. Constraints from CMB anisotropies and B-modes}, \href{https://doi.org/10.1088/1475-7516/2014/06/053}{J. Cosmol. Astropart. Phys. 06 (2014) 053} [\href{https://arxiv.org/abs/1403.5516}{arXiv: 1403.5516 [astro-ph.CO]}]. 
	
	\bibitem{Fujita:2016} T.~Fujita and R.~Namba, Pre-reheating magnetogenesis in the kinetic coupling model, \href{https://doi.org/10.1103/PhysRevD.94.043523}{Phys. Rev. D \textbf{94}, 043523 (2016)} [\href{https://arxiv.org/abs/1602.05673}{arXiv: 1602.05673 [astro-ph.CO]}]. 
	
	\bibitem{Vilchinskii:2017} S.~Vilchinskii, O.~Sobol, E.V.~Gorbar, and I.~Rudenok,  {Magnetogenesis during inflation and preheating in the Starobinsky model}, \href{https://doi.org/10.1103/PhysRevD.95.083509}{Phys. Rev. D \textbf{95}, 083509 (2017)} [\href{https://arxiv.org/abs/1702.02774}{arXiv: 1702.02774 [astro-ph.CO]}]. 
	
	\bibitem{Sharma:2017b} R.~Sharma, S.~Jagannathan, T.R.~Seshadri, and K.~Subramanian,  {Challenges in inflationary magnetogenesis: Constraints from strong coupling, backreaction, and the Schwinger effect}, \href{https://doi.org/10.1103/PhysRevD.96.083511}{Phys. Rev. D \textbf{96}, 083511 (2017)} [\href{https://arxiv.org/abs/1708.08119}{arXiv: 1708.08119 [astro-ph.CO]}]. 
	
	\bibitem{Shtanov:2018} O.~Savchenko and Yu.~Shtanov,  {Magnetogenesis by non-minimal coupling to gravity in the Starobinsky inflationary model}, \href{https://doi.org/10.1088/1475-7516/2018/10/040}{J. Cosmol. Astropart. Phys. 10 (2018) 040} [\href{https://arxiv.org/abs/1808.06193}{arXiv: 1808.06193 [astro-ph.CO]}]. 
	
	
	\bibitem{Sobol:2018} O.O.~Sobol, E.V.~Gorbar, M.~Kamarpour, and S.I.~Vilchinskii,  {Influence of backreaction of electric fields and Schwinger effect on inflationary magnetogenesis}, \href{https://doi.org/10.1103/PhysRevD.98.063534}{Phys. Rev. D \textbf{98}, 063534 (2018)} [\href{https://arxiv.org/abs/1807.09851}{arXiv: 1807.09851 [hep-ph]}]. 
	
	\bibitem{Shtanov:2020} Y.~Shtanov and M.~Pavliuk, Model-independent constraints in inflationary magnetogenesis,
	\href{https://doi.org/10.1088/1475-7516/2020/08/042}{J. Cosmol. Astropart. Phys. \textbf{08} (2020) 042} [\href{https://arxiv.org/abs/2004.00947}{arXiv: 2004.00947 [astro-ph.CO]}]. 
	
	\bibitem{Talebian:2020}	A.~Talebian, A.~Nassiri-Rad and H.~Firouzjahi, Revisiting Magnetogenesis during Inflation, [\href{https://arxiv.org/abs/2007.11066}{arXiv: 2007.11066 [gr-qc]}].	
	
	
	
	\bibitem{Kobayashi:2014} T.~Kobayashi and N.~Afshordi,  {Schwinger effect in 4D de Sitter space and constraints on magnetogenesis in the early universe}, \href{https://doi.org/10.1007/JHEP10(2014)166}{J. High Energy Phys. 10 (2014) 166} [\href{https://arxiv.org/abs/1408.4141}{arXiv: 1408.4141 [hep-th]}]. 
	
	
	\bibitem{Fujita:2015} T.~Fujita, R.~Namba, Y.~Tada, N.~Takeda, and H.~Tashiro, \textit{Consistent generation of magnetic fields in axion inflation models}, \href{https://doi.org/10.1088/1475-7516/2015/05/054}{J. Cosmol. Astropart. Phys. \textbf{05} (2015) 054} [\href{http://arxiv.org/abs/1503.05802}{arXiv: 1503.05802 [astro-ph.CO]}]. 
	
	\bibitem{Notari:2016} A.~Notari and K.~Tywoniuk, \textit{Dissipative axial inflation}, \href{https://doi.org/10.1088/1475-7516/2016/12/038}{J. Cosmol. Astropart. Phys. \textbf{12} (2016) 038} [\href{http://arxiv.org/abs/1608.06223}{arXiv: 1608.06223 [hep-th]}]. 
	
	
	\bibitem{Cuissa:2018} J.R.C.~Cuissa and D.G.~Figueroa, Lattice formulation of axion inflation. Application to preheating,
	\href{https://doi.org/10.1088/1475-7516/2019/06/002}{J. Cosmol. Astropart. Phys. \textbf{06}, 002 (2019)}. [\href{https://arxiv.org/abs/1812.03132}{arXiv: 1812.03132 [astro-ph.CO]}].
	
	

	\bibitem{Sobol:2019} O.O.~Sobol, E.V.~Gorbar, and S.I.~Vilchinskii,  {Backreaction of electromagnetic fields and Schwinger effect in pseudoscalar inflation magnetogenesis}, \href{https://doi.org/10.1103/PhysRevD.100.063523}{Phys. Rev. D \textbf{100}, 063523 (2019)} [\href{https://arxiv.org/abs/1907.10443}{arXiv: 1907.10443 [astro-ph.CO]}]. 
	
	
	
	\bibitem{Bunch:1978} 
	T.S.~Bunch and P.C.W.~Davies, Quantum field theory in de Sitter space: Renormalization by point splitting,
	\href{https://doi.org/10.1098/rspa.1978.0060}{Proc.\ R.\ Soc.\ Lond.\ A {\bf 360}, 117 (1978)}. 
	
	
	
	\bibitem{Starobinsky:1980} A.A.~Starobinsky,  {A new type of isotropic cosmological models without singularity}, \href{https://doi.org/10.1016/0370-2693(80)90670-X}{Phys. Lett. \textbf{91B}, 99 (1980)}. 
	
	\bibitem{Planck:2018-infl} Y.~Akrami  {et al.} (Planck Collaboration),  {Planck 2018 results. X. Constraints on inflation}, \href{https://doi.org/10.1051/0004-6361/201833887}{Astron. Astrophys. \textbf{641}, A10 (2020)} [\href{https://arxiv.org/abs/1807.06211}{arXiv: 1807.06211 [astro-ph.CO]}]. 
	
	
	
	
	
	
	
	\bibitem{Kofman:1997} L.~Kofman, A.~Linde, and A.A.~Starobinsky,  {Towards the theory of reheating after inflation}, \href{https://doi.org/10.1103/PhysRevD.56.3258}{Phys. Rev. D \textbf{56}, 3258 (1997)} [\href{https://arxiv.org/abs/hep-ph/9704452}{arXiv: hep-ph/9704452}]. 
	
	\bibitem{Amin:2015} M.A.~Amin, M.P.~Hertzberg, D.I.~Kaiser, and J.~Karouby,  {Nonperturbative dynamics of reheating after inflation: A review}, \href{https://doi.org/10.1142/S0218271815300037}{Int. J. Mod. Phys. D \textbf{24}, 1530003 (2015)} [\href{https://arxiv.org/abs/1410.3808}{arXiv: 1410.3808 [hep-ph]}]. 
	
	
	
	
	
	\bibitem{Luke:book} Y.~L.~Luke, \textit{The Special Functions and Their Approximations}
	(Academic Press, New York, 1969), Vol. 1.
\end{thebibliography}
\end{document}